\documentclass[superscriptaddress,twocolumn,amsmath,amssymb,footinbib]{revtex4-1}
\usepackage[english]{babel}
\usepackage[pdftex]{graphicx}
\usepackage{gensymb}

\begin{document}

\newcommand{\physicsDept}{Department of Physics, Laboratory of Atomic and Solid State Physics, Cornell University, Ithaca, NY 14853, USA}
\newcommand{\paradim}{Platform for the Accelerated Realization, Analysis, and Discovery of Interface Materials (PARADIM), Cornell University, Ithaca, New York 14853, USA}
\newcommand{\mseDept}{Department of Materials Science and Engineering, Cornell University, Ithaca, NY 14853, USA}
\newcommand{\aepDept}{School of Applied and Engineering Physics, Cornell University, Ithaca, New York 14853, USA}
\newcommand{\kavli}{Kavli Institute at Cornell for Nanoscale Science, Ithaca, NY 14853, USA}
\newcommand{\RO}{RuO$_2$ }
\newcommand{\ROnospace}{RuO$_2$}
\newcommand{\ROA}{RuO$_2$(110) }
\newcommand{\ROAnospace}{RuO$_2$(110)}
\newcommand{\ROB}{RuO$_2$(101) }
\newcommand{\ROBnospace}{RuO$_2$(101)}
\newcommand{\TO}{TiO$_2$ }
\newcommand{\TOnospace}{TiO$_2$}
\newcommand{\TC}{$T_c$ }
\newcommand{\TCnospace}{$T_c$}
\newcommand{\HC}{$H_c$ }
\newcommand{\HCnospace}{$H_c$}
\newcommand{\ang}{\text{\normalfont\AA}}
\newcommand{\dpar}{$d_{||}$ }
\newcommand{\dparnospace}{$d_{||}$}
\newcommand{\dxzyz}{$(d_{\text{xz}}$, $d_{\text{yz}})$ }
\newcommand{\dxzyznospace}{$(d_{\text{xz}}$, $d_{\text{yz}})$}

\title{Strain-stabilized superconductivity}

\author{J. P. Ruf}
\email{jpr239@cornell.edu}
\affiliation{\physicsDept}
\author{H. Paik}
\affiliation{\paradim}
\author{N. J. Schreiber}
\affiliation{\mseDept}
\author{H. P. Nair}
\affiliation{\mseDept}
\author{L. Miao}
\affiliation{\physicsDept}
\author{J. K. Kawasaki}
\affiliation{\physicsDept}
\affiliation{Department of Materials Science and Engineering, University of Wisconsin, Madison WI 53706}
\author{J. N. Nelson}
\affiliation{\physicsDept}
\author{B. D. Faeth}
\affiliation{\physicsDept}
\author{Y. Lee}
\affiliation{\physicsDept}
\author{B. H. Goodge}
\affiliation{\aepDept}
\affiliation{\kavli}
\author{B. Pamuk}
\affiliation{\aepDept}
\author{C. J. Fennie}
\affiliation{\aepDept}
\author{L. F. Kourkoutis}
\affiliation{\aepDept}
\affiliation{\kavli}
\author{D. G. Schlom}
\affiliation{\mseDept}
\affiliation{\kavli}
\author{K. M. Shen}
\email{kmshen@cornell.edu}
\affiliation{\physicsDept}
\affiliation{\kavli}

\date{\today}

\maketitle

\textbf{Superconductivity is among the most fascinating and well-studied quantum states of matter. Despite over 100 years of research, a detailed understanding of how features of the normal-state electronic structure determine superconducting properties has remained elusive.  For instance, the ability to deterministically enhance the superconducting transition temperature by design, rather than by serendipity, has been a long sought-after goal in condensed matter physics and materials science, but achieving this objective may require new tools, techniques and approaches. Here, we report the first instance of the transmutation of a normal metal into a superconductor through the application of epitaxial strain. We demonstrate that synthesizing \RO thin films on (110)-oriented \TO substrates enhances the density of states near the Fermi level, which stabilizes superconductivity under strain, and suggests that a promising strategy to create new transition-metal superconductors is to apply judiciously chosen anisotropic strains that redistribute carriers within the low-energy manifold of $d$ orbitals.}

In typical weak-coupling theories of superconductivity, the effective attraction $V$ between electrons is mediated by the exchange of bosons having a characteristic energy scale $\omega_{\text{B}}$, and superconductivity condenses below a transition temperature \TC parameterized as~\cite{carbotte_properties_1990}:

\begin{equation}
T_{c} \sim \omega_{\text{B}}\exp\left(-\frac{1}{N(E_F)V}\right) = \omega_{\text{B}}\exp\left(-\frac{1 + \lambda}{\lambda - \mu^{*}}\right) 
\end{equation}

\noindent where $N(E_F)$ is the density of states (DOS) near the Fermi level, $\lambda$ is the electron-boson coupling strength, and $\mu^{*}$ is the Coulomb pseudopotential that describes the residual Coulomb repulsion between quasiparticles~\footnote{For simplicity, we assume that all of the non-isotropic $\mathbf{q}$- and $\mathbf{k}$-dependencies that appear in a more realistic formulation of the Cooper pairing problem have been averaged away.  Note that within the range of validity of equation~(1)---\textit{viz.}, $1 >> \lambda > \mu^{*}$---increasing $\lambda$ (increasing $\mu^{*}$) generally enhances (suppresses) \TCnospace, respectively, assuming that superconductivity remains the dominant instability.}.  

Experimental methods that boost \TC are highly desired from a practical perspective. Furthermore, by analyzing how these available knobs couple to the normal-state properties on the right side of equation~(1), one can envisage engineering the electronic structure and electron-boson coupling to optimize \TCnospace.  For example, increasing $N(E_F)$ is a frequently suggested route towards realizing higher \TCnospace, but how to achieve this  for specific materials often remains unclear.

Chemical doping and hydrostatic pressure have been the most common knobs used to manipulate superconductivity.  Unfortunately, doping has the complication of introducing substitutional disorder, whereas pressure studies are incompatible with most probes of electronic structure.  Moreover, because large pressures are usually required to appreciably increase \TCnospace~\cite{hamlin_superconductivity_2015}, pressure-enhanced superconductivity exists transiently---typically in different structural polymorphs than at ambient conditions---rendering it inaccessible for applications. 

An alternative strategy for controlling superconductivity is epitaxial strain engineering.  This approach is static, disorder-free, allows for the use of sophisticated experimental probes~\cite{burganov_strain_2016}, and enables integration with other materials in novel artificial interfaces and device structures~\cite{ohtomo_high-mobility_2004,kawasaki_rutile_2018}.  To date, epitaxial strain has only been used to modulate \TC in known superconductors~\cite{lock_penetration_1951,locquet_doubling_1998,*si_strain_1999,*si_epitaxial-strain-induced_2001, *bozovic_epitaxial_2002}.  In this letter, we report the first example of a new superconductor created through epitaxial strain, starting from a compound, \ROnospace, previously not known to be superconducting~\footnote{In principle, assuming that all Fermi liquids are eventually unstable towards superconductivity at low enough temperatures and magnetic fields (including internal fields arising from magnetic impurities), this is not strictly a change in the ground state of the system. However, for our purposes, extremely low temperatures and fields below what are experimentally achievable can be thought of as effectively zero, justifying the use of phrases such as ``strain-induced superconductivity'' interchangeably with ``huge enhancement of critical temperature''.}.  By comparing the results of angle-resolved photoemission spectroscopy (ARPES) experiments with density functional theory (DFT) calculations, we show that splittings between the effective low-energy $d$ orbital degrees of freedom in \RO respond sensitively to appropriate modes of strain, and we discuss how this approach may open the door to strain tuning of superconductivity in other materials.

\begin{figure*}
\begin{center}
\includegraphics[width=7in]{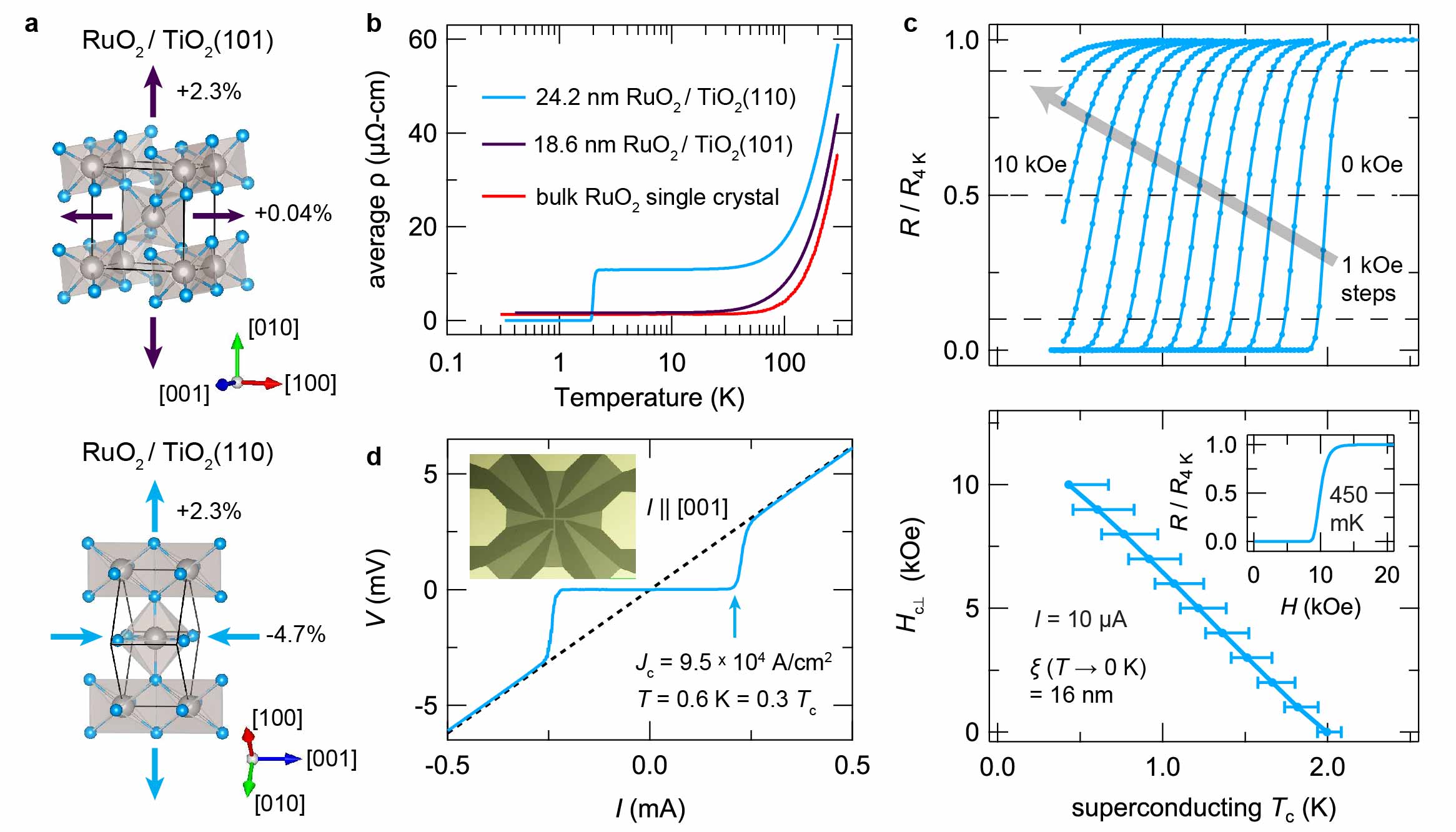}
\end{center}
\caption{\label{figure1} \textbf{Electrical transport behavior of bulk \RO single crystals and epitaxially strained \RO thin films.} \textbf{a,} Schematic diagrams of the crystal structures and in-plane lattice mismatches with \TO of \RO thin films synthesized in (101)- and (110)-orientations. \textbf{b,} Average resistivity versus temperature curves for 24.2 nm thick \ROA and 18.6 nm thick \ROB films and bulk \RO single crystals reproduced from Ref.~\cite{lin_low_2004}. For clarity the results for bulk \RO have been rigidly shifted upward by 1 $\mu\Omega$-cm ($\rho_0 \approx 0.3 \text{ }\mu\Omega$-cm). \textbf{c,} Upper critical magnetic fields $H_{c\perp}$ versus superconducting \TCnospace s extracted from magnetoresistance measurements for the \ROA sample in \textbf{b}, along with (inset) a characteristic $R(H)$ sweep acquired at $T = 0.45 \text{ K}$. Superconducting \TCnospace s are taken as the temperatures at which the resistance drops to $50\%$ of its residual normal-state value $R_{4\,\text{K}}$; error bars on these \TCnospace s delineate drops to $90\%$ and $10\%$ of $R_{4\,\text{K}}$, respectively (\textit{cf.} the horizontal dashed lines in the top panel).  \textbf{d,} $V(I)$ curve measured at $0.6 \text{ K}$ on a $10 \, \mu\text{m}$-wide resistivity bridge lithographically patterned on the \ROA sample from \textbf{b} and \textbf{c}, with the direction of current flow parallel to $[001]_{\text{rutile}}$.  Similarly large critical current densities are obtained with $I||[1\overline{1}0]$ (Supplementary Fig.\,S1).}
\end{figure*}

Bulk \RO crystallizes in the ideal tetragonal rutile structure (space group $\#136$, $P4_2/mnm$) with lattice constants at 295 K of $(a=4.492 \text{ \ang}, c=3.106 \text{ \ang})$~\cite{berlijn_itinerant_2017}.  \RO thin films in distinct epitaxial strain states were synthesized using oxide molecular-beam epitaxy by employing different orientations of isostructural \TO substrates, $(a=4.594 \text{ \ang}, c=2.959 \text{ \ang})$~\cite{burdett_structural-electronic_1987}.  As shown in Fig.\,1a, the surfaces of (101)-oriented substrates are spanned by the $[\overline{1}01]$ and $[010]$ lattice vectors of \TOnospace, which ideally impart in-plane tensile strains on \RO (at 295 K) of $+0.04\%$ and $+2.3\%$, respectively.  On \TOnospace(110), the lattice mismatches with \RO are larger: $-4.7\%$ along $[001]$ and $+2.3\%$ along $[1\overline{1}0]$. Figure 1b shows electrical resistivity $\rho(T)$ measurements for \RO films, along with results for bulk \RO single crystals from Ref.~\cite{lin_low_2004}.  To compare with bulk \ROnospace, for the thin-film samples we plot the geometric mean of $\rho$ along the two in-plane directions; the intrinsic resistivity anisotropy is known to be small (Supplementary Fig.\,S1).

$\rho(T)$ data for the lightly strained \ROnospace/\TOnospace(101) sample---henceforth referred to as \ROBnospace---are nearly indistinguishable from bulk, exhibiting metallic behavior with a low residual resistivity $\rho(0.4\text{ K}) < 2\text{ } \mu\Omega\text{-cm}$.  In contrast, a clear superconducting transition is observed for the more heavily strained \ROnospace/\TOnospace(110) sample---referred to as \ROAnospace---at $T_{c} = 2.0 \pm 0.1$ K. Magnetoresistance measurements with $H_{\perp}$ applied along $[110]$ (the out-of-plane direction) show a suppression of \TC with increasing fields and an extrapolated value of $H_{c\perp}(T \rightarrow 0 \text{ K}) = 13.3 \pm 1.5$ kOe, corresponding to an average in-plane superconducting coherence length of $\xi(T \rightarrow 0 \text{ K}) = 15.8 \pm 0.9$ nm (Supplementary Fig.\,S2).  In Fig.\,1d, we show a $V(I)$ curve measured on a lithographically patterned resistivity bridge at $T/T_{c} = 0.3$, from which we extract a critical current density $J_c = (9.5 \pm 1.2) \times 10^4$ A/$\text{cm}^2$. This large value of $J_c$ (over one order of magnitude larger than values reported on typical elemental superconductors with comparable \TCnospace s) indicates that the superconductivity does not arise from a filamentary network, structural defects, minority phases, or from the substrate-film interface, which would all yield much smaller values of $J_c$. 

\begin{figure*}
\begin{center}
\includegraphics[width=7in]{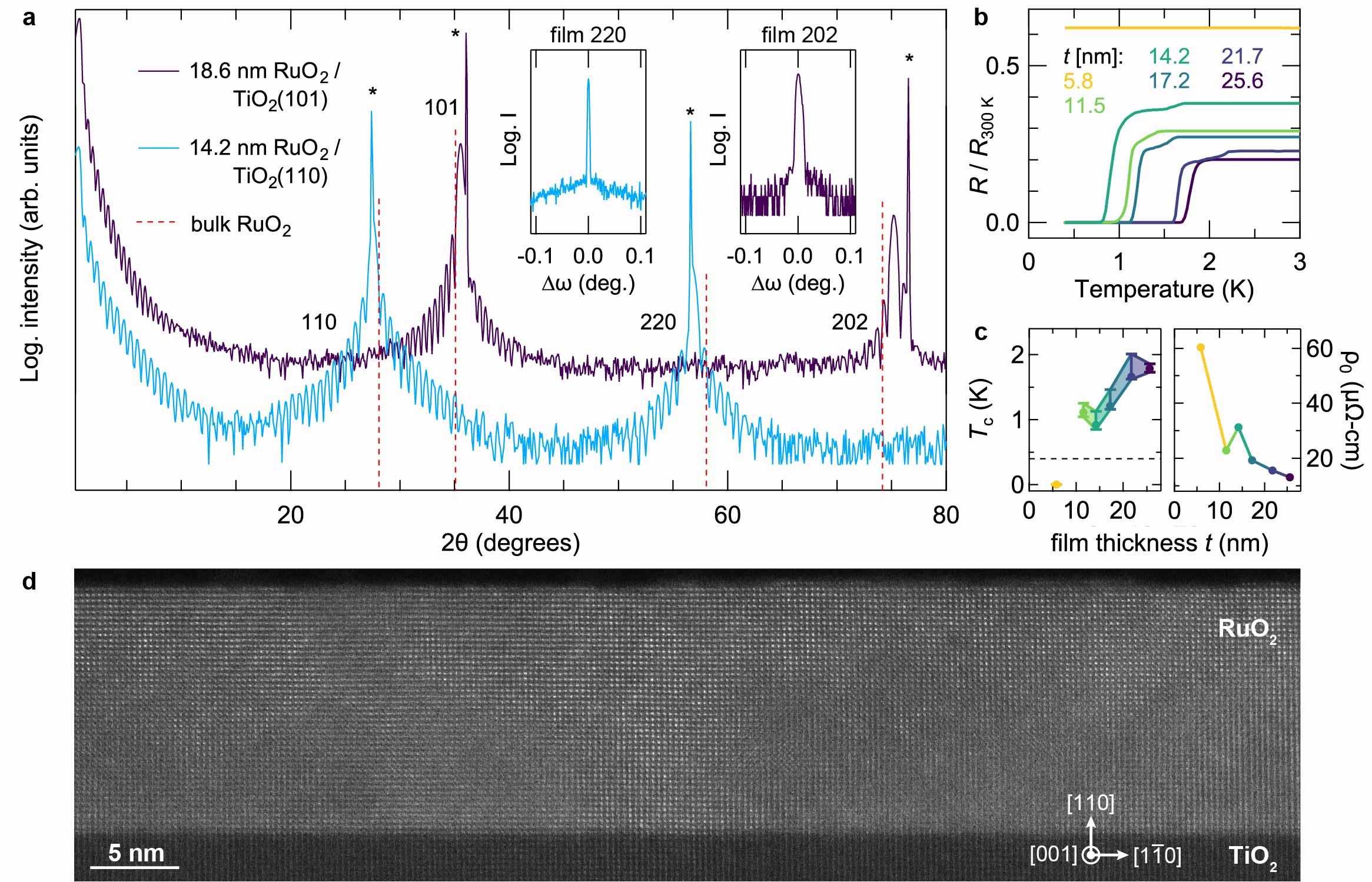}
\end{center}
\caption{\label{figure2} \textbf{Structural characterization of epitaxially strained \RO thin films, and thickness-dependent superconductivity for \ROAnospace.} \textbf{a,}~XRD data acquired with Cu-K$\alpha$ radiation along the specular crystal truncation rods for 18.6 nm thick \ROB and 14.2 nm thick \ROA films.  Bragg peaks arising from the \TO substrates are marked with asterisks, and the peak positions that would be expected for unstrained bulk \RO are indicated by dashed red lines~\cite{berlijn_itinerant_2017}.  Insets display rocking curves with FWHMs $< 0.01^{\circ}$ taken at the $2\theta$ values corresponding to the primary 220 and 202 film peaks.  Here $q_{||}$ is aligned with TiO$_2[1\overline{1}0]$ for the (110)-oriented sample, and with TiO$_2[\overline{1}01]$ for the (101)-oriented sample. \textbf{b,}~Resistance versus temperature curves for \ROA samples with different film thicknesses $t$, normalized to their values at $300\text{ K}$. \textbf{c,}~Superconducting \TCnospace s and residual resistivities $\rho_{0}$ plotted versus film thickness for the \ROA samples from \textbf{b}. Error bars on \TCnospace s have the same meaning as in Fig.\,1.  The horizontal dashed line represents the base temperature attainable in our refrigerator, 0.4 K. \textbf{d,} STEM image of the same 14.2 nm thick \ROA sample from \textbf{a - c}.  More comprehensive structural and electrical characterization of the samples shown here are included in Supplementary Figs.\,S3-S10.}  
\end{figure*}

In order to disentangle the effects of strain from other possible sources of superconductivity, we compare \RO films as functions of strain and film thickness, $t$.  In Fig.\,2a, we plot x-ray diffraction (XRD) data from similar-thickness films of \ROB and \ROAnospace, showing that the bulk-averaged crystal structures of the films are strained as expected along the out-of-plane direction based on their net in-plane lattice mismatches with \TOnospace.  The primary 101 and 202 film peaks of \ROB are shifted to larger angles than bulk \ROnospace, corresponding to a $1.1 \%$ compression of $d_{101}$, while Nelson-Riley analysis of the primary 110, 220 and 330 (not shown) peak positions for \ROA evidence a $2.0 \%$ expansion of $d_{110}$ relative to bulk.  In Fig.\,2b,c, we plot resistivity data showing that reducing $t$ in \ROA decreases \TCnospace, characteristic of numerous families of thin-film superconductors~\cite{pinto_dimensional_2018,meyer_strain-relaxation_2015}, with \TC dropping below our experimental threshold (0.4 K) between $t = 11.5 \text{ nm}$ and $5.8 \text{ nm}$. This suppression of \TC with thickness indicates superconductivity is \emph{not} confined near the substrate-film interface, so possible interfacial modifications of the crystal structure~\cite{gozar_high-temperature_2008}, carrier density~\cite{he_phase_2013}, substrate-film mode coupling~\cite{lee_interfacial_2014}, and non-stoichiometry in the films or substrates~\cite{paik_transport_2015,yoshimatsu_superconductivity_2017} can all be eliminated as potential causes of superconductivity.  These conclusions are also supported by the facts that superconductivity is not observed in \ROB films, nor in bare \TO substrates treated in an identical fashion to the \RO films.  Finally, in Fig.\,2d we show a scanning transmission electron microscopy (STEM) image of a superconducting \ROA sample, which confirms uniform growth of the film over lateral length scales exceeding those expected to be relevant for superconductivity (\textit{e.g.}, $\xi$), with no evidence of minority phases and a chemically abrupt interface between \RO and \TO (Supplementary Figs.\,S5-S7).

We believe the thickness dependence of \TC results primarily from the competition between: 1) an intrinsic strain-induced enhancement of \TC that should be maximized for thinner, commensurately strained \ROA films, versus 2) disorder-induced suppressions of \TC that become amplified in the ultrathin limit (see $\rho_{0}$ versus $t$ in Fig.\,2c).  While the thinnest films experience the largest substrate-imposed strains, stronger disorder scattering (likely from interfacial defects) reduces \TC below our detection threshold.  Films of intermediate thickness ($t \approx 10 - 25\text{ nm}$) have lower residual resistivities and higher \TCnospace s, but do exhibit signatures of partial strain relaxation.  Nevertheless, a detailed analysis of misfit dislocations by STEM and XRD reciprocal-space mapping (Supplementary Figs.\,S3-S10) indicates that these films are largely structurally homogeneous and, on average, much closer to commensurately strained than fully relaxed. Finally, in much thicker samples ($t = 48$ nm) where a more significant volume fraction of the film should be relaxed, the strain is further released by oriented micro-cracks that make these samples spatially inhomogeneous and cause anisotropic distributions of current flow, preventing reliable resistivity measurements (Supplementary Fig.\,S11).

\begin{figure*}
\begin{center}
\includegraphics[width=7in]{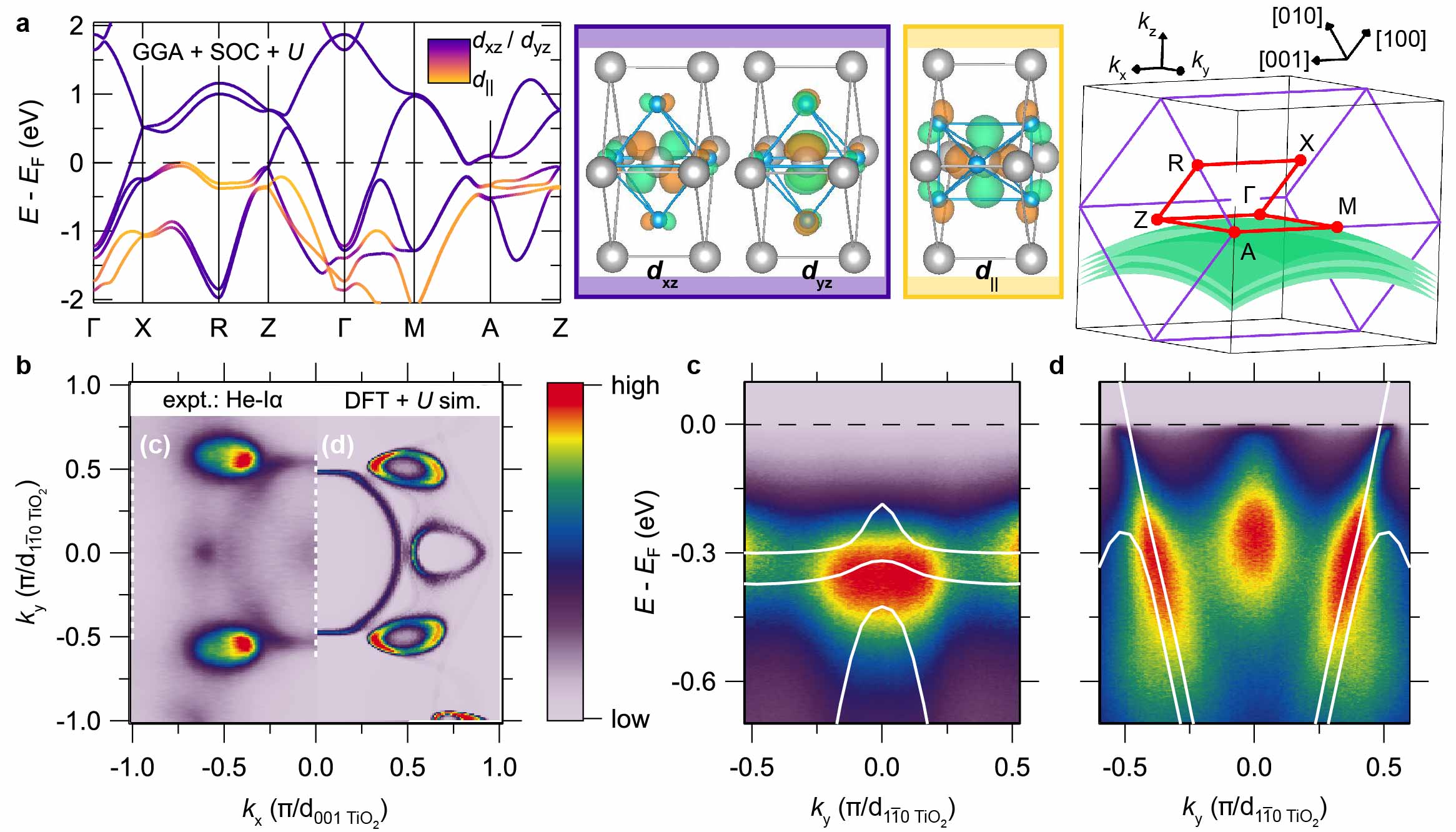}
\end{center}
\caption{\label{figure3} \textbf{Electronic structure of \ROnospace.}  \textbf{a,} Non-magnetic band structure of \ROA according to DFT, calculated within the generalized gradient approximation (GGA) including spin-orbit coupling (SOC) and a static $+U=2 \text{ eV}$ correction on the Ru sites.  The color scale indicates the magnitudes of projections onto Ru-centered Wannier functions with \dpar and \dxzyz orbital characters, which are constructed from the near-$E_F$ Kohn-Sham eigenstates and plotted in the above drawings of the crystal structure.  The schematic in the upper right shows the Brillouin zone of the parent tetragonal rutile structure in purple, the high-symmetry contour for the spaghetti plot in red, and the region probed on (110)-oriented surfaces at the experimental photon energy in green (Supplementary Fig.\,S13).  \textbf{b,} Slice through the Fermi surface measured on a 7 nm thick \RO / \TOnospace(110) thin film (left), compared to the DFT$+U$ Fermi surface (right) projected onto the green region of the Brillouin zone. \textbf{c - d,} $E(k)$ spectra along the one-dimensional cuts indicated in \textbf{b}, showing flat bands with \dpar orbital character and more dispersive bands with \dxzyz character, consistent with DFT$+U$ expectations (solid white lines).}
\end{figure*}

Having established the strain-induced nature of the superconductivity in \ROAnospace, we now explore its underlying origin using a combination of DFT and ARPES.  In Fig.\,3a, we present the electronic structure of commensurately strained \ROA calculated by DFT$+U$ ($U$ = 2 eV), following the methods of Berlijn \textit{et al.}~\cite{berlijn_itinerant_2017}.  Despite being constructed of RuO$_6$ octahedra having the same $4d^4$ electronic configuration as in (Ca,Sr,Ba)RuO$_3$, the electronic structure of \RO is markedly different from that of perovskite-based ruthenates. These distinctions arise from a sizable ligand-field splitting of the $t_{2g}$ orbitals, such that the most natural description of the low-energy electronic structure is in terms of states derived from two distinct types of orbitals:  \dpar and \dxzyznospace, as illustrated by plots of Wannier functions in Fig.\,3~\cite{goodenough_two_1971,eyert_embedded_2000}.  Viewed in the band basis in Fig.\,3a, the differentiation in $\mathbf{k}$-space between these orbitals becomes apparent: the  near-$E_F$ \dpar states form mostly flat bands concentrated around the $k_{001}=\pi/c$ (Z-R-A) plane, whereas the \dxzyz states form more isotropically dispersing bands distributed uniformly throughout the Brillouin zone.  

In many other $d^4$ ruthenates (such as Sr$_2$RuO$_4$ and Ca$_2$RuO$_4$), static mean-field electronic structure calculations (such as DFT$+U$) often predict quantitatively incorrect effective masses~\cite{mravlje_coherence-incoherence_2011, *mackenzie_quantum_1996, *tamai_high-resolution_2019, ricco_situ_2018, *sutter_orbitally_2019} or even qualitatively incorrect ground states~\cite{sutter_hallmarks_2017} because these approaches neglect local atomic-like (Hund's rule) spin correlations that strongly renormalize the low-energy quasiparticle excitations.  Therefore, it is imperative to compare DFT calculations for \RO with experimental data, to establish the reliability of any theoretically predicted dependence of the electronic structure on strain. The left half of Fig.\,3b shows the Fermi surface of \ROA measured with He-I$\alpha$ (21.2 eV) photons at 17 K, which agrees well with a non-magnetic DFT+$U$ simulation of the Fermi surface at a reduced out-of-plane momentum of $k_{110} = -0.2 \pm 0.2 \text{ } \pi/d_{110}$ (right half of Fig.\,3b).  In Fig.\,3c,d, we plot energy versus momentum spectra acquired along the dashed lines in Fig.\,3b: in Fig.\,3c, the spectrum is dominated by the flat \dpar bands around a binding energy of 300 meV, whereas in Fig.\,3d the \dxzyz bands are steeply dispersing and can be tracked down to several hundred meV below $E_F$, both of which are well reproduced by DFT$+U$ calculations. The agreement between the experimental and DFT band velocities is consistent with recent ARPES studies of Ir-doped \RO single crystals~\cite{jovic_dirac_2018} and with earlier specific heat measurements of the Sommerfeld coefficient in bulk \ROnospace, which suggested a modest momentum-averaged mass renormalization of $\gamma_{\text{exp.}} = 1.45\gamma_{\text{DFT}}$~\cite{passenheim_heat_1969, glassford_electron_1994}. The fact that the true electronic structure of \RO can be well accounted for by DFT+$U$ allows us to utilize such calculations to understand how epitaxial strains can be employed to engineer features of the electronic structure to enhance the instability towards superconductivity. 

\begin{figure*}
\begin{center}
\includegraphics[width=7in]{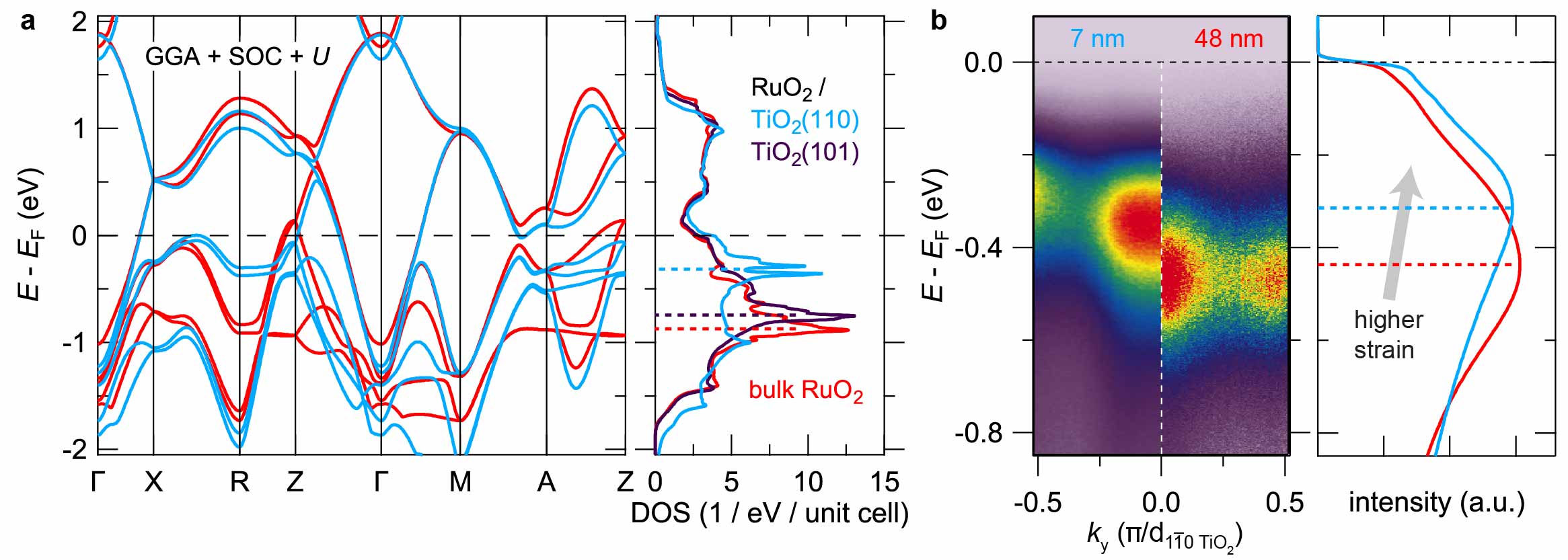}
\end{center}
\caption{\label{figure4} \textbf{Strain-induced changes to the electronic structure of \ROnospace.} \textbf{a,} DFT + $U$ band structures and corresponding density of states (DOS) traces for bulk \RO and epitaxially strained \ROA and \ROBnospace.  The \ROB results are omitted from the spaghetti plot for clarity since they are very similar to bulk.  \textbf{b,} Comparison of $E(k)$ spectra along cut (c) from Fig.\,3 for two different \ROA samples: a highly strained 7 nm thick film (left), and a partially strain-relaxed 48 nm thick film (right).  As an approximate proxy of the full DOS, for these samples we show the energy distribution curves of photoemission intensity averaged over the entire region of $k$-space probed experimentally at $h\nu=21.2\text{ eV}$, which demonstrate that the epitaxial strains imposed by \TOnospace(110) shift \dpar states towards $E_F$ and thereby increase $N(E_F)$.}
\end{figure*}

In Fig.\,4a, we show the strain dependence of the DFT band structure (left panel) and DOS (right panel) for \ROAnospace, \ROBnospace, and bulk \ROnospace. While the results for \ROB are almost identical to bulk, the results for \ROA exhibit significant differences: the large \dparnospace-derived peak in the DOS (centered around a binding energy of 800 meV for bulk) is split into multiple peaks for \ROAnospace, several of which are shifted closer to the Fermi level, thereby increasing $N(E_F)$. In our studies, we found that this strain-dependent trend was robust against details of the DFT calculations, such as whether $U$ was finite (Supplementary Fig.\,S12). In order to determine whether this strain dependence of $N(E_F)$ is realized in experiment, we compared the electronic structure of a thin (7 nm) highly strained \ROA film with a much thicker (48 nm) strain-relaxed \ROA film.  The surface lattice constants of the 48 nm thick film were closer to bulk \RO than the 7 nm thick film (Supplementary Fig.\,S14), so we expect that the surface electronic structure probed by ARPES of the thicker film to be more representative of bulk \ROnospace.  Comparisons between the \ROA and \ROB surfaces are less straightforward, since different parts of the three-dimensional Brillouin zone are sampled by ARPES (Supplementary Fig.\,S15). Figure 4b shows $E(k)$ spectra side-by-side for the 7 nm (left) and 48 nm (right) films of \ROA along the same cut (c) through $\mathbf{k}$-space from Fig.\,3 where the photoemission intensity is dominated by \dpar initial states. The higher levels of strain present at the film surface for the 7 nm thick sample cause a substantial shift of the flat bands towards $E_F$ by $120 \pm 20$ meV relative to the more relaxed 48 nm thick sample.  Integrating the ARPES data over the full measured region of $\mathbf{k}$-space for both samples gives the average energy distribution curves plotted in the right half of Fig.\,4b, which show that spectral weight near $E_F$ is enhanced as the \dpar states move towards $E_F$, in qualitative agreement with the trend predicted by DFT. Our results show that the primary electronic effect of epitaxial strains in \ROA is to alter the relative occupancies of the \dpar and \dxzyz orbitals as compared with bulk, and to push a large number of states with \dpar character closer to $E_F$, which enhances $N(E_F)$ and likely \TCnospace.

Observations of Fermi-liquid-like quasiparticles near $E_F$ that scatter at higher energies primarily via their interaction with phonons~\cite{glassford_electron_1994}, along with the fact that superconductivity in \ROA persists in the dirty limit (Supplementary Fig.\,S9), are both consistent with conventional Cooper pairing, suggesting that calculations assuming an electron-phonon mechanism may be enlightening. We performed DFT-based Migdal-Eliashberg calculations of \TC for bulk \RO and strained \ROA that indeed indicate epitaxial strain can enhance \TC by several orders of magnitude.  For bulk \ROnospace, we find that the empirical Coulomb pseudopotential must satisfy $\mu^{*} > 0.30$ to be compatible with the experimentally measured upper bound on $T_{c} < 0.3 \text{ K}$.  For this range of $\mu^{*}$, \TC for \ROA can be as high as $7 \text{ K}$ (Supplementary Fig.\,S16).  A robust strain-induced enhancement of the electron-phonon coupling $\lambda_{\text{el-ph}}$ boosts \TC by a factor of 20 (for $\mu^{*} = 0.30$), and this ratio becomes even larger for higher values of $\mu^{*}$ (\textit{e.g.} for $\mu^{*} = 0.37$, $T_{c}(110)/T_{c}(\text{bulk}) = 5 \text{ K}/5 \text{ mK}$).  Although these estimations of \TC are broadly consistent with our experimental findings, conventional superconductivity in \RO remains a working hypothesis until measurements of the order parameter are possible.

We believe our results demonstrate that a promising strategy to create new transition-metal superconductors is to apply judiciously chosen anisotropic strains that modulate degeneracies among $d$ orbitals near $E_F$.  Many classic studies of conventional superconductors that have nearly-free-electron $(s,p)$-derived states spanning $E_F$ show \textit{decreases} in \TC under hydrostatic pressure, due to lattice stiffening~\cite{smith_will_1967}.  In a limited number of elements where \TC actually \textit{increases} under pressure, electron transfer between $s \rightarrow d$ orbitals has been suggested as causing the enhanced \TCnospace s~\cite{hamlin_superconductivity_2015}; a drawback of this approach, however, is that large pressures of $\approx 10$ GPa are typically required to, \textit{e.g.}, double \TCnospace.  More recently, measurements on single crystals of the unconventional superconductor Sr$_2$RuO$_4$ have shown that uniaxial pressures of only $\approx 1\text{ GPa}$ can boost \TC by more than a factor of two~\cite{steppke_strong_2017}.  Independent of the underlying mechanism, it appears that anisotropic strains may prove much more effective than hydrostatic pressure for tuning superconductivity in multi-orbital systems, as shown here for \ROnospace, as well as in Sr$_2$RuO$_4$.
 
Sizable coupling between the lattice and electronic degrees of freedom in rutile-like crystal structures has been well established both theoretically~\cite{eyert_embedded_2000} and experimentally in VO$_{2}$, where strain-induced variations in the orbital occupancies can modify the metal-insulator transition by $\delta T_{\text{MIT}} \approx 70 \text{ K}$~\cite{muraoka_metalinsulator_2002,*aetukuri_control_2013}.  Therefore, it may be promising to explore other less strongly correlated (\textit{i.e.}, $4d$ and $5d$) rutile compounds such as MoO$_2$ for strain-stabilized superconductivity, instead of employing chemical doping~\cite{alves_unconventional_2010, *alves_superconductivity_2012, *parker_evidence_2014}. Finally, since \ROnospace/\TOnospace(110) is the first known stoichiometric superconductor within the rutile family, further optimization of the superconductivity may enable the creation of structures that integrate superconductivity with other functional properties that have been extensively studied in other rutile compounds, such as high photocatalytic efficiency, half-metallic ferromagnetism, and large spin Hall conductivities.

\section*{Acknowledgements}  The authors thank Y. Li for assistance with electrical transport measurements. This work was supported through the National Science Foundation [Platform for the Accelerated Realization, Analysis, and Discovery of Interface Materials (PARADIM)] under Cooperative Agreement No. DMR-1539918, NSF DMR-1709255, the Air Force Office of Scientific Research Grant No. FA9550-15-1-0474, and the Department of Energy (Award No. DE-SC0019414). This research is funded in part by the Gordon and Betty Moore Foundation's EPiQS Initiative through Grant No. GBMF3850 to Cornell University. This work made use of the Cornell Center for Materials Research (CCMR) Shared Facilities, which are supported through the NSF MRSEC Program (No. DMR-1719875). The FEI Titan Themis 300 was acquired through NSF-MRI-1429155, with additional support from Cornell University, the Weill Institute, and the Kavli Institute at Cornell.  Device fabrication and substrate preparation were performed in part at the Cornell NanoScale Facility, a member of the National Nanotechnology Coordinated Infrastructure (NNCI), which is supported by the NSF (Grant No. ECCS-1542081). 

\section*{Author contributions} H.P., N.J.S., and H.P.N. synthesized the samples by MBE.  J.P.R., H.P., N.J.S., and H.P.N. characterized the samples by XRD.  J.P.R., L.M., and Y.L. characterized the samples by electrical transport; L.M. patterned resistivity bridges on select films.  J.P.R., J.K.K., J.N.N., and B.D.F. characterized the samples by ARPES and LEED.  B.H.G. characterized the samples by STEM.  J.P.R. performed DFT calculations of the electronic structure and B.P. performed DFT-based calculations of the electron-phonon coupling.  L.F.K., D.G.S., and K.M.S. supervised the various aspects of this project.  J.P.R. and K.M.S. wrote the manuscript with input from all authors.  

\section*{Competing interests}  The authors declare no competing interests.

\section*{Data availability statement}  The datasets generated during and analysed during the current study are available from the corresponding author on reasonable request.

\section*{Methods}

\subsection*{Film synthesis}

Epitaxial thin films of \RO were synthesized on various orientations of rutile \TO substrates using a GEN10 reactive oxide molecular-beam epitaxy (MBE) system (Veeco Instruments).  Prior to growth, \TO substrates (Crystec, GmbH) were cleaned with organic solvents, etched in acid, and annealed in air to produce starting surfaces with step-terrace morphology, following the methods in Ref.~\cite{yamamoto_preparation_2005}.  Elemental ruthenium (99.99\% purity, ESPI Metals) was evaporated using an electron-beam evaporator in background oxidant partial pressures of $1 \times 10^{-6} - 5 \times 10^{-6}$ Torr of distilled ozone ($\approx$ 80\% O$_3$ + 20\% O$_2$) at substrate temperatures of $250 - 400^{\circ}$ C, as measured by a thermocouple.  Reflection high-energy electron diffraction was used to monitor the surface crystallinity of the films \textit{in situ} and showed characteristic oscillations in intensity during most of the Ru deposition, indicating a layer-by-layer growth mode following the initial nucleation of several-monolayer-thick \RO islands~\cite{he_versatile_2015}. 

\subsection*{Film characterization}

The crystal structures of all \RO thin-film samples were characterized via lab-based x-ray diffraction (XRD) measurements with Cu-K$\alpha$ radiation (Rigaku SmartLab and Malvern Panalytical Empyrean diffractometers).  Four-point probe electrical transport measurements were conducted on greater than 20 \RO films from 300 K down to a base temperature of 0.4 K using a Physical Properties Measurement System equipped with a He-3 refrigerator (Quantum Design).  All \RO / \TOnospace(110) samples were superconducting with \TCnospace s ranging from $0.5 - 2.4$ K, except for ultrathin films with residual resistivities $\rho_0 \gtrapprox 40 \text{ } \mu\Omega\text{-cm}$, as described in the main text and supplemental information.   A systematic investigation of how the crystal structures, structural defects and superconducting properties (\textit{e.g.} \TC and \HCnospace) of the films are correlated---and how these vary under different film synthesis conditions---is beyond the scope of the present work and will be published elsewhere.
 
A subset of films studied by XRD and transport were also characterized \textit{in situ} by angle-resolved photoemission spectroscopy (ARPES) and low-energy electron diffraction (LEED).  For these measurements, films were transferred under ultrahigh vacuum immediately following growth to an analysis chamber with a base pressure of $5\times10^{-11}$ Torr equipped with a helium plasma discharge lamp, a hemispherical electron analyzer (VG Scienta R4000), and a four-grid LEED optics (SPECS ErLEED 150). 

A subset of films studied by XRD and transport were also imaged using cross-sectional scanning transmission electron microscopy.  Cross-sectional specimens were prepared using the standard focused ion beam (FIB) lift-out process on a Thermo Scientific Helios G4 X FIB. High-angle annular dark-field scanning transmission electron microscopy (HAADF-STEM) images were acquired on an aberration-corrected FEI Titan Themis at 300 keV with a probe convergence semi-angle of 21.4 mrad and inner and outer collection angles of 68 and 340 mrad.  

\subsection*{Electronic structure calculations}

Non-magnetic density functional theory (DFT) calculations for the electronic structure of \RO were performed using the {\sc Quantum ESPRESSO} software package~\cite{giannozzi_quantum_2009, *giannozzi_advanced_2017} with fully relativistic ultrasoft pseudopotentials for Ru and O~\cite{dal_corso_pseudopotentials_2014}.  We represented the Kohn-Sham wavefunctions in a basis set of plane waves extending up to a kinetic energy cutoff of 60 Ry, and used a cutoff of 400 Ry for representing the charge density.   Brillouin zone integrations were carried out on an $8\times8\times12$ $k$-mesh with 70 meV of Gaussian smearing.  Perdew, Burke and Ernzerhof's parametrization of the generalized gradient approximation was employed as the exchange-correlation functional~\cite{PBE_1996}, supplemented by an on-site correction of $+U_{\text{eff}} = U - J = 2$ eV within spheres surrounding the Ru sites, following Ref.~\cite{berlijn_itinerant_2017}.

After obtaining self-consistent Kohn-Sham eigenstates via DFT, we used the pw2wannier and Wannier90 codes~\cite{Wannier90_2014} to construct 20 Wannier functions spanning the manifold of eigenstates surrounding $E_F$ (20 = 10 Kramers-degenerate $d$-orbitals per Ru atom $\times$ 2 Ru atoms per unit cell).  Following Ref.~\cite{Eyert_2002}, to account for the non-symmorphic space group symmetries of rutile crystal structures, we referenced the trial orbitals employed in the Wannierisation routine to locally rotated coordinate systems centered on the two Ru sites within each unit cell.  Orbital designations employed in the main text such as \dpar and \dxzyz refer to projections onto this basis of Wannier functions.  The more computationally efficient Wannier basis was used to calculate quantities that required dense $k$ meshes to be properly converged, such as the the projected Fermi surface in Fig.\,3 ($51 \times 51 \times 51$ $k$-mesh) and the near-$E_F$ density of states traces in Fig.\,4 ($32 \times 32 \times 48$ $k$-meshes).

Because the \RO samples studied in this work are thin films subject to biaxial epitaxial strains imposed by differently oriented rutile TiO$_2$ substrates, we performed DFT + Wannier calculations of the electronic structure for several different crystal structures of \RO as summarized in the supplemental information (Supplementary Table 1).  We used the ISOTROPY software package~\cite{isotropy} to study distortions of the parent tetragonal rutile crystal structure that are induced in biaxially strained thin films. Crystal structures and Wannier functions were visualized using the VESTA software package~\cite{momma_vesta_2011}. 

\subsection*{Electron-phonon coupling calculations}

To generate the inputs required for the electron-phonon coupling calculations described below, first-principles electronic structure and phonon calculations were performed using the {\sc Quantum ESPRESSO} software package with norm-conserving pseudopotentials and plane-wave basis sets~\cite{giannozzi_quantum_2009, *giannozzi_advanced_2017}.  Here we employed a kinetic energy cutoff of 160 Ry, an electronic momentum $k$-point mesh of $16\times16\times24$, $20 \text{ meV}$ of Methfessel-Paxton smearing for the occupation of the electronic states, and a tolerance of $10^{-10} \text{  eV}$ for the total energy convergence.  The generalized gradient approximation as implemented in the PBEsol functional~\cite{perdew_restoring_2008} was employed as the exchange-correlation functional.  For the Wannier interpolation, we used an interpolating electron-momentum mesh of $8\times8\times12$ and a phonon-momentum mesh of $2\times2\times3$.  Results for bulk \RO were calculated using the crystal structure that minimizes the DFT-computed total energy with the PBEsol functional: $(a=4.464 \text{ \ang}, c=3.093 \text{ \ang})$ and $x_{\text{oxygen}} = 0.3062$.  Results for strained \ROA were calculated by changing the lattice constants of this ``bulk'' crystal structure by $+2.3\%$ along $[1\overline{1}0]$, $-4.7\%$ along $[001]$, $+2.2\%$ along $[110]$, and setting $x_{\text{oxygen}} = y_{\text{oxygen}} = 0.2996$.  The lattice constant along $[110]$ and internal coordinates of this simulated \ROA structure were determined by allowing the structure to relax so as to (locally) minimize the DFT-computed total energy. 

Electron-phonon coupling calculations were performed using the EPW code~\cite{ponce_epw_2016}, using an interpolated electron-momentum mesh of $32\times32\times48$ and an interpolated phonon-momentum mesh of $8\times8\times12$.  The isotropic Eliashberg spectral function $\alpha^2 F(\omega)$ and total electron-phonon coupling constant $\lambda_{\text{el-ph}}$ (integrated over all phonon modes and wavevectors) were calculated with a phonon smearing of 0.2 meV.  From the calculated $\alpha^2 F(\omega)$ and $\lambda_{\text{el-ph}}$, we estimated the superconducting transition temperature using the semi-empirical McMillan-Allen-Dynes formula~\cite{mcmillan_transition_1968,*allen_transition_1975}:

\begin{equation}
T_c = \frac{\omega_{\log}}{1.2} \exp\left[ - \frac{1.04 (1 + \lambda_{\text{el-ph}} ) } { \lambda_{\text{el-ph}} - \mu^{*} (1 + 0.62 \lambda_{\text{el-ph}} ) } \right]  
\end{equation}


\begin{thebibliography}{54}%
\makeatletter
\providecommand \@ifxundefined [1]{%
 \@ifx{#1\undefined}
}%
\providecommand \@ifnum [1]{%
 \ifnum #1\expandafter \@firstoftwo
 \else \expandafter \@secondoftwo
 \fi
}%
\providecommand \@ifx [1]{%
 \ifx #1\expandafter \@firstoftwo
 \else \expandafter \@secondoftwo
 \fi
}%
\providecommand \natexlab [1]{#1}%
\providecommand \enquote  [1]{``#1''}%
\providecommand \bibnamefont  [1]{#1}%
\providecommand \bibfnamefont [1]{#1}%
\providecommand \citenamefont [1]{#1}%
\providecommand \href@noop [0]{\@secondoftwo}%
\providecommand \href [0]{\begingroup \@sanitize@url \@href}%
\providecommand \@href[1]{\@@startlink{#1}\@@href}%
\providecommand \@@href[1]{\endgroup#1\@@endlink}%
\providecommand \@sanitize@url [0]{\catcode `\\12\catcode `\$12\catcode
  `\&12\catcode `\#12\catcode `\^12\catcode `\_12\catcode `\%12\relax}%
\providecommand \@@startlink[1]{}%
\providecommand \@@endlink[0]{}%
\providecommand \url  [0]{\begingroup\@sanitize@url \@url }%
\providecommand \@url [1]{\endgroup\@href {#1}{\urlprefix }}%
\providecommand \urlprefix  [0]{URL }%
\providecommand \Eprint [0]{\href }%
\providecommand \doibase [0]{http://dx.doi.org/}%
\providecommand \selectlanguage [0]{\@gobble}%
\providecommand \bibinfo  [0]{\@secondoftwo}%
\providecommand \bibfield  [0]{\@secondoftwo}%
\providecommand \translation [1]{[#1]}%
\providecommand \BibitemOpen [0]{}%
\providecommand \bibitemStop [0]{}%
\providecommand \bibitemNoStop [0]{.\EOS\space}%
\providecommand \EOS [0]{\spacefactor3000\relax}%
\providecommand \BibitemShut  [1]{\csname bibitem#1\endcsname}%
\let\auto@bib@innerbib\@empty
%</preamble>
\bibitem [{\citenamefont {Carbotte}(1990)}]{carbotte_properties_1990}%
  \BibitemOpen
  \bibfield  {author} {\bibinfo {author} {\bibfnamefont {J.~P.}\ \bibnamefont
  {Carbotte}},\ }\href {\doibase 10.1103/RevModPhys.62.1027} {\bibfield
  {journal} {\bibinfo  {journal} {Reviews of Modern Physics}\ }\textbf
  {\bibinfo {volume} {62}},\ \bibinfo {pages} {1027} (\bibinfo {year}
  {1990})}\BibitemShut {NoStop}%
\bibitem [{Note1()}]{Note1}%
  \BibitemOpen
  \bibinfo {note} {For simplicity, we assume that all of the non-isotropic
  $\protect \mathbf {q}$- and $\protect \mathbf {k}$-dependencies that appear
  in a more realistic formulation of the Cooper pairing problem have been
  averaged away. Note that within the range of validity of
  equation~(1)---\protect \textit {viz.}, $1 >> \lambda > \mu
  ^{*}$---increasing $\lambda $ (increasing $\mu ^{*}$) generally enhances
  (suppresses) $T_c$, respectively, assuming that superconductivity remains the
  dominant instability.}\BibitemShut {Stop}%
\bibitem [{\citenamefont {Hamlin}(2015)}]{hamlin_superconductivity_2015}%
  \BibitemOpen
  \bibfield  {author} {\bibinfo {author} {\bibfnamefont {J.~J.}\ \bibnamefont
  {Hamlin}},\ }\href {\doibase 10.1016/j.physc.2015.02.032} {\bibfield
  {journal} {\bibinfo  {journal} {Physica C: Superconductivity and its
  Applications}\ }\bibinfo {series} {Superconducting {Materials}:
  {Conventional}, {Unconventional} and {Undetermined}},\ \textbf {\bibinfo
  {volume} {514}},\ \bibinfo {pages} {59} (\bibinfo {year} {2015})}\BibitemShut
  {NoStop}%
\bibitem [{\citenamefont {Burganov}\ \emph {et~al.}(2016)\citenamefont
  {Burganov}, \citenamefont {Adamo}, \citenamefont {Mulder}, \citenamefont
  {Uchida}, \citenamefont {King}, \citenamefont {Harter}, \citenamefont {Shai},
  \citenamefont {Gibbs}, \citenamefont {Mackenzie}, \citenamefont {Uecker},
  \citenamefont {Bruetzam}, \citenamefont {Beasley}, \citenamefont {Fennie},
  \citenamefont {Schlom},\ and\ \citenamefont {Shen}}]{burganov_strain_2016}%
  \BibitemOpen
  \bibfield  {author} {\bibinfo {author} {\bibfnamefont {B.}~\bibnamefont
  {Burganov}}, \bibinfo {author} {\bibfnamefont {C.}~\bibnamefont {Adamo}},
  \bibinfo {author} {\bibfnamefont {A.}~\bibnamefont {Mulder}}, \bibinfo
  {author} {\bibfnamefont {M.}~\bibnamefont {Uchida}}, \bibinfo {author}
  {\bibfnamefont {P.}~\bibnamefont {King}}, \bibinfo {author} {\bibfnamefont
  {J.}~\bibnamefont {Harter}}, \bibinfo {author} {\bibfnamefont
  {D.}~\bibnamefont {Shai}}, \bibinfo {author} {\bibfnamefont {A.}~\bibnamefont
  {Gibbs}}, \bibinfo {author} {\bibfnamefont {A.}~\bibnamefont {Mackenzie}},
  \bibinfo {author} {\bibfnamefont {R.}~\bibnamefont {Uecker}}, \bibinfo
  {author} {\bibfnamefont {M.}~\bibnamefont {Bruetzam}}, \bibinfo {author}
  {\bibfnamefont {M.}~\bibnamefont {Beasley}}, \bibinfo {author} {\bibfnamefont
  {C.}~\bibnamefont {Fennie}}, \bibinfo {author} {\bibfnamefont
  {D.}~\bibnamefont {Schlom}}, \ and\ \bibinfo {author} {\bibfnamefont
  {K.}~\bibnamefont {Shen}},\ }\href {\doibase 10.1103/PhysRevLett.116.197003}
  {\bibfield  {journal} {\bibinfo  {journal} {Physical Review Letters}\
  }\textbf {\bibinfo {volume} {116}},\ \bibinfo {pages} {197003} (\bibinfo
  {year} {2016})}\BibitemShut {NoStop}%
\bibitem [{\citenamefont {Ohtomo}\ and\ \citenamefont
  {Hwang}(2004)}]{ohtomo_high-mobility_2004}%
  \BibitemOpen
  \bibfield  {author} {\bibinfo {author} {\bibfnamefont {A.}~\bibnamefont
  {Ohtomo}}\ and\ \bibinfo {author} {\bibfnamefont {H.~Y.}\ \bibnamefont
  {Hwang}},\ }\href {\doibase 10.1038/nature02308} {\bibfield  {journal}
  {\bibinfo  {journal} {Nature}\ }\textbf {\bibinfo {volume} {427}},\ \bibinfo
  {pages} {423} (\bibinfo {year} {2004})}\BibitemShut {NoStop}%
\bibitem [{\citenamefont {Kawasaki}\ \emph {et~al.}(2018)\citenamefont
  {Kawasaki}, \citenamefont {Baek}, \citenamefont {Paik}, \citenamefont {Nair},
  \citenamefont {Kourkoutis}, \citenamefont {Schlom},\ and\ \citenamefont
  {Shen}}]{kawasaki_rutile_2018}%
  \BibitemOpen
  \bibfield  {author} {\bibinfo {author} {\bibfnamefont {J.~K.}\ \bibnamefont
  {Kawasaki}}, \bibinfo {author} {\bibfnamefont {D.}~\bibnamefont {Baek}},
  \bibinfo {author} {\bibfnamefont {H.}~\bibnamefont {Paik}}, \bibinfo {author}
  {\bibfnamefont {H.~P.}\ \bibnamefont {Nair}}, \bibinfo {author}
  {\bibfnamefont {L.~F.}\ \bibnamefont {Kourkoutis}}, \bibinfo {author}
  {\bibfnamefont {D.~G.}\ \bibnamefont {Schlom}}, \ and\ \bibinfo {author}
  {\bibfnamefont {K.~M.}\ \bibnamefont {Shen}},\ }\href {\doibase
  10.1103/PhysRevMaterials.2.054206} {\bibfield  {journal} {\bibinfo  {journal}
  {Physical Review Materials}\ }\textbf {\bibinfo {volume} {2}},\ \bibinfo
  {pages} {054206} (\bibinfo {year} {2018})}\BibitemShut {NoStop}%
\bibitem [{\citenamefont {{Lock J. M.}}\ and\ \citenamefont {{Bragg William
  Lawrence}}(1951)}]{lock_penetration_1951}%
  \BibitemOpen
  \bibfield  {author} {\bibinfo {author} {\bibnamefont {{Lock J. M.}}}\ and\
  \bibinfo {author} {\bibnamefont {{Bragg William Lawrence}}},\ }\href
  {\doibase 10.1098/rspa.1951.0169} {\bibfield  {journal} {\bibinfo  {journal}
  {Proceedings of the Royal Society of London. Series A. Mathematical and
  Physical Sciences}\ }\textbf {\bibinfo {volume} {208}},\ \bibinfo {pages}
  {391} (\bibinfo {year} {1951})}\BibitemShut {NoStop}%
\bibitem [{\citenamefont {Locquet}\ \emph {et~al.}(1998)\citenamefont
  {Locquet}, \citenamefont {Perret}, \citenamefont {Fompeyrine}, \citenamefont
  {Mächler}, \citenamefont {Seo},\ and\ \citenamefont
  {Tendeloo}}]{locquet_doubling_1998}%
  \BibitemOpen
  \bibfield  {author} {\bibinfo {author} {\bibfnamefont {J.-P.}\ \bibnamefont
  {Locquet}}, \bibinfo {author} {\bibfnamefont {J.}~\bibnamefont {Perret}},
  \bibinfo {author} {\bibfnamefont {J.}~\bibnamefont {Fompeyrine}}, \bibinfo
  {author} {\bibfnamefont {E.}~\bibnamefont {Mächler}}, \bibinfo {author}
  {\bibfnamefont {J.~W.}\ \bibnamefont {Seo}}, \ and\ \bibinfo {author}
  {\bibfnamefont {G.~V.}\ \bibnamefont {Tendeloo}},\ }\href {\doibase
  10.1038/28810} {\bibfield  {journal} {\bibinfo  {journal} {Nature}\ }\textbf
  {\bibinfo {volume} {394}},\ \bibinfo {pages} {453} (\bibinfo {year}
  {1998})}\BibitemShut {NoStop}%
\bibitem [{\citenamefont {Si}\ \emph {et~al.}(1999)\citenamefont {Si},
  \citenamefont {Li},\ and\ \citenamefont {Xi}}]{si_strain_1999}%
  \BibitemOpen
  \bibfield  {author} {\bibinfo {author} {\bibfnamefont {W.}~\bibnamefont
  {Si}}, \bibinfo {author} {\bibfnamefont {H.-C.}\ \bibnamefont {Li}}, \ and\
  \bibinfo {author} {\bibfnamefont {X.~X.}\ \bibnamefont {Xi}},\ }\href
  {\doibase 10.1063/1.124031} {\bibfield  {journal} {\bibinfo  {journal}
  {Applied Physics Letters}\ }\textbf {\bibinfo {volume} {74}},\ \bibinfo
  {pages} {2839} (\bibinfo {year} {1999})}\BibitemShut {NoStop}%
\bibitem [{\citenamefont {Si}\ and\ \citenamefont
  {Xi}(2001)}]{si_epitaxial-strain-induced_2001}%
  \BibitemOpen
  \bibfield  {author} {\bibinfo {author} {\bibfnamefont {W.}~\bibnamefont
  {Si}}\ and\ \bibinfo {author} {\bibfnamefont {X.~X.}\ \bibnamefont {Xi}},\
  }\href {\doibase 10.1063/1.1338966} {\bibfield  {journal} {\bibinfo
  {journal} {Applied Physics Letters}\ }\textbf {\bibinfo {volume} {78}},\
  \bibinfo {pages} {240} (\bibinfo {year} {2001})}\BibitemShut {NoStop}%
\bibitem [{\citenamefont {Bozovic}\ \emph {et~al.}(2002)\citenamefont
  {Bozovic}, \citenamefont {Logvenov}, \citenamefont {Belca}, \citenamefont
  {Narimbetov},\ and\ \citenamefont {Sveklo}}]{bozovic_epitaxial_2002}%
  \BibitemOpen
  \bibfield  {author} {\bibinfo {author} {\bibfnamefont {I.}~\bibnamefont
  {Bozovic}}, \bibinfo {author} {\bibfnamefont {G.}~\bibnamefont {Logvenov}},
  \bibinfo {author} {\bibfnamefont {I.}~\bibnamefont {Belca}}, \bibinfo
  {author} {\bibfnamefont {B.}~\bibnamefont {Narimbetov}}, \ and\ \bibinfo
  {author} {\bibfnamefont {I.}~\bibnamefont {Sveklo}},\ }\href {\doibase
  10.1103/PhysRevLett.89.107001} {\bibfield  {journal} {\bibinfo  {journal}
  {Physical Review Letters}\ }\textbf {\bibinfo {volume} {89}},\ \bibinfo
  {pages} {107001} (\bibinfo {year} {2002})}\BibitemShut {NoStop}%
\bibitem [{Note2()}]{Note2}%
  \BibitemOpen
  \bibinfo {note} {In principle, assuming that all Fermi liquids are eventually
  unstable towards superconductivity at low enough temperatures and magnetic
  fields (including internal fields arising from magnetic impurities), this is
  not strictly a change in the ground state of the system. However, for our
  purposes, extremely low temperatures and fields below what are experimentally
  achievable can be thought of as effectively zero, justifying the use of
  phrases such as ``strain-induced superconductivity'' interchangeably with
  ``huge enhancement of critical temperature''.}\BibitemShut {Stop}%
\bibitem [{\citenamefont {Lin}\ \emph {et~al.}(2004)\citenamefont {Lin},
  \citenamefont {Huang}, \citenamefont {Lin}, \citenamefont {Lee},
  \citenamefont {Liu}, \citenamefont {Zhang}, \citenamefont {Chen},\ and\
  \citenamefont {Huang}}]{lin_low_2004}%
  \BibitemOpen
  \bibfield  {author} {\bibinfo {author} {\bibfnamefont {J.~J.}\ \bibnamefont
  {Lin}}, \bibinfo {author} {\bibfnamefont {S.~M.}\ \bibnamefont {Huang}},
  \bibinfo {author} {\bibfnamefont {Y.~H.}\ \bibnamefont {Lin}}, \bibinfo
  {author} {\bibfnamefont {T.~C.}\ \bibnamefont {Lee}}, \bibinfo {author}
  {\bibfnamefont {H.}~\bibnamefont {Liu}}, \bibinfo {author} {\bibfnamefont
  {X.~X.}\ \bibnamefont {Zhang}}, \bibinfo {author} {\bibfnamefont {R.~S.}\
  \bibnamefont {Chen}}, \ and\ \bibinfo {author} {\bibfnamefont {Y.~S.}\
  \bibnamefont {Huang}},\ }\href {\doibase 10.1088/0953-8984/16/45/025}
  {\bibfield  {journal} {\bibinfo  {journal} {Journal of Physics: Condensed
  Matter}\ }\textbf {\bibinfo {volume} {16}},\ \bibinfo {pages} {8035}
  (\bibinfo {year} {2004})}\BibitemShut {NoStop}%
\bibitem [{\citenamefont {Berlijn}\ \emph {et~al.}(2017)\citenamefont
  {Berlijn}, \citenamefont {Snijders}, \citenamefont {Delaire}, \citenamefont
  {Zhou}, \citenamefont {Maier}, \citenamefont {Cao}, \citenamefont {Chi},
  \citenamefont {Matsuda}, \citenamefont {Wang}, \citenamefont {Koehler},
  \citenamefont {Kent},\ and\ \citenamefont
  {Weitering}}]{berlijn_itinerant_2017}%
  \BibitemOpen
  \bibfield  {author} {\bibinfo {author} {\bibfnamefont {T.}~\bibnamefont
  {Berlijn}}, \bibinfo {author} {\bibfnamefont {P.}~\bibnamefont {Snijders}},
  \bibinfo {author} {\bibfnamefont {O.}~\bibnamefont {Delaire}}, \bibinfo
  {author} {\bibfnamefont {H.-D.}\ \bibnamefont {Zhou}}, \bibinfo {author}
  {\bibfnamefont {T.}~\bibnamefont {Maier}}, \bibinfo {author} {\bibfnamefont
  {H.-B.}\ \bibnamefont {Cao}}, \bibinfo {author} {\bibfnamefont {S.-X.}\
  \bibnamefont {Chi}}, \bibinfo {author} {\bibfnamefont {M.}~\bibnamefont
  {Matsuda}}, \bibinfo {author} {\bibfnamefont {Y.}~\bibnamefont {Wang}},
  \bibinfo {author} {\bibfnamefont {M.}~\bibnamefont {Koehler}}, \bibinfo
  {author} {\bibfnamefont {P.}~\bibnamefont {Kent}}, \ and\ \bibinfo {author}
  {\bibfnamefont {H.}~\bibnamefont {Weitering}},\ }\href {\doibase
  10.1103/PhysRevLett.118.077201} {\bibfield  {journal} {\bibinfo  {journal}
  {Physical Review Letters}\ }\textbf {\bibinfo {volume} {118}},\ \bibinfo
  {pages} {077201} (\bibinfo {year} {2017})}\BibitemShut {NoStop}%
\bibitem [{\citenamefont {Burdett}\ \emph {et~al.}(1987)\citenamefont
  {Burdett}, \citenamefont {Hughbanks}, \citenamefont {Miller}, \citenamefont
  {Richardson},\ and\ \citenamefont
  {Smith}}]{burdett_structural-electronic_1987}%
  \BibitemOpen
  \bibfield  {author} {\bibinfo {author} {\bibfnamefont {J.~K.}\ \bibnamefont
  {Burdett}}, \bibinfo {author} {\bibfnamefont {T.}~\bibnamefont {Hughbanks}},
  \bibinfo {author} {\bibfnamefont {G.~J.}\ \bibnamefont {Miller}}, \bibinfo
  {author} {\bibfnamefont {J.~W.}\ \bibnamefont {Richardson}}, \ and\ \bibinfo
  {author} {\bibfnamefont {J.~V.}\ \bibnamefont {Smith}},\ }\href {\doibase
  10.1021/ja00246a021} {\bibfield  {journal} {\bibinfo  {journal} {Journal of
  the American Chemical Society}\ }\textbf {\bibinfo {volume} {109}},\ \bibinfo
  {pages} {3639} (\bibinfo {year} {1987})}\BibitemShut {NoStop}%
\bibitem [{\citenamefont {Pinto}\ \emph {et~al.}(2018)\citenamefont {Pinto},
  \citenamefont {Rezvani}, \citenamefont {Perali}, \citenamefont {Flammia},
  \citenamefont {Milošević}, \citenamefont {Fretto}, \citenamefont
  {Cassiago},\ and\ \citenamefont {Leo}}]{pinto_dimensional_2018}%
  \BibitemOpen
  \bibfield  {author} {\bibinfo {author} {\bibfnamefont {N.}~\bibnamefont
  {Pinto}}, \bibinfo {author} {\bibfnamefont {S.~J.}\ \bibnamefont {Rezvani}},
  \bibinfo {author} {\bibfnamefont {A.}~\bibnamefont {Perali}}, \bibinfo
  {author} {\bibfnamefont {L.}~\bibnamefont {Flammia}}, \bibinfo {author}
  {\bibfnamefont {M.~V.}\ \bibnamefont {Milošević}}, \bibinfo {author}
  {\bibfnamefont {M.}~\bibnamefont {Fretto}}, \bibinfo {author} {\bibfnamefont
  {C.}~\bibnamefont {Cassiago}}, \ and\ \bibinfo {author} {\bibfnamefont
  {N.~D.}\ \bibnamefont {Leo}},\ }\href {\doibase 10.1038/s41598-018-22983-6}
  {\bibfield  {journal} {\bibinfo  {journal} {Scientific Reports}\ }\textbf
  {\bibinfo {volume} {8}},\ \bibinfo {pages} {1} (\bibinfo {year}
  {2018})}\BibitemShut {NoStop}%
\bibitem [{\citenamefont {Meyer}\ \emph {et~al.}(2015)\citenamefont {Meyer},
  \citenamefont {Jiang}, \citenamefont {Park}, \citenamefont {Egami},\ and\
  \citenamefont {Lee}}]{meyer_strain-relaxation_2015}%
  \BibitemOpen
  \bibfield  {author} {\bibinfo {author} {\bibfnamefont {T.~L.}\ \bibnamefont
  {Meyer}}, \bibinfo {author} {\bibfnamefont {L.}~\bibnamefont {Jiang}},
  \bibinfo {author} {\bibfnamefont {S.}~\bibnamefont {Park}}, \bibinfo {author}
  {\bibfnamefont {T.}~\bibnamefont {Egami}}, \ and\ \bibinfo {author}
  {\bibfnamefont {H.~N.}\ \bibnamefont {Lee}},\ }\href {\doibase
  10.1063/1.4937170} {\bibfield  {journal} {\bibinfo  {journal} {APL
  Materials}\ }\textbf {\bibinfo {volume} {3}},\ \bibinfo {pages} {126102}
  (\bibinfo {year} {2015})}\BibitemShut {NoStop}%
\bibitem [{\citenamefont {Gozar}\ \emph {et~al.}(2008)\citenamefont {Gozar},
  \citenamefont {Logvenov}, \citenamefont {Kourkoutis}, \citenamefont
  {Bollinger}, \citenamefont {Giannuzzi}, \citenamefont {Muller},\ and\
  \citenamefont {Bozovic}}]{gozar_high-temperature_2008}%
  \BibitemOpen
  \bibfield  {author} {\bibinfo {author} {\bibfnamefont {A.}~\bibnamefont
  {Gozar}}, \bibinfo {author} {\bibfnamefont {G.}~\bibnamefont {Logvenov}},
  \bibinfo {author} {\bibfnamefont {L.~F.}\ \bibnamefont {Kourkoutis}},
  \bibinfo {author} {\bibfnamefont {A.~T.}\ \bibnamefont {Bollinger}}, \bibinfo
  {author} {\bibfnamefont {L.~A.}\ \bibnamefont {Giannuzzi}}, \bibinfo {author}
  {\bibfnamefont {D.~A.}\ \bibnamefont {Muller}}, \ and\ \bibinfo {author}
  {\bibfnamefont {I.}~\bibnamefont {Bozovic}},\ }\href {\doibase
  10.1038/nature07293} {\bibfield  {journal} {\bibinfo  {journal} {Nature}\
  }\textbf {\bibinfo {volume} {455}},\ \bibinfo {pages} {782} (\bibinfo {year}
  {2008})}\BibitemShut {NoStop}%
\bibitem [{\citenamefont {He}\ \emph {et~al.}(2013)\citenamefont {He},
  \citenamefont {He}, \citenamefont {Zhang}, \citenamefont {Zhao},
  \citenamefont {Liu}, \citenamefont {Liu}, \citenamefont {Mou}, \citenamefont
  {Ou}, \citenamefont {Wang}, \citenamefont {Li}, \citenamefont {Wang},
  \citenamefont {Peng}, \citenamefont {Liu}, \citenamefont {Chen},
  \citenamefont {Yu}, \citenamefont {Liu}, \citenamefont {Dong}, \citenamefont
  {Zhang}, \citenamefont {Chen}, \citenamefont {Xu}, \citenamefont {Chen},
  \citenamefont {Ma}, \citenamefont {Xue},\ and\ \citenamefont
  {Zhou}}]{he_phase_2013}%
  \BibitemOpen
  \bibfield  {author} {\bibinfo {author} {\bibfnamefont {S.}~\bibnamefont
  {He}}, \bibinfo {author} {\bibfnamefont {J.}~\bibnamefont {He}}, \bibinfo
  {author} {\bibfnamefont {W.}~\bibnamefont {Zhang}}, \bibinfo {author}
  {\bibfnamefont {L.}~\bibnamefont {Zhao}}, \bibinfo {author} {\bibfnamefont
  {D.}~\bibnamefont {Liu}}, \bibinfo {author} {\bibfnamefont {X.}~\bibnamefont
  {Liu}}, \bibinfo {author} {\bibfnamefont {D.}~\bibnamefont {Mou}}, \bibinfo
  {author} {\bibfnamefont {Y.-B.}\ \bibnamefont {Ou}}, \bibinfo {author}
  {\bibfnamefont {Q.-Y.}\ \bibnamefont {Wang}}, \bibinfo {author}
  {\bibfnamefont {Z.}~\bibnamefont {Li}}, \bibinfo {author} {\bibfnamefont
  {L.}~\bibnamefont {Wang}}, \bibinfo {author} {\bibfnamefont {Y.}~\bibnamefont
  {Peng}}, \bibinfo {author} {\bibfnamefont {Y.}~\bibnamefont {Liu}}, \bibinfo
  {author} {\bibfnamefont {C.}~\bibnamefont {Chen}}, \bibinfo {author}
  {\bibfnamefont {L.}~\bibnamefont {Yu}}, \bibinfo {author} {\bibfnamefont
  {G.}~\bibnamefont {Liu}}, \bibinfo {author} {\bibfnamefont {X.}~\bibnamefont
  {Dong}}, \bibinfo {author} {\bibfnamefont {J.}~\bibnamefont {Zhang}},
  \bibinfo {author} {\bibfnamefont {C.}~\bibnamefont {Chen}}, \bibinfo {author}
  {\bibfnamefont {Z.}~\bibnamefont {Xu}}, \bibinfo {author} {\bibfnamefont
  {X.}~\bibnamefont {Chen}}, \bibinfo {author} {\bibfnamefont {X.}~\bibnamefont
  {Ma}}, \bibinfo {author} {\bibfnamefont {Q.}~\bibnamefont {Xue}}, \ and\
  \bibinfo {author} {\bibfnamefont {X.~J.}\ \bibnamefont {Zhou}},\ }\href
  {\doibase 10.1038/nmat3648} {\bibfield  {journal} {\bibinfo  {journal}
  {Nature Materials}\ }\textbf {\bibinfo {volume} {12}},\ \bibinfo {pages}
  {605} (\bibinfo {year} {2013})}\BibitemShut {NoStop}%
\bibitem [{\citenamefont {Lee}\ \emph {et~al.}(2014)\citenamefont {Lee},
  \citenamefont {Schmitt}, \citenamefont {Moore}, \citenamefont {Johnston},
  \citenamefont {Cui}, \citenamefont {Li}, \citenamefont {Yi}, \citenamefont
  {Liu}, \citenamefont {Hashimoto}, \citenamefont {Zhang}, \citenamefont {Lu},
  \citenamefont {Devereaux}, \citenamefont {Lee},\ and\ \citenamefont
  {Shen}}]{lee_interfacial_2014}%
  \BibitemOpen
  \bibfield  {author} {\bibinfo {author} {\bibfnamefont {J.~J.}\ \bibnamefont
  {Lee}}, \bibinfo {author} {\bibfnamefont {F.~T.}\ \bibnamefont {Schmitt}},
  \bibinfo {author} {\bibfnamefont {R.~G.}\ \bibnamefont {Moore}}, \bibinfo
  {author} {\bibfnamefont {S.}~\bibnamefont {Johnston}}, \bibinfo {author}
  {\bibfnamefont {Y.-T.}\ \bibnamefont {Cui}}, \bibinfo {author} {\bibfnamefont
  {W.}~\bibnamefont {Li}}, \bibinfo {author} {\bibfnamefont {M.}~\bibnamefont
  {Yi}}, \bibinfo {author} {\bibfnamefont {Z.~K.}\ \bibnamefont {Liu}},
  \bibinfo {author} {\bibfnamefont {M.}~\bibnamefont {Hashimoto}}, \bibinfo
  {author} {\bibfnamefont {Y.}~\bibnamefont {Zhang}}, \bibinfo {author}
  {\bibfnamefont {D.~H.}\ \bibnamefont {Lu}}, \bibinfo {author} {\bibfnamefont
  {T.~P.}\ \bibnamefont {Devereaux}}, \bibinfo {author} {\bibfnamefont {D.-H.}\
  \bibnamefont {Lee}}, \ and\ \bibinfo {author} {\bibfnamefont {Z.-X.}\
  \bibnamefont {Shen}},\ }\href {\doibase 10.1038/nature13894} {\bibfield
  {journal} {\bibinfo  {journal} {Nature}\ }\textbf {\bibinfo {volume} {515}},\
  \bibinfo {pages} {245} (\bibinfo {year} {2014})}\BibitemShut {NoStop}%
\bibitem [{\citenamefont {Paik}\ \emph {et~al.}(2015)\citenamefont {Paik},
  \citenamefont {Moyer}, \citenamefont {Spila}, \citenamefont {Tashman},
  \citenamefont {Mundy}, \citenamefont {Freeman}, \citenamefont {Shukla},
  \citenamefont {Lapano}, \citenamefont {Engel-Herbert}, \citenamefont
  {Zander}, \citenamefont {Schubert}, \citenamefont {Muller}, \citenamefont
  {Datta}, \citenamefont {Schiffer},\ and\ \citenamefont
  {Schlom}}]{paik_transport_2015}%
  \BibitemOpen
  \bibfield  {author} {\bibinfo {author} {\bibfnamefont {H.}~\bibnamefont
  {Paik}}, \bibinfo {author} {\bibfnamefont {J.~A.}\ \bibnamefont {Moyer}},
  \bibinfo {author} {\bibfnamefont {T.}~\bibnamefont {Spila}}, \bibinfo
  {author} {\bibfnamefont {J.~W.}\ \bibnamefont {Tashman}}, \bibinfo {author}
  {\bibfnamefont {J.~A.}\ \bibnamefont {Mundy}}, \bibinfo {author}
  {\bibfnamefont {E.}~\bibnamefont {Freeman}}, \bibinfo {author} {\bibfnamefont
  {N.}~\bibnamefont {Shukla}}, \bibinfo {author} {\bibfnamefont {J.~M.}\
  \bibnamefont {Lapano}}, \bibinfo {author} {\bibfnamefont {R.}~\bibnamefont
  {Engel-Herbert}}, \bibinfo {author} {\bibfnamefont {W.}~\bibnamefont
  {Zander}}, \bibinfo {author} {\bibfnamefont {J.}~\bibnamefont {Schubert}},
  \bibinfo {author} {\bibfnamefont {D.~A.}\ \bibnamefont {Muller}}, \bibinfo
  {author} {\bibfnamefont {S.}~\bibnamefont {Datta}}, \bibinfo {author}
  {\bibfnamefont {P.}~\bibnamefont {Schiffer}}, \ and\ \bibinfo {author}
  {\bibfnamefont {D.~G.}\ \bibnamefont {Schlom}},\ }\href {\doibase
  10.1063/1.4932123} {\bibfield  {journal} {\bibinfo  {journal} {Applied
  Physics Letters}\ }\textbf {\bibinfo {volume} {107}},\ \bibinfo {pages}
  {163101} (\bibinfo {year} {2015})}\BibitemShut {NoStop}%
\bibitem [{\citenamefont {Yoshimatsu}\ \emph {et~al.}(2017)\citenamefont
  {Yoshimatsu}, \citenamefont {Sakata},\ and\ \citenamefont
  {Ohtomo}}]{yoshimatsu_superconductivity_2017}%
  \BibitemOpen
  \bibfield  {author} {\bibinfo {author} {\bibfnamefont {K.}~\bibnamefont
  {Yoshimatsu}}, \bibinfo {author} {\bibfnamefont {O.}~\bibnamefont {Sakata}},
  \ and\ \bibinfo {author} {\bibfnamefont {A.}~\bibnamefont {Ohtomo}},\ }\href
  {\doibase 10.1038/s41598-017-12815-4} {\bibfield  {journal} {\bibinfo
  {journal} {Scientific Reports}\ }\textbf {\bibinfo {volume} {7}},\ \bibinfo
  {pages} {12544} (\bibinfo {year} {2017})}\BibitemShut {NoStop}%
\bibitem [{\citenamefont {Goodenough}(1971)}]{goodenough_two_1971}%
  \BibitemOpen
  \bibfield  {author} {\bibinfo {author} {\bibfnamefont {J.~B.}\ \bibnamefont
  {Goodenough}},\ }\href {\doibase 10.1016/0022-4596(71)90091-0} {\bibfield
  {journal} {\bibinfo  {journal} {Journal of Solid State Chemistry}\ }\textbf
  {\bibinfo {volume} {3}},\ \bibinfo {pages} {490} (\bibinfo {year}
  {1971})}\BibitemShut {NoStop}%
\bibitem [{\citenamefont {Eyert}\ \emph {et~al.}(2000)\citenamefont {Eyert},
  \citenamefont {Horny}, \citenamefont {Höck},\ and\ \citenamefont
  {Horn}}]{eyert_embedded_2000}%
  \BibitemOpen
  \bibfield  {author} {\bibinfo {author} {\bibfnamefont {V.}~\bibnamefont
  {Eyert}}, \bibinfo {author} {\bibfnamefont {R.}~\bibnamefont {Horny}},
  \bibinfo {author} {\bibfnamefont {K.-H.}\ \bibnamefont {Höck}}, \ and\
  \bibinfo {author} {\bibfnamefont {S.}~\bibnamefont {Horn}},\ }\href {\doibase
  10.1088/0953-8984/12/23/303} {\bibfield  {journal} {\bibinfo  {journal}
  {Journal of Physics: Condensed Matter}\ }\textbf {\bibinfo {volume} {12}},\
  \bibinfo {pages} {4923} (\bibinfo {year} {2000})}\BibitemShut {NoStop}%
\bibitem [{\citenamefont {Mravlje}\ \emph {et~al.}(2011)\citenamefont
  {Mravlje}, \citenamefont {Aichhorn}, \citenamefont {Miyake}, \citenamefont
  {Haule}, \citenamefont {Kotliar},\ and\ \citenamefont
  {Georges}}]{mravlje_coherence-incoherence_2011}%
  \BibitemOpen
  \bibfield  {author} {\bibinfo {author} {\bibfnamefont {J.}~\bibnamefont
  {Mravlje}}, \bibinfo {author} {\bibfnamefont {M.}~\bibnamefont {Aichhorn}},
  \bibinfo {author} {\bibfnamefont {T.}~\bibnamefont {Miyake}}, \bibinfo
  {author} {\bibfnamefont {K.}~\bibnamefont {Haule}}, \bibinfo {author}
  {\bibfnamefont {G.}~\bibnamefont {Kotliar}}, \ and\ \bibinfo {author}
  {\bibfnamefont {A.}~\bibnamefont {Georges}},\ }\href {\doibase
  10.1103/PhysRevLett.106.096401} {\bibfield  {journal} {\bibinfo  {journal}
  {Physical Review Letters}\ }\textbf {\bibinfo {volume} {106}},\ \bibinfo
  {pages} {096401} (\bibinfo {year} {2011})}\BibitemShut {NoStop}%
\bibitem [{\citenamefont {Mackenzie}(1996)}]{mackenzie_quantum_1996}%
  \BibitemOpen
  \bibfield  {author} {\bibinfo {author} {\bibfnamefont {A.~P.}\ \bibnamefont
  {Mackenzie}},\ }\href {\doibase 10.1103/PhysRevLett.76.3786} {\bibfield
  {journal} {\bibinfo  {journal} {Physical Review Letters}\ }\textbf {\bibinfo
  {volume} {76}},\ \bibinfo {pages} {3786} (\bibinfo {year}
  {1996})}\BibitemShut {NoStop}%
\bibitem [{\citenamefont {Tamai}\ \emph {et~al.}(2019)\citenamefont {Tamai},
  \citenamefont {Zingl}, \citenamefont {Rozbicki}, \citenamefont {Cappelli},
  \citenamefont {Ricco}, \citenamefont {de~la Torre}, \citenamefont
  {McKeown~Walker}, \citenamefont {Bruno}, \citenamefont {King}, \citenamefont
  {Meevasana}, \citenamefont {Shi}, \citenamefont {Radovic}, \citenamefont
  {Plumb}, \citenamefont {Gibbs}, \citenamefont {Mackenzie}, \citenamefont
  {Berthod}, \citenamefont {Strand}, \citenamefont {Kim}, \citenamefont
  {Georges},\ and\ \citenamefont {Baumberger}}]{tamai_high-resolution_2019}%
  \BibitemOpen
  \bibfield  {author} {\bibinfo {author} {\bibfnamefont {A.}~\bibnamefont
  {Tamai}}, \bibinfo {author} {\bibfnamefont {M.}~\bibnamefont {Zingl}},
  \bibinfo {author} {\bibfnamefont {E.}~\bibnamefont {Rozbicki}}, \bibinfo
  {author} {\bibfnamefont {E.}~\bibnamefont {Cappelli}}, \bibinfo {author}
  {\bibfnamefont {S.}~\bibnamefont {Ricco}}, \bibinfo {author} {\bibfnamefont
  {A.}~\bibnamefont {de~la Torre}}, \bibinfo {author} {\bibfnamefont
  {S.}~\bibnamefont {McKeown~Walker}}, \bibinfo {author} {\bibfnamefont
  {F.}~\bibnamefont {Bruno}}, \bibinfo {author} {\bibfnamefont
  {P.}~\bibnamefont {King}}, \bibinfo {author} {\bibfnamefont {W.}~\bibnamefont
  {Meevasana}}, \bibinfo {author} {\bibfnamefont {M.}~\bibnamefont {Shi}},
  \bibinfo {author} {\bibfnamefont {M.}~\bibnamefont {Radovic}}, \bibinfo
  {author} {\bibfnamefont {N.}~\bibnamefont {Plumb}}, \bibinfo {author}
  {\bibfnamefont {A.}~\bibnamefont {Gibbs}}, \bibinfo {author} {\bibfnamefont
  {A.}~\bibnamefont {Mackenzie}}, \bibinfo {author} {\bibfnamefont
  {C.}~\bibnamefont {Berthod}}, \bibinfo {author} {\bibfnamefont
  {H.}~\bibnamefont {Strand}}, \bibinfo {author} {\bibfnamefont
  {M.}~\bibnamefont {Kim}}, \bibinfo {author} {\bibfnamefont {A.}~\bibnamefont
  {Georges}}, \ and\ \bibinfo {author} {\bibfnamefont {F.}~\bibnamefont
  {Baumberger}},\ }\href {\doibase 10.1103/PhysRevX.9.021048} {\bibfield
  {journal} {\bibinfo  {journal} {Physical Review X}\ }\textbf {\bibinfo
  {volume} {9}},\ \bibinfo {pages} {021048} (\bibinfo {year}
  {2019})}\BibitemShut {NoStop}%
\bibitem [{\citenamefont {Ricco}\ \emph {et~al.}(2018)\citenamefont {Ricco},
  \citenamefont {Kim}, \citenamefont {Tamai}, \citenamefont {Walker},
  \citenamefont {Bruno}, \citenamefont {Cucchi}, \citenamefont {Cappelli},
  \citenamefont {Besnard}, \citenamefont {Kim}, \citenamefont {Dudin},
  \citenamefont {Hoesch}, \citenamefont {Gutmann}, \citenamefont {Georges},
  \citenamefont {Perry},\ and\ \citenamefont {Baumberger}}]{ricco_situ_2018}%
  \BibitemOpen
  \bibfield  {author} {\bibinfo {author} {\bibfnamefont {S.}~\bibnamefont
  {Ricco}}, \bibinfo {author} {\bibfnamefont {M.}~\bibnamefont {Kim}}, \bibinfo
  {author} {\bibfnamefont {A.}~\bibnamefont {Tamai}}, \bibinfo {author}
  {\bibfnamefont {S.~M.}\ \bibnamefont {Walker}}, \bibinfo {author}
  {\bibfnamefont {F.~Y.}\ \bibnamefont {Bruno}}, \bibinfo {author}
  {\bibfnamefont {I.}~\bibnamefont {Cucchi}}, \bibinfo {author} {\bibfnamefont
  {E.}~\bibnamefont {Cappelli}}, \bibinfo {author} {\bibfnamefont
  {C.}~\bibnamefont {Besnard}}, \bibinfo {author} {\bibfnamefont {T.~K.}\
  \bibnamefont {Kim}}, \bibinfo {author} {\bibfnamefont {P.}~\bibnamefont
  {Dudin}}, \bibinfo {author} {\bibfnamefont {M.}~\bibnamefont {Hoesch}},
  \bibinfo {author} {\bibfnamefont {M.~J.}\ \bibnamefont {Gutmann}}, \bibinfo
  {author} {\bibfnamefont {A.}~\bibnamefont {Georges}}, \bibinfo {author}
  {\bibfnamefont {R.~S.}\ \bibnamefont {Perry}}, \ and\ \bibinfo {author}
  {\bibfnamefont {F.}~\bibnamefont {Baumberger}},\ }\href {\doibase
  10.1038/s41467-018-06945-0} {\bibfield  {journal} {\bibinfo  {journal}
  {Nature Communications}\ }\textbf {\bibinfo {volume} {9}},\ \bibinfo {pages}
  {4535} (\bibinfo {year} {2018})}\BibitemShut {NoStop}%
\bibitem [{\citenamefont {Sutter}\ \emph {et~al.}(2019)\citenamefont {Sutter},
  \citenamefont {Kim}, \citenamefont {Matt}, \citenamefont {Horio},
  \citenamefont {Fittipaldi}, \citenamefont {Vecchione}, \citenamefont
  {Granata}, \citenamefont {Hauser}, \citenamefont {Sassa}, \citenamefont
  {Gatti}, \citenamefont {Grioni}, \citenamefont {Hoesch}, \citenamefont {Kim},
  \citenamefont {Rienks}, \citenamefont {Plumb}, \citenamefont {Shi},
  \citenamefont {Neupert}, \citenamefont {Georges},\ and\ \citenamefont
  {Chang}}]{sutter_orbitally_2019}%
  \BibitemOpen
  \bibfield  {author} {\bibinfo {author} {\bibfnamefont {D.}~\bibnamefont
  {Sutter}}, \bibinfo {author} {\bibfnamefont {M.}~\bibnamefont {Kim}},
  \bibinfo {author} {\bibfnamefont {C.~E.}\ \bibnamefont {Matt}}, \bibinfo
  {author} {\bibfnamefont {M.}~\bibnamefont {Horio}}, \bibinfo {author}
  {\bibfnamefont {R.}~\bibnamefont {Fittipaldi}}, \bibinfo {author}
  {\bibfnamefont {A.}~\bibnamefont {Vecchione}}, \bibinfo {author}
  {\bibfnamefont {V.}~\bibnamefont {Granata}}, \bibinfo {author} {\bibfnamefont
  {K.}~\bibnamefont {Hauser}}, \bibinfo {author} {\bibfnamefont
  {Y.}~\bibnamefont {Sassa}}, \bibinfo {author} {\bibfnamefont
  {G.}~\bibnamefont {Gatti}}, \bibinfo {author} {\bibfnamefont
  {M.}~\bibnamefont {Grioni}}, \bibinfo {author} {\bibfnamefont
  {M.}~\bibnamefont {Hoesch}}, \bibinfo {author} {\bibfnamefont {T.~K.}\
  \bibnamefont {Kim}}, \bibinfo {author} {\bibfnamefont {E.}~\bibnamefont
  {Rienks}}, \bibinfo {author} {\bibfnamefont {N.~C.}\ \bibnamefont {Plumb}},
  \bibinfo {author} {\bibfnamefont {M.}~\bibnamefont {Shi}}, \bibinfo {author}
  {\bibfnamefont {T.}~\bibnamefont {Neupert}}, \bibinfo {author} {\bibfnamefont
  {A.}~\bibnamefont {Georges}}, \ and\ \bibinfo {author} {\bibfnamefont
  {J.}~\bibnamefont {Chang}},\ }\href {\doibase 10.1103/PhysRevB.99.121115}
  {\bibfield  {journal} {\bibinfo  {journal} {Physical Review B}\ }\textbf
  {\bibinfo {volume} {99}},\ \bibinfo {pages} {121115} (\bibinfo {year}
  {2019})}\BibitemShut {NoStop}%
\bibitem [{\citenamefont {Sutter}\ \emph {et~al.}(2017)\citenamefont {Sutter},
  \citenamefont {Fatuzzo}, \citenamefont {Moser}, \citenamefont {Kim},
  \citenamefont {Fittipaldi}, \citenamefont {Vecchione}, \citenamefont
  {Granata}, \citenamefont {Sassa}, \citenamefont {Cossalter}, \citenamefont
  {Gatti}, \citenamefont {Grioni}, \citenamefont {Rønnow}, \citenamefont
  {Plumb}, \citenamefont {Matt}, \citenamefont {Shi}, \citenamefont {Hoesch},
  \citenamefont {Kim}, \citenamefont {Chang}, \citenamefont {Jeng},
  \citenamefont {Jozwiak}, \citenamefont {Bostwick}, \citenamefont {Rotenberg},
  \citenamefont {Georges}, \citenamefont {Neupert},\ and\ \citenamefont
  {Chang}}]{sutter_hallmarks_2017}%
  \BibitemOpen
  \bibfield  {author} {\bibinfo {author} {\bibfnamefont {D.}~\bibnamefont
  {Sutter}}, \bibinfo {author} {\bibfnamefont {C.~G.}\ \bibnamefont {Fatuzzo}},
  \bibinfo {author} {\bibfnamefont {S.}~\bibnamefont {Moser}}, \bibinfo
  {author} {\bibfnamefont {M.}~\bibnamefont {Kim}}, \bibinfo {author}
  {\bibfnamefont {R.}~\bibnamefont {Fittipaldi}}, \bibinfo {author}
  {\bibfnamefont {A.}~\bibnamefont {Vecchione}}, \bibinfo {author}
  {\bibfnamefont {V.}~\bibnamefont {Granata}}, \bibinfo {author} {\bibfnamefont
  {Y.}~\bibnamefont {Sassa}}, \bibinfo {author} {\bibfnamefont
  {F.}~\bibnamefont {Cossalter}}, \bibinfo {author} {\bibfnamefont
  {G.}~\bibnamefont {Gatti}}, \bibinfo {author} {\bibfnamefont
  {M.}~\bibnamefont {Grioni}}, \bibinfo {author} {\bibfnamefont {H.~M.}\
  \bibnamefont {Rønnow}}, \bibinfo {author} {\bibfnamefont {N.~C.}\
  \bibnamefont {Plumb}}, \bibinfo {author} {\bibfnamefont {C.~E.}\ \bibnamefont
  {Matt}}, \bibinfo {author} {\bibfnamefont {M.}~\bibnamefont {Shi}}, \bibinfo
  {author} {\bibfnamefont {M.}~\bibnamefont {Hoesch}}, \bibinfo {author}
  {\bibfnamefont {T.~K.}\ \bibnamefont {Kim}}, \bibinfo {author} {\bibfnamefont
  {T.-R.}\ \bibnamefont {Chang}}, \bibinfo {author} {\bibfnamefont {H.-T.}\
  \bibnamefont {Jeng}}, \bibinfo {author} {\bibfnamefont {C.}~\bibnamefont
  {Jozwiak}}, \bibinfo {author} {\bibfnamefont {A.}~\bibnamefont {Bostwick}},
  \bibinfo {author} {\bibfnamefont {E.}~\bibnamefont {Rotenberg}}, \bibinfo
  {author} {\bibfnamefont {A.}~\bibnamefont {Georges}}, \bibinfo {author}
  {\bibfnamefont {T.}~\bibnamefont {Neupert}}, \ and\ \bibinfo {author}
  {\bibfnamefont {J.}~\bibnamefont {Chang}},\ }\href {\doibase
  10.1038/ncomms15176} {\bibfield  {journal} {\bibinfo  {journal} {Nature
  Communications}\ }\textbf {\bibinfo {volume} {8}},\ \bibinfo {pages} {15176}
  (\bibinfo {year} {2017})}\BibitemShut {NoStop}%
\bibitem [{\citenamefont {Jovic}\ \emph {et~al.}(2018)\citenamefont {Jovic},
  \citenamefont {Koch}, \citenamefont {Panda}, \citenamefont {Berger},
  \citenamefont {Bugnon}, \citenamefont {Magrez}, \citenamefont {Smith},
  \citenamefont {Biermann}, \citenamefont {Jozwiak}, \citenamefont {Bostwick},
  \citenamefont {Rotenberg},\ and\ \citenamefont {Moser}}]{jovic_dirac_2018}%
  \BibitemOpen
  \bibfield  {author} {\bibinfo {author} {\bibfnamefont {V.}~\bibnamefont
  {Jovic}}, \bibinfo {author} {\bibfnamefont {R.~J.}\ \bibnamefont {Koch}},
  \bibinfo {author} {\bibfnamefont {S.~K.}\ \bibnamefont {Panda}}, \bibinfo
  {author} {\bibfnamefont {H.}~\bibnamefont {Berger}}, \bibinfo {author}
  {\bibfnamefont {P.}~\bibnamefont {Bugnon}}, \bibinfo {author} {\bibfnamefont
  {A.}~\bibnamefont {Magrez}}, \bibinfo {author} {\bibfnamefont {K.~E.}\
  \bibnamefont {Smith}}, \bibinfo {author} {\bibfnamefont {S.}~\bibnamefont
  {Biermann}}, \bibinfo {author} {\bibfnamefont {C.}~\bibnamefont {Jozwiak}},
  \bibinfo {author} {\bibfnamefont {A.}~\bibnamefont {Bostwick}}, \bibinfo
  {author} {\bibfnamefont {E.}~\bibnamefont {Rotenberg}}, \ and\ \bibinfo
  {author} {\bibfnamefont {S.}~\bibnamefont {Moser}},\ }\href {\doibase
  10.1103/PhysRevB.98.241101} {\bibfield  {journal} {\bibinfo  {journal}
  {Physical Review B}\ }\textbf {\bibinfo {volume} {98}},\ \bibinfo {pages}
  {241101} (\bibinfo {year} {2018})}\BibitemShut {NoStop}%
\bibitem [{\citenamefont {Passenheim}\ and\ \citenamefont
  {McCollum}(1969)}]{passenheim_heat_1969}%
  \BibitemOpen
  \bibfield  {author} {\bibinfo {author} {\bibfnamefont {B.~C.}\ \bibnamefont
  {Passenheim}}\ and\ \bibinfo {author} {\bibfnamefont {D.~C.}\ \bibnamefont
  {McCollum}},\ }\href {\doibase 10.1063/1.1671725} {\bibfield  {journal}
  {\bibinfo  {journal} {The Journal of Chemical Physics}\ }\textbf {\bibinfo
  {volume} {51}},\ \bibinfo {pages} {320} (\bibinfo {year} {1969})}\BibitemShut
  {NoStop}%
\bibitem [{\citenamefont {Glassford}\ and\ \citenamefont
  {Chelikowsky}(1994)}]{glassford_electron_1994}%
  \BibitemOpen
  \bibfield  {author} {\bibinfo {author} {\bibfnamefont {K.~M.}\ \bibnamefont
  {Glassford}}\ and\ \bibinfo {author} {\bibfnamefont {J.~R.}\ \bibnamefont
  {Chelikowsky}},\ }\href {\doibase 10.1103/PhysRevB.49.7107} {\bibfield
  {journal} {\bibinfo  {journal} {Physical Review B}\ }\textbf {\bibinfo
  {volume} {49}},\ \bibinfo {pages} {7107} (\bibinfo {year}
  {1994})}\BibitemShut {NoStop}%
\bibitem [{\citenamefont {Smith}\ and\ \citenamefont
  {Chu}(1967)}]{smith_will_1967}%
  \BibitemOpen
  \bibfield  {author} {\bibinfo {author} {\bibfnamefont {T.~F.}\ \bibnamefont
  {Smith}}\ and\ \bibinfo {author} {\bibfnamefont {C.~W.}\ \bibnamefont
  {Chu}},\ }\href {\doibase 10.1103/PhysRev.159.353} {\bibfield  {journal}
  {\bibinfo  {journal} {Physical Review}\ }\textbf {\bibinfo {volume} {159}},\
  \bibinfo {pages} {353} (\bibinfo {year} {1967})}\BibitemShut {NoStop}%
\bibitem [{\citenamefont {Steppke}\ \emph {et~al.}(2017)\citenamefont
  {Steppke}, \citenamefont {Zhao}, \citenamefont {Barber}, \citenamefont
  {Scaffidi}, \citenamefont {Jerzembeck}, \citenamefont {Rosner}, \citenamefont
  {Gibbs}, \citenamefont {Maeno}, \citenamefont {Simon}, \citenamefont
  {Mackenzie},\ and\ \citenamefont {Hicks}}]{steppke_strong_2017}%
  \BibitemOpen
  \bibfield  {author} {\bibinfo {author} {\bibfnamefont {A.}~\bibnamefont
  {Steppke}}, \bibinfo {author} {\bibfnamefont {L.}~\bibnamefont {Zhao}},
  \bibinfo {author} {\bibfnamefont {M.~E.}\ \bibnamefont {Barber}}, \bibinfo
  {author} {\bibfnamefont {T.}~\bibnamefont {Scaffidi}}, \bibinfo {author}
  {\bibfnamefont {F.}~\bibnamefont {Jerzembeck}}, \bibinfo {author}
  {\bibfnamefont {H.}~\bibnamefont {Rosner}}, \bibinfo {author} {\bibfnamefont
  {A.~S.}\ \bibnamefont {Gibbs}}, \bibinfo {author} {\bibfnamefont
  {Y.}~\bibnamefont {Maeno}}, \bibinfo {author} {\bibfnamefont {S.~H.}\
  \bibnamefont {Simon}}, \bibinfo {author} {\bibfnamefont {A.~P.}\ \bibnamefont
  {Mackenzie}}, \ and\ \bibinfo {author} {\bibfnamefont {C.~W.}\ \bibnamefont
  {Hicks}},\ }\href {\doibase 10.1126/science.aaf9398} {\bibfield  {journal}
  {\bibinfo  {journal} {Science}\ }\textbf {\bibinfo {volume} {355}},\ \bibinfo
  {pages} {eaaf9398} (\bibinfo {year} {2017})}\BibitemShut {NoStop}%
\bibitem [{\citenamefont {Muraoka}\ and\ \citenamefont
  {Hiroi}(2002)}]{muraoka_metalinsulator_2002}%
  \BibitemOpen
  \bibfield  {author} {\bibinfo {author} {\bibfnamefont {Y.}~\bibnamefont
  {Muraoka}}\ and\ \bibinfo {author} {\bibfnamefont {Z.}~\bibnamefont
  {Hiroi}},\ }\href {\doibase 10.1063/1.1446215} {\bibfield  {journal}
  {\bibinfo  {journal} {Applied Physics Letters}\ }\textbf {\bibinfo {volume}
  {80}},\ \bibinfo {pages} {583} (\bibinfo {year} {2002})}\BibitemShut
  {NoStop}%
\bibitem [{\citenamefont {Aetukuri}\ \emph {et~al.}(2013)\citenamefont
  {Aetukuri}, \citenamefont {Gray}, \citenamefont {Drouard}, \citenamefont
  {Cossale}, \citenamefont {Gao}, \citenamefont {Reid}, \citenamefont
  {Kukreja}, \citenamefont {Ohldag}, \citenamefont {Jenkins}, \citenamefont
  {Arenholz}, \citenamefont {Roche}, \citenamefont {Dürr}, \citenamefont
  {Samant},\ and\ \citenamefont {Parkin}}]{aetukuri_control_2013}%
  \BibitemOpen
  \bibfield  {author} {\bibinfo {author} {\bibfnamefont {N.~B.}\ \bibnamefont
  {Aetukuri}}, \bibinfo {author} {\bibfnamefont {A.~X.}\ \bibnamefont {Gray}},
  \bibinfo {author} {\bibfnamefont {M.}~\bibnamefont {Drouard}}, \bibinfo
  {author} {\bibfnamefont {M.}~\bibnamefont {Cossale}}, \bibinfo {author}
  {\bibfnamefont {L.}~\bibnamefont {Gao}}, \bibinfo {author} {\bibfnamefont
  {A.~H.}\ \bibnamefont {Reid}}, \bibinfo {author} {\bibfnamefont
  {R.}~\bibnamefont {Kukreja}}, \bibinfo {author} {\bibfnamefont
  {H.}~\bibnamefont {Ohldag}}, \bibinfo {author} {\bibfnamefont {C.~A.}\
  \bibnamefont {Jenkins}}, \bibinfo {author} {\bibfnamefont {E.}~\bibnamefont
  {Arenholz}}, \bibinfo {author} {\bibfnamefont {K.~P.}\ \bibnamefont {Roche}},
  \bibinfo {author} {\bibfnamefont {H.~A.}\ \bibnamefont {Dürr}}, \bibinfo
  {author} {\bibfnamefont {M.~G.}\ \bibnamefont {Samant}}, \ and\ \bibinfo
  {author} {\bibfnamefont {S.~S.~P.}\ \bibnamefont {Parkin}},\ }\href {\doibase
  10.1038/nphys2733} {\bibfield  {journal} {\bibinfo  {journal} {Nature
  Physics}\ }\textbf {\bibinfo {volume} {9}},\ \bibinfo {pages} {661} (\bibinfo
  {year} {2013})}\BibitemShut {NoStop}%
\bibitem [{\citenamefont {Alves}\ \emph {et~al.}(2010)\citenamefont {Alves},
  \citenamefont {Damasceno}, \citenamefont {dos Santos}, \citenamefont
  {Bortolozo}, \citenamefont {Suzuki}, \citenamefont {Izario~Filho},
  \citenamefont {Machado},\ and\ \citenamefont
  {Fisk}}]{alves_unconventional_2010}%
  \BibitemOpen
  \bibfield  {author} {\bibinfo {author} {\bibfnamefont {L.~M.~S.}\
  \bibnamefont {Alves}}, \bibinfo {author} {\bibfnamefont {V.~I.}\ \bibnamefont
  {Damasceno}}, \bibinfo {author} {\bibfnamefont {C.~A.~M.}\ \bibnamefont {dos
  Santos}}, \bibinfo {author} {\bibfnamefont {A.~D.}\ \bibnamefont
  {Bortolozo}}, \bibinfo {author} {\bibfnamefont {P.~A.}\ \bibnamefont
  {Suzuki}}, \bibinfo {author} {\bibfnamefont {H.~J.}\ \bibnamefont
  {Izario~Filho}}, \bibinfo {author} {\bibfnamefont {A.~J.~S.}\ \bibnamefont
  {Machado}}, \ and\ \bibinfo {author} {\bibfnamefont {Z.}~\bibnamefont
  {Fisk}},\ }\href {\doibase 10.1103/PhysRevB.81.174532} {\bibfield  {journal}
  {\bibinfo  {journal} {Physical Review B}\ }\textbf {\bibinfo {volume} {81}},\
  \bibinfo {pages} {174532} (\bibinfo {year} {2010})}\BibitemShut {NoStop}%
\bibitem [{\citenamefont {Alves}\ \emph {et~al.}(2012)\citenamefont {Alves},
  \citenamefont {dos Santos}, \citenamefont {Benaion}, \citenamefont {Machado},
  \citenamefont {de~Lima}, \citenamefont {Neumeier}, \citenamefont {Marques},
  \citenamefont {Aguiar}, \citenamefont {Mossanek},\ and\ \citenamefont
  {Abbate}}]{alves_superconductivity_2012}%
  \BibitemOpen
  \bibfield  {author} {\bibinfo {author} {\bibfnamefont {L.~M.~S.}\
  \bibnamefont {Alves}}, \bibinfo {author} {\bibfnamefont {C.~a.~M.}\
  \bibnamefont {dos Santos}}, \bibinfo {author} {\bibfnamefont {S.~S.}\
  \bibnamefont {Benaion}}, \bibinfo {author} {\bibfnamefont {A.~J.~S.}\
  \bibnamefont {Machado}}, \bibinfo {author} {\bibfnamefont {B.~S.}\
  \bibnamefont {de~Lima}}, \bibinfo {author} {\bibfnamefont {J.~J.}\
  \bibnamefont {Neumeier}}, \bibinfo {author} {\bibfnamefont {M.~D.~R.}\
  \bibnamefont {Marques}}, \bibinfo {author} {\bibfnamefont {J.~A.}\
  \bibnamefont {Aguiar}}, \bibinfo {author} {\bibfnamefont {R.~J.~O.}\
  \bibnamefont {Mossanek}}, \ and\ \bibinfo {author} {\bibfnamefont
  {M.}~\bibnamefont {Abbate}},\ }\href {\doibase 10.1063/1.4757003} {\bibfield
  {journal} {\bibinfo  {journal} {Journal of Applied Physics}\ }\textbf
  {\bibinfo {volume} {112}},\ \bibinfo {pages} {073923} (\bibinfo {year}
  {2012})}\BibitemShut {NoStop}%
\bibitem [{\citenamefont {Parker}\ \emph {et~al.}(2014)\citenamefont {Parker},
  \citenamefont {Idrobo}, \citenamefont {Cantoni},\ and\ \citenamefont
  {Sefat}}]{parker_evidence_2014}%
  \BibitemOpen
  \bibfield  {author} {\bibinfo {author} {\bibfnamefont {D.}~\bibnamefont
  {Parker}}, \bibinfo {author} {\bibfnamefont {J.~C.}\ \bibnamefont {Idrobo}},
  \bibinfo {author} {\bibfnamefont {C.}~\bibnamefont {Cantoni}}, \ and\
  \bibinfo {author} {\bibfnamefont {A.~S.}\ \bibnamefont {Sefat}},\ }\href
  {\doibase 10.1103/PhysRevB.90.054505} {\bibfield  {journal} {\bibinfo
  {journal} {Physical Review B}\ }\textbf {\bibinfo {volume} {90}},\ \bibinfo
  {pages} {054505} (\bibinfo {year} {2014})}\BibitemShut {NoStop}%
\bibitem [{\citenamefont {Yamamoto}\ \emph {et~al.}(2005)\citenamefont
  {Yamamoto}, \citenamefont {Nakajima}, \citenamefont {Ohsawa}, \citenamefont
  {Matsumoto},\ and\ \citenamefont {Koinuma}}]{yamamoto_preparation_2005}%
  \BibitemOpen
  \bibfield  {author} {\bibinfo {author} {\bibfnamefont {Y.}~\bibnamefont
  {Yamamoto}}, \bibinfo {author} {\bibfnamefont {K.}~\bibnamefont {Nakajima}},
  \bibinfo {author} {\bibfnamefont {T.}~\bibnamefont {Ohsawa}}, \bibinfo
  {author} {\bibfnamefont {Y.}~\bibnamefont {Matsumoto}}, \ and\ \bibinfo
  {author} {\bibfnamefont {H.}~\bibnamefont {Koinuma}},\ }\href {\doibase
  10.1143/JJAP.44.L511} {\bibfield  {journal} {\bibinfo  {journal} {Japanese
  Journal of Applied Physics}\ }\textbf {\bibinfo {volume} {44}},\ \bibinfo
  {pages} {L511} (\bibinfo {year} {2005})}\BibitemShut {NoStop}%
\bibitem [{\citenamefont {He}\ \emph {et~al.}(2015)\citenamefont {He},
  \citenamefont {Langsdorf}, \citenamefont {Li},\ and\ \citenamefont
  {Over}}]{he_versatile_2015}%
  \BibitemOpen
  \bibfield  {author} {\bibinfo {author} {\bibfnamefont {Y.}~\bibnamefont
  {He}}, \bibinfo {author} {\bibfnamefont {D.}~\bibnamefont {Langsdorf}},
  \bibinfo {author} {\bibfnamefont {L.}~\bibnamefont {Li}}, \ and\ \bibinfo
  {author} {\bibfnamefont {H.}~\bibnamefont {Over}},\ }\href {\doibase
  10.1021/jp5121405} {\bibfield  {journal} {\bibinfo  {journal} {The Journal of
  Physical Chemistry C}\ }\textbf {\bibinfo {volume} {119}},\ \bibinfo {pages}
  {2692} (\bibinfo {year} {2015})}\BibitemShut {NoStop}%
\bibitem [{\citenamefont {Giannozzi}\ \emph {et~al.}(2009)\citenamefont
  {Giannozzi}, \citenamefont {Baroni}, \citenamefont {Bonini}, \citenamefont
  {Calandra}, \citenamefont {Car}, \citenamefont {Cavazzoni}, \citenamefont
  {Ceresoli}, \citenamefont {Chiarotti}, \citenamefont {Cococcioni},
  \citenamefont {Dabo}, \citenamefont {Corso}, \citenamefont {Gironcoli},
  \citenamefont {Fabris}, \citenamefont {Fratesi}, \citenamefont {Gebauer},
  \citenamefont {Gerstmann}, \citenamefont {Gougoussis}, \citenamefont
  {Kokalj}, \citenamefont {Lazzeri}, \citenamefont {Martin-Samos},
  \citenamefont {Marzari}, \citenamefont {Mauri}, \citenamefont {Mazzarello},
  \citenamefont {Paolini}, \citenamefont {Pasquarello}, \citenamefont
  {Paulatto}, \citenamefont {Sbraccia}, \citenamefont {Scandolo}, \citenamefont
  {Sclauzero}, \citenamefont {Seitsonen}, \citenamefont {Smogunov},
  \citenamefont {Umari},\ and\ \citenamefont
  {Wentzcovitch}}]{giannozzi_quantum_2009}%
  \BibitemOpen
  \bibfield  {author} {\bibinfo {author} {\bibfnamefont {P.}~\bibnamefont
  {Giannozzi}}, \bibinfo {author} {\bibfnamefont {S.}~\bibnamefont {Baroni}},
  \bibinfo {author} {\bibfnamefont {N.}~\bibnamefont {Bonini}}, \bibinfo
  {author} {\bibfnamefont {M.}~\bibnamefont {Calandra}}, \bibinfo {author}
  {\bibfnamefont {R.}~\bibnamefont {Car}}, \bibinfo {author} {\bibfnamefont
  {C.}~\bibnamefont {Cavazzoni}}, \bibinfo {author} {\bibfnamefont
  {D.}~\bibnamefont {Ceresoli}}, \bibinfo {author} {\bibfnamefont {G.~L.}\
  \bibnamefont {Chiarotti}}, \bibinfo {author} {\bibfnamefont {M.}~\bibnamefont
  {Cococcioni}}, \bibinfo {author} {\bibfnamefont {I.}~\bibnamefont {Dabo}},
  \bibinfo {author} {\bibfnamefont {A.~D.}\ \bibnamefont {Corso}}, \bibinfo
  {author} {\bibfnamefont {S.~d.}\ \bibnamefont {Gironcoli}}, \bibinfo {author}
  {\bibfnamefont {S.}~\bibnamefont {Fabris}}, \bibinfo {author} {\bibfnamefont
  {G.}~\bibnamefont {Fratesi}}, \bibinfo {author} {\bibfnamefont
  {R.}~\bibnamefont {Gebauer}}, \bibinfo {author} {\bibfnamefont
  {U.}~\bibnamefont {Gerstmann}}, \bibinfo {author} {\bibfnamefont
  {C.}~\bibnamefont {Gougoussis}}, \bibinfo {author} {\bibfnamefont
  {A.}~\bibnamefont {Kokalj}}, \bibinfo {author} {\bibfnamefont
  {M.}~\bibnamefont {Lazzeri}}, \bibinfo {author} {\bibfnamefont
  {L.}~\bibnamefont {Martin-Samos}}, \bibinfo {author} {\bibfnamefont
  {N.}~\bibnamefont {Marzari}}, \bibinfo {author} {\bibfnamefont
  {F.}~\bibnamefont {Mauri}}, \bibinfo {author} {\bibfnamefont
  {R.}~\bibnamefont {Mazzarello}}, \bibinfo {author} {\bibfnamefont
  {S.}~\bibnamefont {Paolini}}, \bibinfo {author} {\bibfnamefont
  {A.}~\bibnamefont {Pasquarello}}, \bibinfo {author} {\bibfnamefont
  {L.}~\bibnamefont {Paulatto}}, \bibinfo {author} {\bibfnamefont
  {C.}~\bibnamefont {Sbraccia}}, \bibinfo {author} {\bibfnamefont
  {S.}~\bibnamefont {Scandolo}}, \bibinfo {author} {\bibfnamefont
  {G.}~\bibnamefont {Sclauzero}}, \bibinfo {author} {\bibfnamefont {A.~P.}\
  \bibnamefont {Seitsonen}}, \bibinfo {author} {\bibfnamefont {A.}~\bibnamefont
  {Smogunov}}, \bibinfo {author} {\bibfnamefont {P.}~\bibnamefont {Umari}}, \
  and\ \bibinfo {author} {\bibfnamefont {R.~M.}\ \bibnamefont {Wentzcovitch}},\
  }\href {\doibase 10.1088/0953-8984/21/39/395502} {\bibfield  {journal}
  {\bibinfo  {journal} {Journal of Physics: Condensed Matter}\ }\textbf
  {\bibinfo {volume} {21}},\ \bibinfo {pages} {395502} (\bibinfo {year}
  {2009})}\BibitemShut {NoStop}%
\bibitem [{\citenamefont {Giannozzi}\ \emph {et~al.}(2017)\citenamefont
  {Giannozzi}, \citenamefont {Andreussi}, \citenamefont {Brumme}, \citenamefont
  {Bunau}, \citenamefont {Nardelli}, \citenamefont {Calandra}, \citenamefont
  {Car}, \citenamefont {Cavazzoni}, \citenamefont {Ceresoli}, \citenamefont
  {Cococcioni}, \citenamefont {Colonna}, \citenamefont {Carnimeo},
  \citenamefont {Corso}, \citenamefont {Gironcoli}, \citenamefont {Delugas},
  \citenamefont {DiStasio}, \citenamefont {Ferretti}, \citenamefont {Floris},
  \citenamefont {Fratesi}, \citenamefont {Fugallo}, \citenamefont {Gebauer},
  \citenamefont {Gerstmann}, \citenamefont {Giustino}, \citenamefont {Gorni},
  \citenamefont {Jia}, \citenamefont {Kawamura}, \citenamefont {Ko},
  \citenamefont {Kokalj}, \citenamefont {Küçükbenli}, \citenamefont
  {Lazzeri}, \citenamefont {Marsili}, \citenamefont {Marzari}, \citenamefont
  {Mauri}, \citenamefont {Nguyen}, \citenamefont {Nguyen}, \citenamefont
  {Otero-de-la Roza}, \citenamefont {Paulatto}, \citenamefont {Poncé},
  \citenamefont {Rocca}, \citenamefont {Sabatini}, \citenamefont {Santra},
  \citenamefont {Schlipf}, \citenamefont {Seitsonen}, \citenamefont {Smogunov},
  \citenamefont {Timrov}, \citenamefont {Thonhauser}, \citenamefont {Umari},
  \citenamefont {Vast}, \citenamefont {Wu},\ and\ \citenamefont
  {Baroni}}]{giannozzi_advanced_2017}%
  \BibitemOpen
  \bibfield  {author} {\bibinfo {author} {\bibfnamefont {P.}~\bibnamefont
  {Giannozzi}}, \bibinfo {author} {\bibfnamefont {O.}~\bibnamefont
  {Andreussi}}, \bibinfo {author} {\bibfnamefont {T.}~\bibnamefont {Brumme}},
  \bibinfo {author} {\bibfnamefont {O.}~\bibnamefont {Bunau}}, \bibinfo
  {author} {\bibfnamefont {M.~B.}\ \bibnamefont {Nardelli}}, \bibinfo {author}
  {\bibfnamefont {M.}~\bibnamefont {Calandra}}, \bibinfo {author}
  {\bibfnamefont {R.}~\bibnamefont {Car}}, \bibinfo {author} {\bibfnamefont
  {C.}~\bibnamefont {Cavazzoni}}, \bibinfo {author} {\bibfnamefont
  {D.}~\bibnamefont {Ceresoli}}, \bibinfo {author} {\bibfnamefont
  {M.}~\bibnamefont {Cococcioni}}, \bibinfo {author} {\bibfnamefont
  {N.}~\bibnamefont {Colonna}}, \bibinfo {author} {\bibfnamefont
  {I.}~\bibnamefont {Carnimeo}}, \bibinfo {author} {\bibfnamefont {A.~D.}\
  \bibnamefont {Corso}}, \bibinfo {author} {\bibfnamefont {S.~d.}\ \bibnamefont
  {Gironcoli}}, \bibinfo {author} {\bibfnamefont {P.}~\bibnamefont {Delugas}},
  \bibinfo {author} {\bibfnamefont {R.~A.}\ \bibnamefont {DiStasio}}, \bibinfo
  {author} {\bibfnamefont {A.}~\bibnamefont {Ferretti}}, \bibinfo {author}
  {\bibfnamefont {A.}~\bibnamefont {Floris}}, \bibinfo {author} {\bibfnamefont
  {G.}~\bibnamefont {Fratesi}}, \bibinfo {author} {\bibfnamefont
  {G.}~\bibnamefont {Fugallo}}, \bibinfo {author} {\bibfnamefont
  {R.}~\bibnamefont {Gebauer}}, \bibinfo {author} {\bibfnamefont
  {U.}~\bibnamefont {Gerstmann}}, \bibinfo {author} {\bibfnamefont
  {F.}~\bibnamefont {Giustino}}, \bibinfo {author} {\bibfnamefont
  {T.}~\bibnamefont {Gorni}}, \bibinfo {author} {\bibfnamefont
  {J.}~\bibnamefont {Jia}}, \bibinfo {author} {\bibfnamefont {M.}~\bibnamefont
  {Kawamura}}, \bibinfo {author} {\bibfnamefont {H.-Y.}\ \bibnamefont {Ko}},
  \bibinfo {author} {\bibfnamefont {A.}~\bibnamefont {Kokalj}}, \bibinfo
  {author} {\bibfnamefont {E.}~\bibnamefont {Küçükbenli}}, \bibinfo {author}
  {\bibfnamefont {M.}~\bibnamefont {Lazzeri}}, \bibinfo {author} {\bibfnamefont
  {M.}~\bibnamefont {Marsili}}, \bibinfo {author} {\bibfnamefont
  {N.}~\bibnamefont {Marzari}}, \bibinfo {author} {\bibfnamefont
  {F.}~\bibnamefont {Mauri}}, \bibinfo {author} {\bibfnamefont {N.~L.}\
  \bibnamefont {Nguyen}}, \bibinfo {author} {\bibfnamefont {H.-V.}\
  \bibnamefont {Nguyen}}, \bibinfo {author} {\bibfnamefont {A.}~\bibnamefont
  {Otero-de-la Roza}}, \bibinfo {author} {\bibfnamefont {L.}~\bibnamefont
  {Paulatto}}, \bibinfo {author} {\bibfnamefont {S.}~\bibnamefont {Poncé}},
  \bibinfo {author} {\bibfnamefont {D.}~\bibnamefont {Rocca}}, \bibinfo
  {author} {\bibfnamefont {R.}~\bibnamefont {Sabatini}}, \bibinfo {author}
  {\bibfnamefont {B.}~\bibnamefont {Santra}}, \bibinfo {author} {\bibfnamefont
  {M.}~\bibnamefont {Schlipf}}, \bibinfo {author} {\bibfnamefont {A.~P.}\
  \bibnamefont {Seitsonen}}, \bibinfo {author} {\bibfnamefont {A.}~\bibnamefont
  {Smogunov}}, \bibinfo {author} {\bibfnamefont {I.}~\bibnamefont {Timrov}},
  \bibinfo {author} {\bibfnamefont {T.}~\bibnamefont {Thonhauser}}, \bibinfo
  {author} {\bibfnamefont {P.}~\bibnamefont {Umari}}, \bibinfo {author}
  {\bibfnamefont {N.}~\bibnamefont {Vast}}, \bibinfo {author} {\bibfnamefont
  {X.}~\bibnamefont {Wu}}, \ and\ \bibinfo {author} {\bibfnamefont
  {S.}~\bibnamefont {Baroni}},\ }\href {\doibase 10.1088/1361-648X/aa8f79}
  {\bibfield  {journal} {\bibinfo  {journal} {Journal of Physics: Condensed
  Matter}\ }\textbf {\bibinfo {volume} {29}},\ \bibinfo {pages} {465901}
  (\bibinfo {year} {2017})}\BibitemShut {NoStop}%
\bibitem [{\citenamefont {Dal~Corso}(2014)}]{dal_corso_pseudopotentials_2014}%
  \BibitemOpen
  \bibfield  {author} {\bibinfo {author} {\bibfnamefont {A.}~\bibnamefont
  {Dal~Corso}},\ }\href {\doibase 10.1016/j.commatsci.2014.07.043} {\bibfield
  {journal} {\bibinfo  {journal} {Computational Materials Science}\ }\textbf
  {\bibinfo {volume} {95}},\ \bibinfo {pages} {337} (\bibinfo {year}
  {2014})}\BibitemShut {NoStop}%
\bibitem [{\citenamefont {Perdew}\ \emph {et~al.}(1996)\citenamefont {Perdew},
  \citenamefont {Burke},\ and\ \citenamefont {Ernzerhof}}]{PBE_1996}%
  \BibitemOpen
  \bibfield  {author} {\bibinfo {author} {\bibfnamefont {J.~P.}\ \bibnamefont
  {Perdew}}, \bibinfo {author} {\bibfnamefont {K.}~\bibnamefont {Burke}}, \
  and\ \bibinfo {author} {\bibfnamefont {M.}~\bibnamefont {Ernzerhof}},\ }\href
  {\doibase 10.1103/PhysRevLett.77.3865} {\bibfield  {journal} {\bibinfo
  {journal} {Physical Review Letters}\ }\textbf {\bibinfo {volume} {77}},\
  \bibinfo {pages} {3865} (\bibinfo {year} {1996})}\BibitemShut {NoStop}%
\bibitem [{\citenamefont {Mostofi}\ \emph {et~al.}(2014)\citenamefont
  {Mostofi}, \citenamefont {Yates}, \citenamefont {Pizzi}, \citenamefont {Lee},
  \citenamefont {Souza}, \citenamefont {Vanderbilt},\ and\ \citenamefont
  {Marzari}}]{Wannier90_2014}%
  \BibitemOpen
  \bibfield  {author} {\bibinfo {author} {\bibfnamefont {A.~A.}\ \bibnamefont
  {Mostofi}}, \bibinfo {author} {\bibfnamefont {J.~R.}\ \bibnamefont {Yates}},
  \bibinfo {author} {\bibfnamefont {G.}~\bibnamefont {Pizzi}}, \bibinfo
  {author} {\bibfnamefont {Y.-S.}\ \bibnamefont {Lee}}, \bibinfo {author}
  {\bibfnamefont {I.}~\bibnamefont {Souza}}, \bibinfo {author} {\bibfnamefont
  {D.}~\bibnamefont {Vanderbilt}}, \ and\ \bibinfo {author} {\bibfnamefont
  {N.}~\bibnamefont {Marzari}},\ }\href {\doibase 10.1016/j.cpc.2014.05.003}
  {\bibfield  {journal} {\bibinfo  {journal} {Computer Physics Communications}\
  }\textbf {\bibinfo {volume} {185}},\ \bibinfo {pages} {2309} (\bibinfo {year}
  {2014})}\BibitemShut {NoStop}%
\bibitem [{\citenamefont {Eyert}(2002)}]{Eyert_2002}%
  \BibitemOpen
  \bibfield  {author} {\bibinfo {author} {\bibfnamefont {V.}~\bibnamefont
  {Eyert}},\ }\href {\doibase 10.1002/1521-3889} {\bibfield  {journal}
  {\bibinfo  {journal} {Annalen der Physik}\ }\textbf {\bibinfo {volume}
  {11}},\ \bibinfo {pages} {650} (\bibinfo {year} {2002})}\BibitemShut
  {NoStop}%
\bibitem [{\citenamefont {Stokes}\ \emph {et~al.}()\citenamefont {Stokes},
  \citenamefont {Hatch},\ and\ \citenamefont {Campbell}}]{isotropy}%
  \BibitemOpen
  \bibfield  {author} {\bibinfo {author} {\bibfnamefont {H.~T.}\ \bibnamefont
  {Stokes}}, \bibinfo {author} {\bibfnamefont {D.~M.}\ \bibnamefont {Hatch}}, \
  and\ \bibinfo {author} {\bibfnamefont {B.~J.}\ \bibnamefont {Campbell}},\
  }\href {iso.byu.edu} {\enquote {\bibinfo {title} {Isotropy software suite},}\
  }\BibitemShut {NoStop}%
\bibitem [{\citenamefont {Momma}\ and\ \citenamefont
  {Izumi}(2011)}]{momma_vesta_2011}%
  \BibitemOpen
  \bibfield  {author} {\bibinfo {author} {\bibfnamefont {K.}~\bibnamefont
  {Momma}}\ and\ \bibinfo {author} {\bibfnamefont {F.}~\bibnamefont {Izumi}},\
  }\href {\doibase 10.1107/S0021889811038970} {\bibfield  {journal} {\bibinfo
  {journal} {Journal of Applied Crystallography}\ }\textbf {\bibinfo {volume}
  {44}},\ \bibinfo {pages} {1272} (\bibinfo {year} {2011})}\BibitemShut
  {NoStop}%
\bibitem [{\citenamefont {Perdew}\ \emph {et~al.}(2008)\citenamefont {Perdew},
  \citenamefont {Ruzsinszky}, \citenamefont {Csonka}, \citenamefont {Vydrov},
  \citenamefont {Scuseria}, \citenamefont {Constantin}, \citenamefont {Zhou},\
  and\ \citenamefont {Burke}}]{perdew_restoring_2008}%
  \BibitemOpen
  \bibfield  {author} {\bibinfo {author} {\bibfnamefont {J.~P.}\ \bibnamefont
  {Perdew}}, \bibinfo {author} {\bibfnamefont {A.}~\bibnamefont {Ruzsinszky}},
  \bibinfo {author} {\bibfnamefont {G.~I.}\ \bibnamefont {Csonka}}, \bibinfo
  {author} {\bibfnamefont {O.~A.}\ \bibnamefont {Vydrov}}, \bibinfo {author}
  {\bibfnamefont {G.~E.}\ \bibnamefont {Scuseria}}, \bibinfo {author}
  {\bibfnamefont {L.~A.}\ \bibnamefont {Constantin}}, \bibinfo {author}
  {\bibfnamefont {X.}~\bibnamefont {Zhou}}, \ and\ \bibinfo {author}
  {\bibfnamefont {K.}~\bibnamefont {Burke}},\ }\href {\doibase
  10.1103/PhysRevLett.100.136406} {\bibfield  {journal} {\bibinfo  {journal}
  {Physical Review Letters}\ }\textbf {\bibinfo {volume} {100}},\ \bibinfo
  {pages} {136406} (\bibinfo {year} {2008})}\BibitemShut {NoStop}%
\bibitem [{\citenamefont {Poncé}\ \emph {et~al.}(2016)\citenamefont {Poncé},
  \citenamefont {Margine}, \citenamefont {Verdi},\ and\ \citenamefont
  {Giustino}}]{ponce_epw_2016}%
  \BibitemOpen
  \bibfield  {author} {\bibinfo {author} {\bibfnamefont {S.}~\bibnamefont
  {Poncé}}, \bibinfo {author} {\bibfnamefont {E.~R.}\ \bibnamefont {Margine}},
  \bibinfo {author} {\bibfnamefont {C.}~\bibnamefont {Verdi}}, \ and\ \bibinfo
  {author} {\bibfnamefont {F.}~\bibnamefont {Giustino}},\ }\href {\doibase
  10.1016/j.cpc.2016.07.028} {\bibfield  {journal} {\bibinfo  {journal}
  {Computer Physics Communications}\ }\textbf {\bibinfo {volume} {209}},\
  \bibinfo {pages} {116} (\bibinfo {year} {2016})}\BibitemShut {NoStop}%
\bibitem [{\citenamefont {McMillan}(1968)}]{mcmillan_transition_1968}%
  \BibitemOpen
  \bibfield  {author} {\bibinfo {author} {\bibfnamefont {W.~L.}\ \bibnamefont
  {McMillan}},\ }\href {\doibase 10.1103/PhysRev.167.331} {\bibfield  {journal}
  {\bibinfo  {journal} {Physical Review}\ }\textbf {\bibinfo {volume} {167}},\
  \bibinfo {pages} {331} (\bibinfo {year} {1968})}\BibitemShut {NoStop}%
\bibitem [{\citenamefont {Allen}\ and\ \citenamefont
  {Dynes}(1975)}]{allen_transition_1975}%
  \BibitemOpen
  \bibfield  {author} {\bibinfo {author} {\bibfnamefont {P.~B.}\ \bibnamefont
  {Allen}}\ and\ \bibinfo {author} {\bibfnamefont {R.~C.}\ \bibnamefont
  {Dynes}},\ }\href {\doibase 10.1103/PhysRevB.12.905} {\bibfield  {journal}
  {\bibinfo  {journal} {Physical Review B}\ }\textbf {\bibinfo {volume} {12}},\
  \bibinfo {pages} {905} (\bibinfo {year} {1975})}\BibitemShut {NoStop}%
\end{thebibliography}
\end{document}

\newcommand{\physicsDept}{Department of Physics, Laboratory of Atomic and Solid State Physics, Cornell University, Ithaca, NY 14853, USA}
\newcommand{\paradim}{Platform for the Accelerated Realization, Analysis, and Discovery of Interface Materials (PARADIM), Cornell University, Ithaca, New York 14853, USA}
\newcommand{\mseDept}{Department of Materials Science and Engineering, Cornell University, Ithaca, NY 14853, USA}
\newcommand{\aepDept}{School of Applied and Engineering Physics, Cornell University, Ithaca, New York 14853, USA}
\newcommand{\kavli}{Kavli Institute at Cornell for Nanoscale Science, Ithaca, NY 14853, USA}
\newcommand{\RO}{RuO$_2$ }
\newcommand{\ROnospace}{RuO$_2$}
\newcommand{\ROA}{RuO$_2$(110) }
\newcommand{\ROAnospace}{RuO$_2$(110)}
\newcommand{\ROB}{RuO$_2$(101) }
\newcommand{\ROBnospace}{RuO$_2$(101)}
\newcommand{\TO}{TiO$_2$ }
\newcommand{\TOnospace}{TiO$_2$}
\newcommand{\TC}{$T_c$ }
\newcommand{\TCnospace}{$T_c$}
\newcommand{\HC}{$H_c$ }
\newcommand{\HCnospace}{$H_c$}
\newcommand{\ang}{\text{\normalfont\AA}}
\newcommand{\dpar}{$d_{||}$ }
\newcommand{\dparnospace}{$d_{||}$}
\newcommand{\dxzyz}{$(d_{\text{xz}}$, $d_{\text{yz}})$ }
\newcommand{\dxzyznospace}{$(d_{\text{xz}}$, $d_{\text{yz}})$}

\title{Strain-stabilized superconductivity \\ ~ \\SUPPLEMENTAL INFORMATION}

\author{J. P. Ruf}
\affiliation{\physicsDept}
\author{H. Paik}
\affiliation{\paradim}
\author{N. J. Schreiber}
\affiliation{\mseDept}
\author{H. P. Nair}
\affiliation{\mseDept}
\author{L. Miao}
\affiliation{\physicsDept}
\author{J. K. Kawasaki}
\affiliation{\physicsDept}
\affiliation{Department of Materials Science and Engineering, University of Wisconsin, Madison WI 53706}
\author{J. N. Nelson}
\affiliation{\physicsDept}
\author{B. D. Faeth}
\affiliation{\physicsDept}
\author{Y. Lee}
\affiliation{\physicsDept}
\author{B. H. Goodge}
\affiliation{\aepDept}
\affiliation{\kavli}
\author{B. Pamuk}
\affiliation{\aepDept}
\author{C. J. Fennie}
\affiliation{\aepDept}
\author{L. F. Kourkoutis}
\affiliation{\aepDept}
\affiliation{\kavli}
\author{D. G. Schlom}
\affiliation{\mseDept}
\affiliation{\kavli}
\author{K. M. Shen}
\affiliation{\physicsDept}
\affiliation{\kavli}

\date{\today}

\maketitle

\section{Electrical transport measurements on patterned resistivity bridges}

\begin{figure*}
\begin{center}
\includegraphics[width=7in]{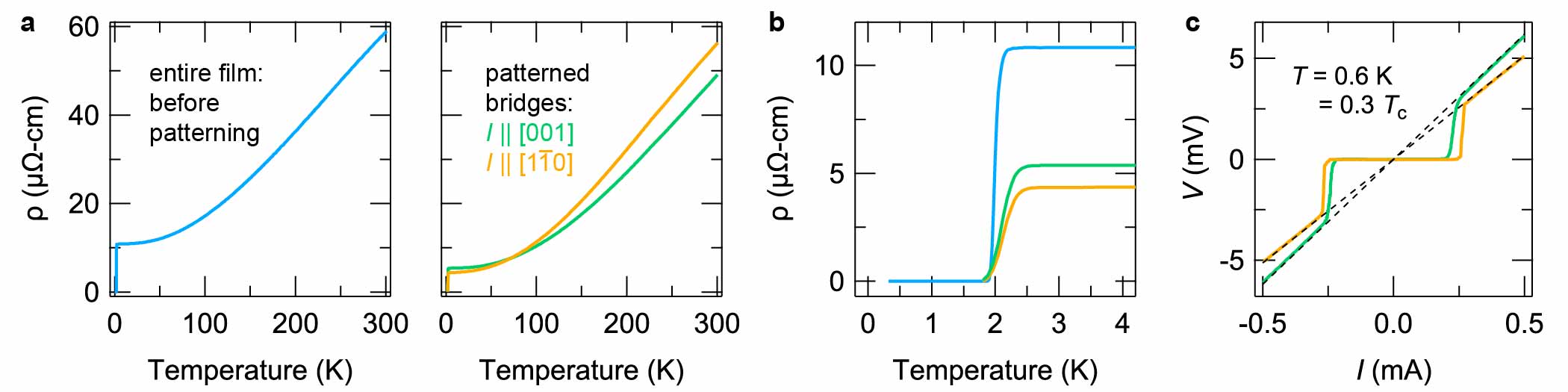}
\end{center}
\caption{\textbf{Electrical transport measurements exploring the anisotropy of the resistivity and superconductivity of a 24.2 nm thick \RO / \TOnospace(110) sample.} \textbf{a - b,} Zero-field $\rho(T)$ data measured on the entire as-grown film (blue) and after lithographically patterning four-point resistivity bridges (green and orange).  \textbf{c,} Superconducting $V(I)$ curves measured on patterned resistivity bridges with the directions of current flow parallel to [001] and $[1\overline{1}0]$.}
\end{figure*}

In Fig.\,S1, we show how the electrical transport properties of an \RO / \TOnospace(110) sample depend on the direction of current flow in the film when it is confined to flow along the orthogonal in-plane crystallographic axes, $[001]$ and $[1\overline{1}0]$.  Prior to lithographically patterning resistivity bridges on the film, we measured the resistance versus temperature of the entire 10~mm $\times$ 10~mm $\times$ 24.2~nm thick film by wire bonding four contacts directly to the surface of the sample in an in-line contact geometry.  Such a contact geometry probes the geometric mean of the two diagonal components of the in-plane resistivity tensor, $\sqrt{\rho_{001}\rho_{1\overline{1}0}}$, neglecting small finite-size corrections that depend on how the contacts are oriented relative to the edges of the wafer~\cite{miccoli_2015}.  The results of these measurements are shown by the blue traces in Fig.\,S1a,b; these are the same data plotted on a logarithmic temperature scale in Fig.\,1b of the main text.

Since \RO has a tetragonal crystal structure in bulk (and orthorhombic or perhaps monoclinic in (110)-oriented films), $\rho_{001}$ and $\rho_{1\overline{1}0}$ are not guaranteed by symmetry to be equal.  The intrinsic transport anisotropy in bulk \RO is known to be small, with differences between $\rho_{100}$ and $\rho_{001}$ that are less than 10\% at 300 K~\cite{ryden_temperature_1968, glassford_electron_1994}; however, in thin films it is common for highly oriented structural defects---\textit{e.g.}, those nucleated at step edges on the substrate---to induce sizable extrinsic anisotropies in the different in-plane components of $\rho$.  To investigate this possibility, we used standard lithographic techniques to pattern the same \RO / \TOnospace(110) sample into four-point resistivity bridges with dimensions 55~$\mu$m (length) $\times$ 10~$\mu$m (width) $\times$ 24.2~nm (thickness), where the direction of current flow is confined (via lithography) to be aligned with specific crystallographic directions.  In the course of performing the lithography, we noticed that the \TO substrates became mildly conducting, possibly due to oxygen vacancies formed during ion milling, as has been reported to occur for SrTiO$_3$~\cite{schneider_microlithography_2006}.  Therefore, we annealed the wafer containing the patterned resistivity bridges in air at elevated temperatures until the substrate again read open-circuit two-point resistances ($> 100 \text{ M}\Omega$); 2 hours at 500$^\circ \text{ C}$ was found to be sufficient. 

The results of electrical measurements on these patterned resistivity bridges are shown by the green and orange traces in Fig.\,S1a-c.  The temperature dependence of $\rho(T)$ is qualitatively consistent with the control measurements performed on the entire film before patterning, and the absolute magnitude of the resistivity anisotropies at 300 K and 4 K are both $< 20\%$.  Furthermore, the superconducting $\rho(T)$ and $V(I)$ behavior does not depend strongly on the direction of current flow; this is contrary to what would be expected if the superconductivity arose purely from oriented structural defects.  In Fig.\,S1b, we ascribe the substantial decrease in low-temperature resistivities observed in the patterned resistivity bridge data relative to the entire film data to the aforementioned annealing involved in preparing the bridges.  We confirmed on other \RO / \TOnospace(110) samples not containing bridges that post-growth annealing in air generically causes the low-temperature values of $\rho$ to drop, by as much as a factor of four.  Because of these complications and additional uncertainties involved in lithographically patterning resistivity bridges on films on TiO$_2$ substrates, all other electrical transport data presented in the main text and in the supplemental information were acquired by wire bonding directly to the surfaces of as-grown samples that were not subject to any post-growth annealing treatments.     

\section{Fitting and extrapolation of superconducting upper critical fields versus temperature}

\begin{figure*}
\begin{center}
\includegraphics[width=7in]{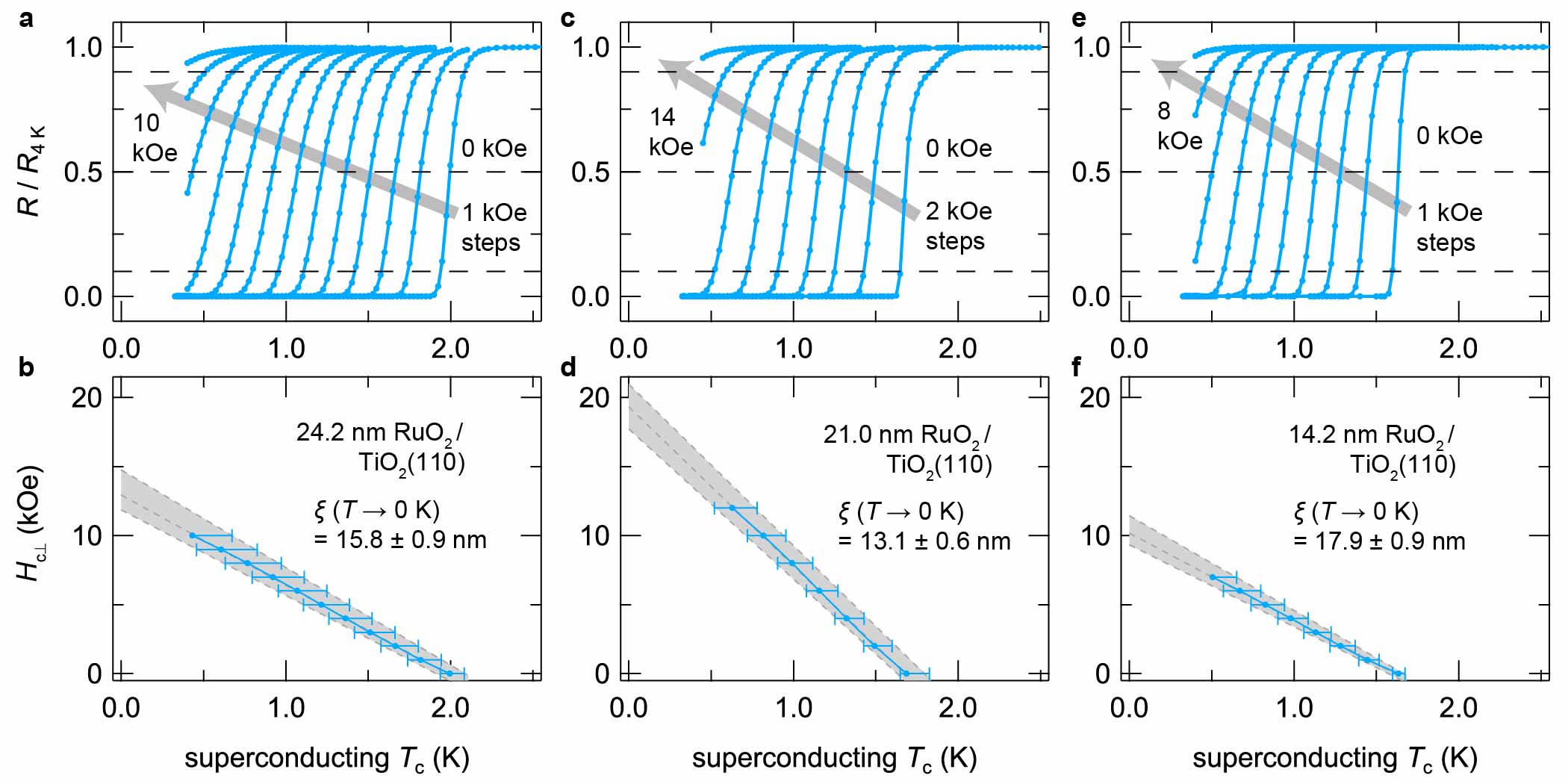}
\end{center}
\caption{\textbf{Magnetoresistance measurements for three superconducting \RO / \TOnospace(110) samples with different film thicknesses.}  All raw data traces in \textbf{a, c, e} are normalized to a common value $R_{\text{4 K}} \equiv R(T = \text{ 4 K}, H = \text{ 0 kOe})$ for ease of visualization and analysis. The extracted scaling behavior of the upper critical fields versus superconducting $T_c$ are plotted in \textbf{b, d, f}, along with the superconducting coherence length $\xi$ corresponding to the extrapolated zero-temperature $H_{c\perp}$. }
\end{figure*}

In Fig.\,S2, we present the results of magnetoresistance measurements for three \RO / \TOnospace(110) samples with different film thicknesses; the data in Fig.\,S2a,b are reproduced from Fig.\,1c of the main text.  Each $R(T)$ trace was acquired at a discrete value of the externally applied magnetic field $H_{\perp}$ (applied perpendicular to the surfaces of the films, along [110]) upon warming the samples up from base temperature through the superconducting transitions.  All resistances are normalized to their zero-field values at 4 K, well above the superconducting transitions; since the normal-state $R(T, H)$ behavior of \RO / \TOnospace(110) in the absence of superconductivity is negligible in this regime of low temperatures and fields, the choice of a single normalization factor $R_{\text{4 K}}$ for all data does not appreciably affect any of the results that follow.  Because percolation effects imply that resistive measurements of critical fields inherently contain some ambiguity about the definition and meaning of $H_{c\perp}$ relative to truly bulk-sensitive measurements of superconductivity~\cite{hsu_superconducting_1992}, here we adopt the same convention employed in the main text:  the temperature at which $R$ drops to 50\% of $R_{\text{4 K}}$ is taken as $T_c$ for the given $H_{c\perp}$, and the error bars on the extracted $T_c$ are the temperatures at which $R$ drops to 90\% and 10\% of $R_{\text{4 K}}$, respectively~\cite{kim_intrinsic_2012}.   

While there are considerable quantitative discrepancies in the values of $H_{c\perp}$ and $T_c$ for the different-thickness samples in Fig.\,S2, the $H_{c\perp}(T_c)$ scaling behavior is remarkably linear for all samples, with no signs of $H_{c\perp}$ saturation down to reduced temperatures $T/T_c \approx 0.2 - 0.3$, unlike what is expected in, \textit{e.g.}, Werthamer-Helfand-Hohenberg (WHH) theory~\cite{werthamer_temperature_1966}.  For example, evaluating the right side of the WHH expression

\begin{equation}
H_{c\perp}(T\rightarrow 0 \text{ K}) \leq -0.693 \left. \frac{dH_{c\perp}}{dT} \right \rvert_{T=T_c} T_c
\end{equation}

\noindent places upper bounds of 9.0, 13.5 and 7.1 kOe on $H_{c\perp}(T\rightarrow 0 \text{ K})$ for these three samples; however, the experimentally measured critical fields at 0.45 K (\textit{i.e.}, $T/T_c =$ 0.23, 0.27, 0.28) are already larger than these bounds: 10.0, 13.7 and 7.4 kOe, respectively.  Therefore, to extrapolate $H_{c\perp}$ down to zero temperature, we performed linear Ginzburg-Landau-type fits to all available data and propagated the systematic uncertainties in the definition of $H_{c\perp}$ according to the gray dashed lines.  The quoted zero-temperature values of the average in-plane superconducting coherence lengths $\xi(T \rightarrow \text{0 K})$ are obtained from the relation

\begin{equation}
\xi(T\rightarrow 0 \text{ K}) = \sqrt{ \frac{\Phi_0}{2 \pi \mu_0 H_{c\perp}(T\rightarrow 0 \text{ K})} }\;\;,
\end{equation}

\noindent where $\Phi_0$ is the superconducting flux quantum and $\mu_0$ is the magnetic permeability of free space.

Notably, these values of $\xi(T \rightarrow \text{0 K})$ are less than values reported for traditional elemental superconductors with comparable \TCnospace s by almost an order of magnitude, corresponding to critical fields that are $\approx 1 - 2$ orders of magnitude greater.  While an explanation and understanding of these sizable critical field enhancements are beyond the scope of the present work, they are internally self-consistent with the large critical current densities noted in the main text and in Fig.\,S1.  These results may motivate future real-space measurements of the superconducting condensate by scanning-probe techniques.  In particular, an interesting question to address is whether the structural defects in \ROA act as pinning sites for the vortices that form under applied fields, similar to what has been observed in numerous other thin-film superconductors~\cite{wissberg_large-scale_2017}, or whether the defects host regions of enhanced superfluid density that effectively act as barriers to vortex motion, akin to twin boundaries in bulk single crystals of iron-based superconductors~\cite{kalisky_stripes_2010, kalisky_behavior_2011}.  

\section{Structural and electrical characterization data for films synthesized on differently oriented substrates}

\begin{figure*}
\begin{center}
\includegraphics[width=7in]{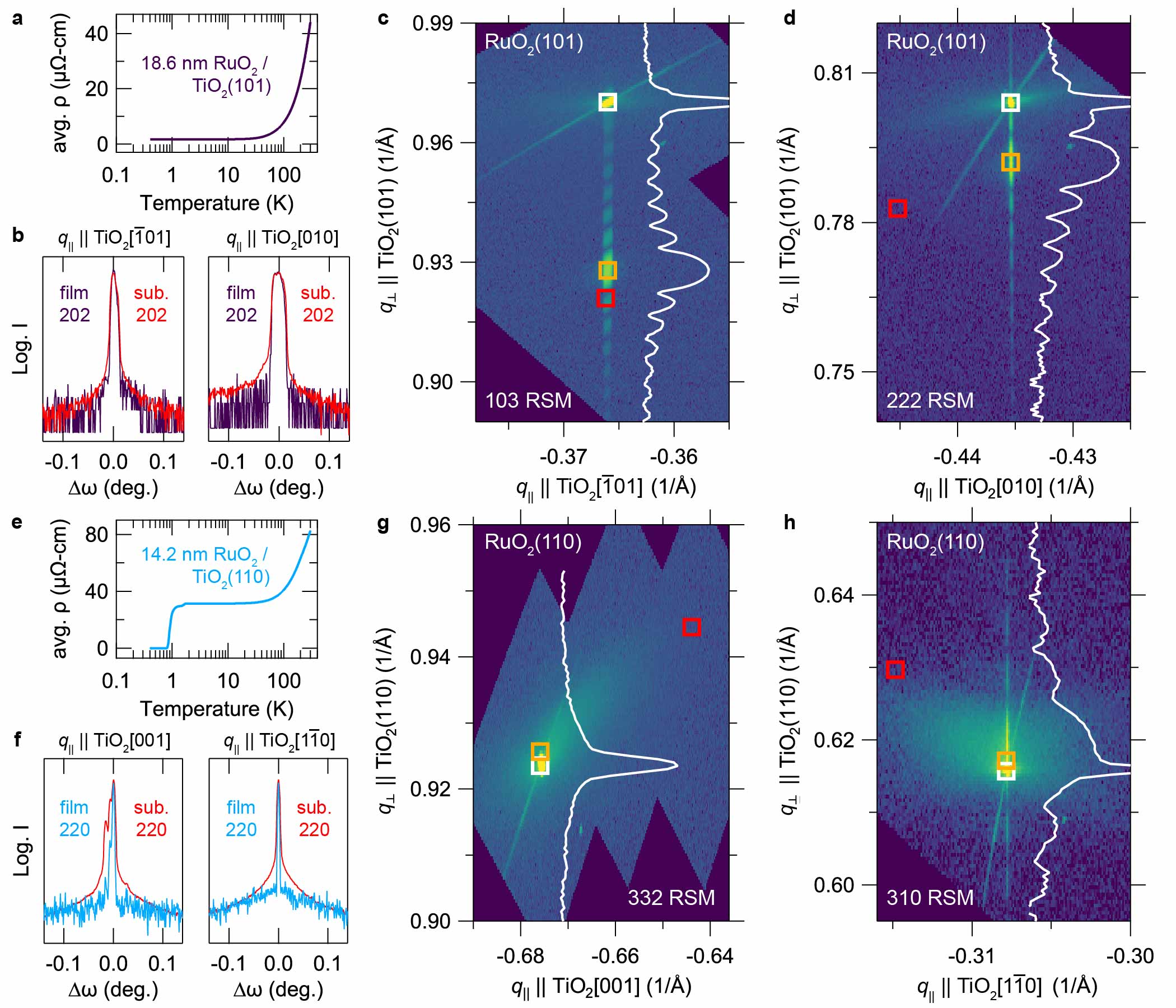}
\end{center}
\caption{\textbf{Electrical and structural characterization of (101)- and (110)-oriented \RO thin films.} \textbf{a,} Zero-field $\rho(T)$ (geometric mean) data for a non-superconducting 18.6 nm thick \ROB sample.  \textbf{b,} Rocking curves for this sample, taken at $2\theta$ values corresponding to the primary film and substrate 202 reflections.  The FWHMs are $0.0081^{\circ}$ with $q_{||}$ aligned along $[\overline{1}01]$ and $0.021^{\circ}$ with $q_{||}$ along [010].  \textbf{c - d,} RSMs for this sample near the 103 and 222 reflections.  Solid white lines are the scattering profiles along the CTRs.  White, red, and orange squares represent the central peak positions expected for bulk \TOnospace, bulk \RO and commensurately strained \RO thin films, respectively.  The in-plane lattice mismatches of \TO with bulk \RO can be read off directly from the lateral offsets of the white and red squares: +0.04\% (tensile) along $[\overline{1}01]$ in \textbf{c} and +2.3\% along $[010]$ in \textbf{d}. \textbf{e - h,} Analogous electrical and structural data for a superconducting 14.2 nm thick \ROA sample.  The rocking curve FWHMs at 220 are $0.0042^{\circ}$ with $q_{||}$ aligned along $[001]$ (although there are clearly multiple peaks discernible in both the substrate and film curves) and $0.0036^{\circ}$ with $q_{||}$ along $[1\overline{1}0]$.  RSMs for \ROA at this film thickness show clear signatures of partial strain relaxation, because of the larger absolute levels of in-plane lattice mismatch with \TOnospace: -4.7\% along $[001]$ in \textbf{g} and +2.3\% in \textbf{h}.}
\end{figure*}

In Fig.\,S3, we include electrical characterization and more comprehensive lab-based x-ray diffraction (XRD) measurements for the \ROB and \ROA films of comparable thickness shown in Fig.\,2 of the main text.  Figure S3a,e show the zero-field $\rho(T)$ behavior for the two films: the 18.6 nm thick \ROB film is non-superconducting down to $< 0.4 \text{ K}$ with a residual resistivity $\rho_0 < 1.7 \text{ } \mu\Omega\text{-cm}$, whereas the 14.2 nm thick \ROA film is superconducting at $T_c = 0.92 \pm \text{ } ^{0.21}_{0.07} \text{ K}$ with a residual resistivity $\rho_0 < 32 \text{ } \mu\Omega\text{-cm}$.  Figure S3b,f show rocking curves for the films overlaid on rocking curves for the \TO substrates they were synthesized on: in all cases the coherent components of the film peaks exhibit narrow full width at half maximum (FWHM) values that are limited by the underlying substrate FWHM, as expected for isostructural film growths.  In our studies we found that the rocking curve shapes and widths of the \TO substrates supplied by CrysTec, GmbH can vary significantly depending on how the in-plane momentum transfer $q_{||}$ is oriented relative to the crystal axes of a given wafer, which may be due to the Verneuil process used to synthesize the crystals; to give some idea of the magnitude of this asymmetric mosaic spread, we show scans with $q_{||}$ oriented along azimuths separated by 90$^\circ$ for each sample.  

In Fig.\,S3c,d and Fig.\,S3g,h we show off-specular $(q_{||}, q_{\perp})$ reciprocal space maps (RSMs) for both samples in regions surrounding $HKL$ Bragg peaks that have $q_{||}$ purely aligned with the crystallographic directions indicated in the labels on the horizontal axes.  For reference, the peak positions that would be expected for bulk \RO and bulk \TO at 295 K~\cite{berlijn_itinerant_2017, burdett_structural-electronic_1987} are shown as red and white squares, respectively; the orange squares represent the central peak positions expected for commensurately strained \RO thin films calculated using appropriately constrained density functional theory structural relaxations.  To give a more quantitative sense of the logarithmic false color scale used here, the solid white lines overlaid on each plot represent the scattered intensity along the crystal truncation rods (CTRs)---\textit{i.e.}, the one-dimensional cuts through the RSMs with $q_{||}$ equal to that of the substrate.  These results show that the 18.6 nm thick \ROB film is coherently strained to the substrate along both in-plane directions, within the $\approx$ 0.1\% resolution of the measurements.  The variable widths of the CTRs versus $q_{||}$ in different RSMs are an artifact of instrumental resolution effects---namely, the ``tall'' incident beam profile convolved with the scattering geometries used to measure each RSM---which we do not attempt to correct for in this work.  On the other hand, the 14.2 nm thick \ROA film is partially strain-relaxed, as evidenced by the more diffuse distribution of scattered intensity versus $q_{||}$ and less prominent finite-thickness fringes versus $q_{\perp}$ along the CTRs.  The diminished (or non-existent) contrast between thickness fringes in the CTRs for \ROnospace(110) is likely a manifestation of crystalline disorder in the film interplanar spacings, since all (110)-oriented films have abrupt bounding interfaces, \textit{cf.} the x-ray reflectivity data in Fig.\,S9. 

\begin{figure*}
\begin{center}
\includegraphics[width=7in]{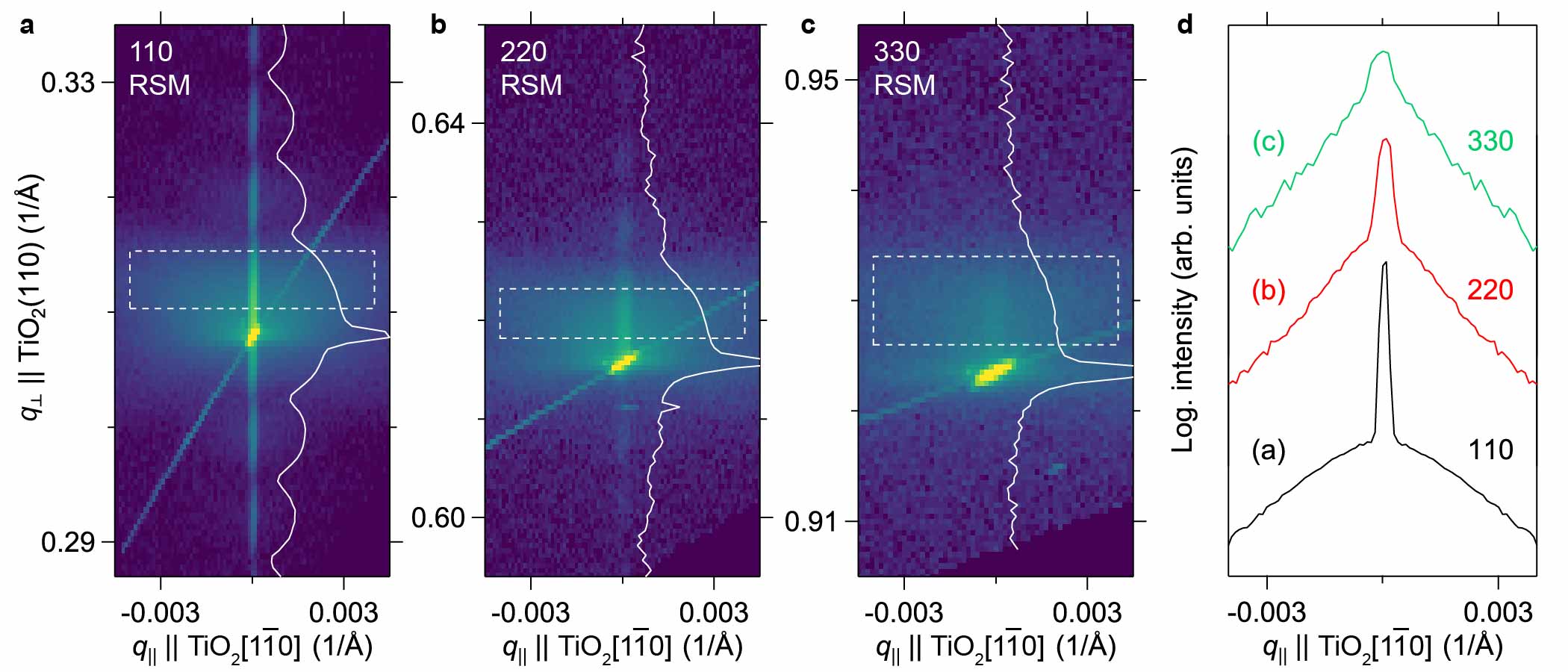}
\end{center}
\caption{\textbf{RSMs and rocking curves for a partially strain-relaxed 14.2 nm thick \ROA sample.}  \textbf{a - c,} RSMs measured near the 110, 220, and 330 Bragg reflections.  Solid white lines are the scattering profiles along the specular CTR.  \textbf{d,} Line cuts of the intensities averaged over the dashed boxes in \textbf{a - c} show rocking curves with a two-component narrow plus broad structure characteristic of partially strain-relaxed epitaxial thin films.  The in-plane momentum transfer $q_{||}$ is aligned with \TOnospace$[1\overline{1}0]$ in all panels; similar results are obtained with $q_{||}$ along \TOnospace$[001]$ (\textit{cf.} Figs.\,S9-S10).}
\end{figure*}

To further substantiate the partial strain relaxation observed in \RO / \TOnospace(110) samples, we measured RSMs around several Bragg peaks along the specular CTR.  Figure S4 summarizes the results of such measurements for the same 14.2 nm thick \ROA sample for which off-specular RSMs are shown in Fig.\,S3, which was also characterized by XRD and scanning transmission electron microscopy in Fig.\,2 of the main text.  By taking line cuts averaged over the dashed boxes---which span ranges of $q_{\perp}$ where the measured intensities are predominantly due to scattering from the film---we obtained the three rocking curves plotted in Fig.\,S4d.  Each rocking curve shows a sharp central peak that is resolution-limited in width (or substrate-limited, \textit{cf.}\,Fig.\,S3), superimposed on a much broader, nearly Lorentzian (FWHM = $0.003 - 0.005$ \ang$^{-1}$), component of the scattering that is also centered at $q_{||} = 0$.  Furthermore, the integrated intensity of the former coherent component of the scattering decays relative to that of the diffuse component as the magnitude of $|\mathbf{q}| = q_{\perp}$ increases in progressing from (a) to (c).

The non-vanishing intensity of the diffuse component in the film rocking curves, and the scaling behavior of how the total integrated intensity is distributed between the coherent and diffuse components as $|\mathbf{q}|$ is varied, are both completely consistent with published data for numerous epitaxial thin films grown on lattice-mismatched substrates where the films are thick enough to exhibit some form of strain relaxation~\cite{miceli_x-ray_1995, miceli_specular_1996, kortan_structure_1999, barabash_x-ray_2001, biegalski_critical_2008, wang_critical_2013}.  In principle, by analyzing the diffuse scattering profiles around multiple Bragg peaks with $\mathbf{q}$ that project differently onto the Burgers vectors of the relevant misfit dislocations that relax the strain, one can obtain quantitative information on the types of dislocations that exist, the dislocation densities, \textit{etc.}~\cite{kaganer_x-ray_1997, csiszar_diffraction_2005}.  We leave a more systematic analysis of this type to future synchrotron XRD studies, where the measurement noise floor is significantly lower and the strongly $\mathbf{q}$-dependent instrumental resolution effects observed here are mitigated by having a more point-like incident beam profile.  We note, however, that the similar FWHM values of the diffuse scattering versus $q_{||}$ around the 110, 220, and 330 peaks imply that the structural defects responsible for this scattering are more translational in nature than rotational (which in typical mosaic crystals, produce rocking curves of constant \textit{angular} widths)~\cite{miceli_x-ray_1995, miceli_specular_1996}.  Whether the inverses of these FWHM values for the fitted Lorentzians can be directly interpreted as the Fourier transform of a real-space correlation length ($200 - 300$ \ang) depends on whether the film is in the limit of \textit{weak disorder}, in the formalism of Refs.\cite{miceli_specular_1996, kaganer_x-ray_1997}.

\begin{figure*}
\begin{center}
\includegraphics[width=7in]{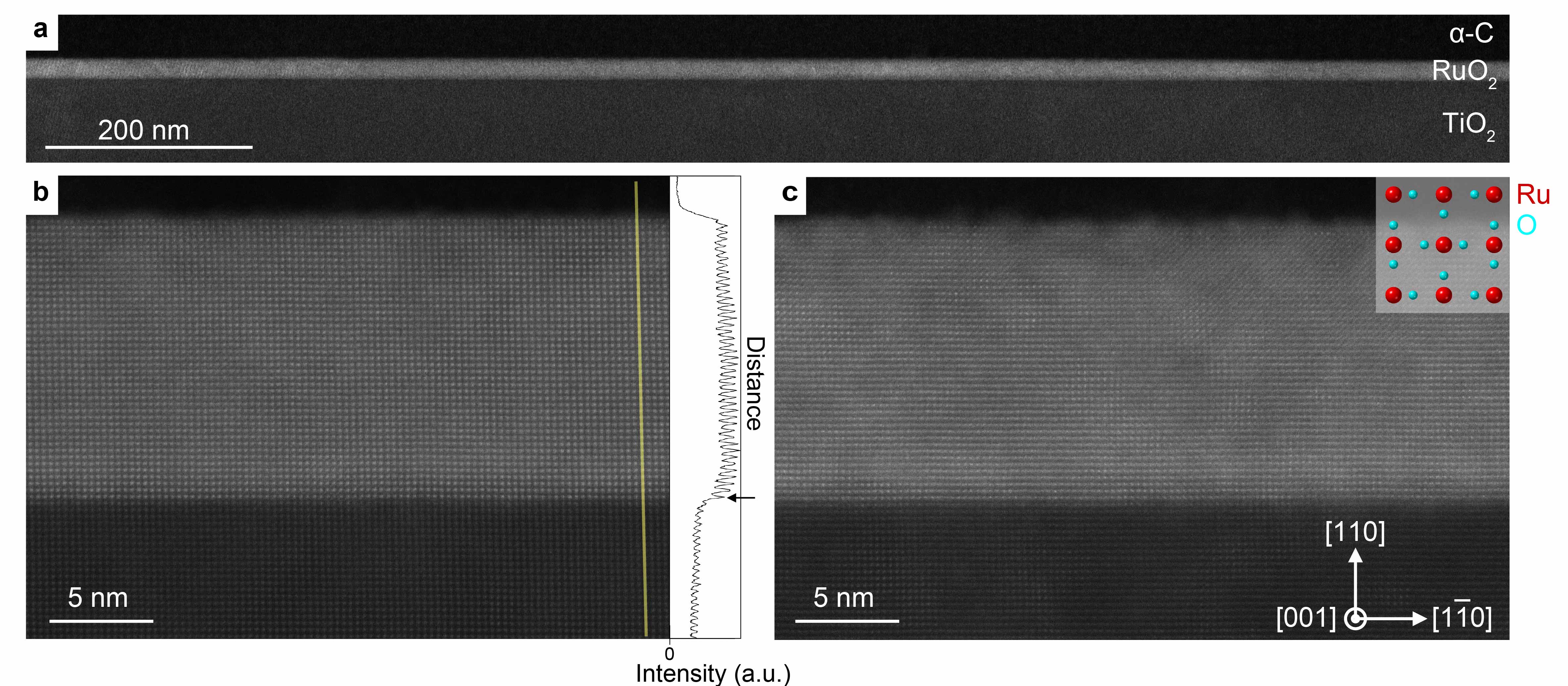}
\end{center}
\caption{\textbf{Variation in local crystalline quality across a superconducting 14.2 nm thick \ROA sample.} \textbf{a,}~Wide field of view image confirming continuous film growth over micron length scales.  \textbf{b,}~A crystallographically coherent region of the \ROA film.  A line cut of the measured intensity across a continuous column of atoms along the growth direction (yellow line) shows an abrupt interface between \TO and \ROnospace, indicated by the black arrow. \textbf{c,}~A relatively less-ordered region of the same film shown in the same projection, perpendicular to $[001]$. Inset shows the expected structure for this projection (not to scale).}
\end{figure*}

\begin{figure*}
\begin{center}
\includegraphics[width=7in]{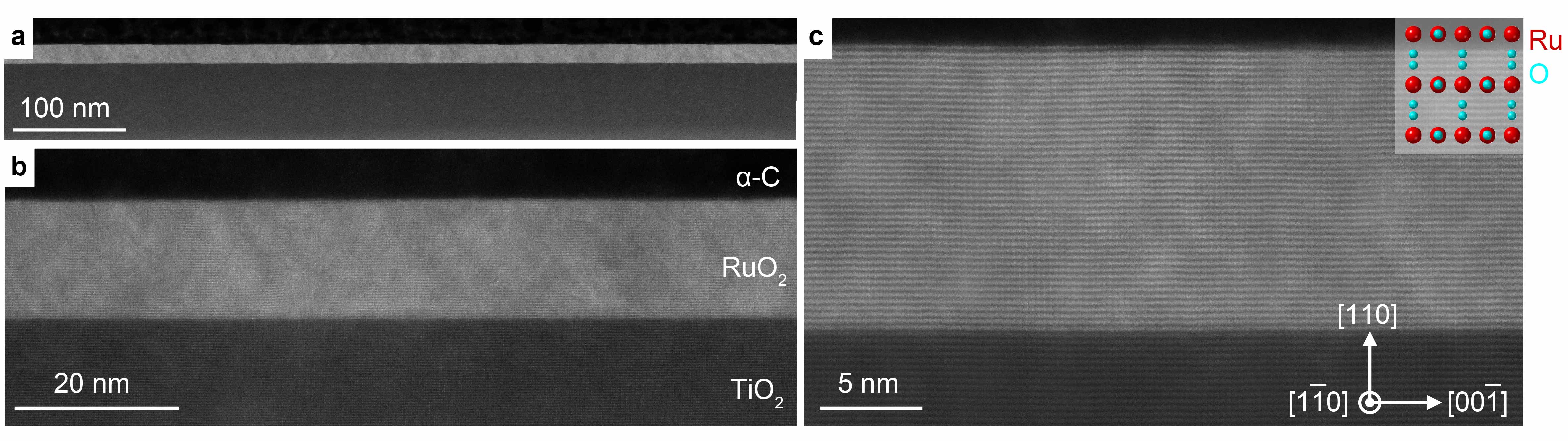}
\end{center}
\caption{\textbf{HAADF-STEM structural characterization of a superconducting 14.2 nm thick \ROA sample in the higher strain direction.} Z-contrast STEM images acquired in $[1\overline{1}0]$ projection demonstrate the effects of the 4.7\% compressive strain applied by the $[001]$ axis of the \TO substrate. \textbf{a,}~Continuous film growth is observed across the full length of the STEM lamella, shown here without interruption over several hundreds of nanometers.  \textbf{b,}~Epitaxial film growth, as observed in the orthogonal projection (Fig.\,S5), is again confirmed. \textbf{c,}~Atomic-resolution image shows the crystalline quality of the strained \RO film. Inset shows the expected structure for this projection (not to scale).}
\end{figure*}

\begin{figure*}
\begin{center}
\includegraphics[width=7in]{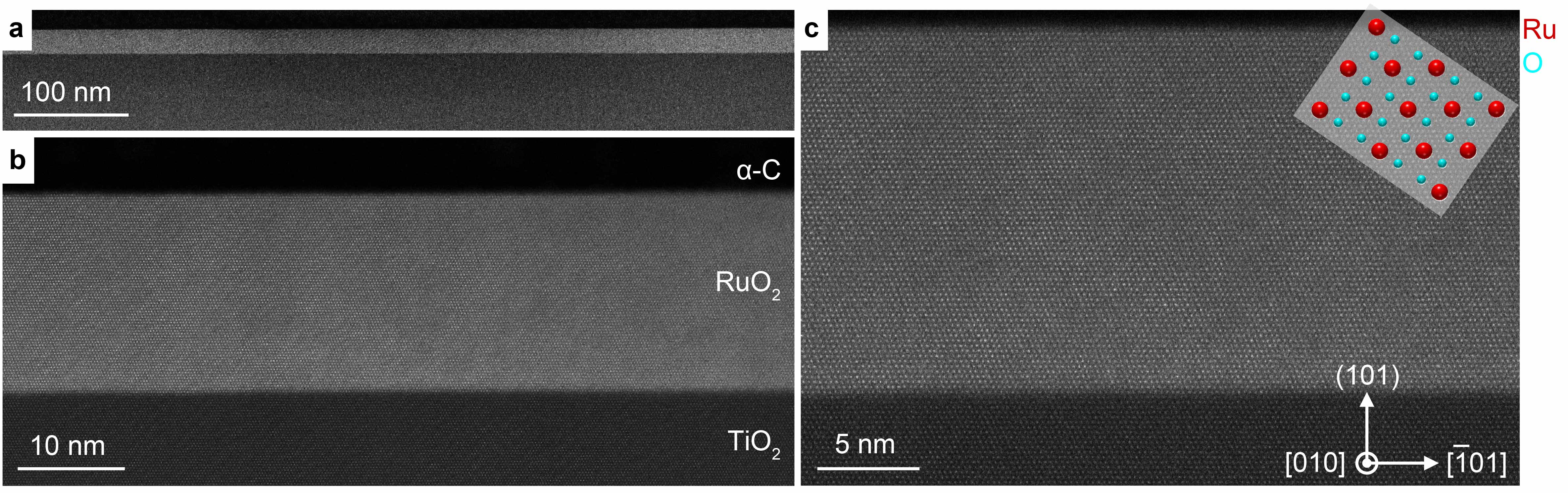}
\end{center}
\caption{\textbf{HAADF-STEM structural characterization of a non-superconducting \ROB sample.} \textbf{a,}~As in the superconducting \ROA samples, continuous film growth is observed across the entire length of the STEM lamella.  \textbf{b,}~Epitaxial growth between the \RO film and \TO substrate is again confirmed. Here, however, the observed contrast is comparatively smooth across the film, without the clear signs of high strain observed in the \ROA sample.  \textbf{c,}~Atomic-resolution STEM image demonstrating the high crystalline quality of the \ROB sample. Inset shows the expected structure for this projection (not to scale).}
\end{figure*}

Figures S5-S6 show additional high-angle annular dark-field scanning transmission electron microscopy (HAADF-STEM) data from the same 14.2 nm thick \ROA sample characterized in Fig.\,2 of the main text and in Figs.\,S3-S4.  Fig.\,S5a shows that the morphology of the \RO film is continuous and epitaxial to the \TO substrate over the largest length scales probed.  An intensity line profile taken along the growth direction (yellow line in Fig.\,S5b) confirms that an abrupt interface exists between \TO and \ROnospace.  In particular, a sharp transition from low intensity peaks in the substrate (Ti: $Z = 22$) to high intensity peaks in the film (Ru: $Z = 44$) occurs over a region thinner than $1 \text{ nm}$ surrounding the black arrow at the substrate-film interface; this indicates that any Ti/Ru chemical interdiffusion is minimal and cannot be the cause of the enhanced superconductivity observed in \ROAnospace.  At the lattice scale, we find that different regions from the same film exhibit varying degrees of crystalline coherence under the epitaxial strain applied by the \TO substrate.  The lateral in-plane direction imaged here is the $[1\overline{1}0]$ axis of the \RO film, subject to $+2.3$\% tensile strain from the \TO substrate.  Some regions, such as that shown in Fig.\,S5b, exhibit exceptionally ``clean'' crystalline quality: all of the atomic columns of Ru stack uniformly in the projection of the STEM image to produce highly ordered atomic contrast.  In other regions of the same film, strain gradients in the film distort the \RO lattice such that the columns of Ru atoms are slightly misaligned to the electron beam projection.  This local misalignment of the lattice causes the apparent blurring and more mottled contrast of the STEM image seen in Fig.\,S5c. 

In Fig.\,S6, the same film is studied with HAADF-STEM imaging in the orthogonal projection direction. This orientation allows us to assess the crystalline response of the \RO film along the [001] direction, which is subject to a larger lattice mismatch with the \TO substrate, $-4.7$\% compressive strain. Again, Fig.\,S6a confirms the continuous and epitaxial growth of the \ROA thin film over the meso- and macroscopic length scales relevant for interpreting the electrical transport data shown elsewhere in the manuscript.  Effects of the large compressive strain along the in-plane direction of this projection are apparent in Fig.\,S6a,b as characteristic V-shaped contrast in the \RO film. Contributions from electron channeling in ADF-STEM imaging produces to this bright/dark contrast in regions of local crystallographic strain; such contrast is a common signature of epitaxial lattice strain in many other oxide systems. Figure S6c shows the same structural response at atomic resolution, where---similar to Fig.\,S5c---the apparent blurring of atomic columns arises from regions where the film lattice has been locally distorted.

Finally, for completeness we also performed HAADF-STEM measurements on the same non-superconducting 18.6 nm thick \ROB film characterized in Fig.\,2 of the main text and in Fig.\,S3.  Z-contrast images of this sample are shown in Fig.\,S7: the film is comparably continuous and epitaxial as the superconducting \ROA films we have studied, without any signatures of extended defects or secondary phase inclusions that might otherwise alter its electrical properties.  In good agreement with the XRD and electrical transport data shown in Fig.\,S3, the \ROB film exhibits more coherent crystalline order than the more drastically strained superconducting \ROA films, even over relatively large fields of view as shown in Fig.\,S7b. Figure S7c shows that the lattice remains largely defect-free down to the atomic scale.  

\begin{figure*}
\begin{center}
\includegraphics[width=7in]{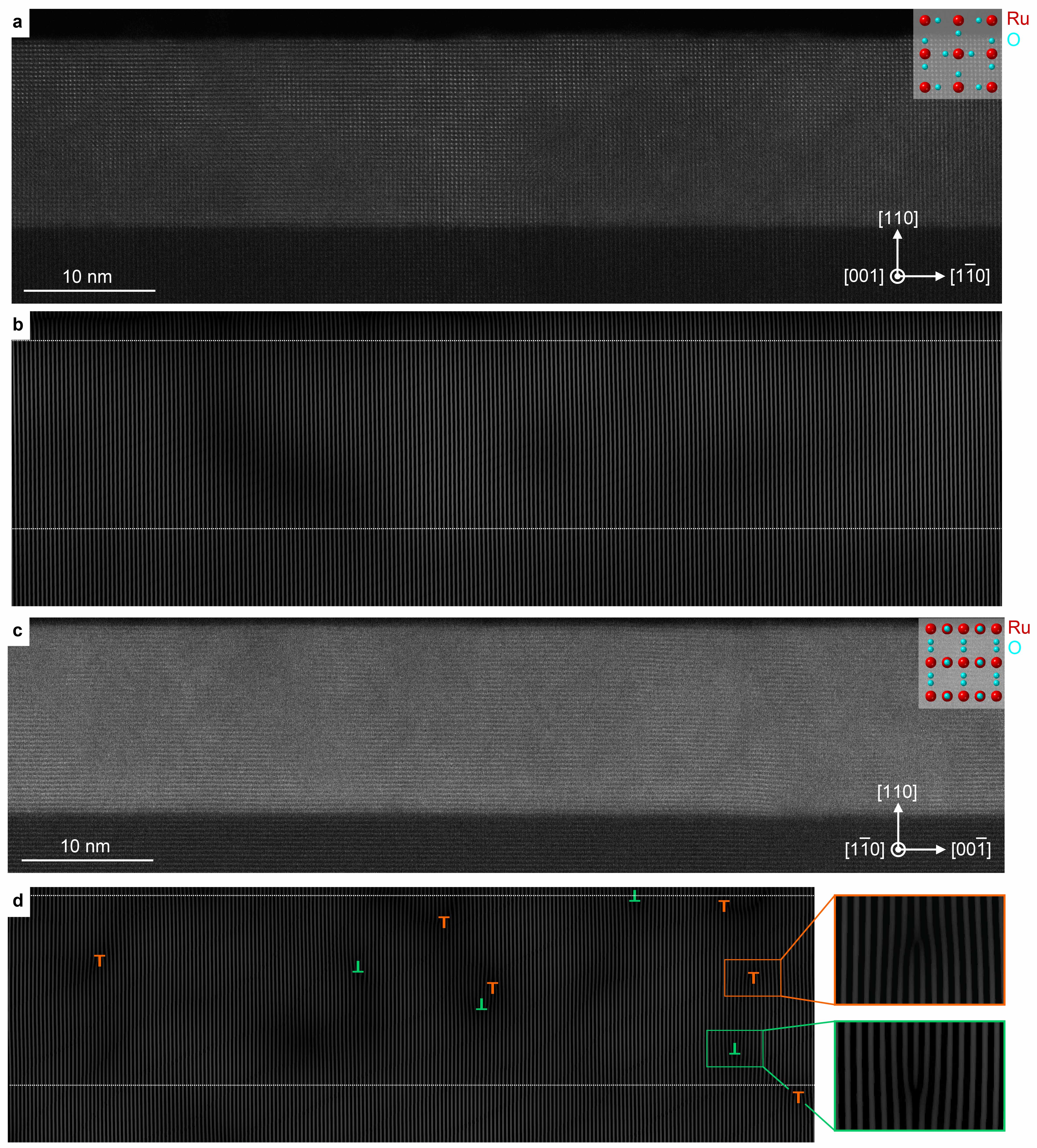}
\end{center}
\caption{\textbf{STEM imaging of edge dislocations in a superconducting \RO / \TOnospace(110) sample.} \textbf{a,}~Atomic-resolution HAADF-STEM image in $[001]$ projection for the same 14.2 nm thick \ROA film characterized elsewhere in the manuscript by STEM, x-ray diffraction, and electrical transport.  \textbf{b,}~Fourier-filtered version of the image in \textbf{a}, keeping only the components having $|q_{||}| \approx 1/d_{1\overline{1}0}$ ($d_{1\overline{1}0} \approx 3.2 \text{ \ang}$).  The horizontal dashed lines represent the boundaries of the film along the out-of-plane direction.  The \TO and \RO lattices are continuously matched in plane over the entire $226$-unit-cell-wide ($73 \text{-nm-wide}$) field of view, without any dislocations. \textbf{c - d,}~Same as in \textbf{a - b}, except in $[1\overline{1}0]$ projection and isolating the Fourier component of the image having $|q_{||}| \approx 1/c$ ($c \approx 3.0 \text{ \ang}$).  Nearly equal numbers of edge dislocations are observed in \textbf{d} with Burgers vectors of $-\mathbf{c} \text{ (orange)}$ and $+\mathbf{c} \text{ (green)}$, respectively, across a $204$-unit-cell-wide ($61 \text{-nm-wide}$) field of view.  The insets on the right side of panel \textbf{d} are magnified by $3 \times$.}
\end{figure*}

The data presented in Figs.\,S3-S6 indicate that the crystal structures of superconducting \ROA films are not commensurately strained to the \TO substrates.  To better visualize how this partial strain relaxation manifests in real space, we employed Fourier filtering of STEM images to find edge dislocations, following the techniques described in Ref.\cite{xie_coherent_2018}.  Specifically, in Fig.\,S8a (Fig.\,S8c) we plot an atomic-resolution HAADF-STEM image acquired in projection along $[001]$ (projection along $[1\overline{1}0]$) where the lateral in-plane direction is aligned with $[1\overline{1}0]$ ($[001]$, respectively).  Figure S8b (Figure S8d) displays the Fourier component of this image with spatial frequencies $|q_{||}| \approx 1/d_{1\overline{1}0}$ ($|q_{||}| \approx 1/c$), which we obtained by computing the fast Fourier transform (FFT) of the image in Fig.\,S8a (Fig.\,S8c), and then taking the inverse FFT with the contributions of all spatial frequencies smoothly masked out except for those in a narrow region of $q$-space surrounding the noted $\pm q_{||}$.  The horizontal dashed lines in Fig.\,S8b,d represent the boundaries of the film along the out-of-plane direction.

In this representation, edge dislocations appear as topological defects in the otherwise continuous vertical streaks appearing in Fig.\,S8b,d.  These vertical streaks are formally the lattice points of the film and substrate crystal structures, blurred into streaks along the out-of-plane direction because we discard any high spatial frequency information about out-of-plane correlations of the electron density (\textit{i.e.}, $q_{\perp} \not\approx 0$) when computing the inverse FFTs.  Hereafter we loosely refer to these streaks as \emph{atomic columns}, since the contrast in these HAADF-STEM data arises predominantly from scattering by the atomic cores with larger $Z$ (\textit{i.e.}, Ti and Ru).  Dislocations indicated by green markers \emph{add} one atomic column to the number of columns that exist in layers beneath it (thus relaxing tensile strain in the lateral direction), whereas dislocations indicated by orange markers \emph{remove} one atomic column to the number of columns that exist in layers beneath it (thus relaxing compressive strain in the lateral direction).  Therefore, a fully strain-relaxed film of \RO / \TOnospace(110) would show a collection of only green (only orange) dislocations accumulated at the substrate-film interface in Fig.\,S8b (Fig.\,S8d, respectively).  The dislocation densities expected in this fully strain-relaxed scenario would be 1 per every $1/0.023 \approx 43$ vertical streaks for green dislocations in Fig.\,S8b, and 1 per every $1/0.047 \approx 21$ vertical streaks for orange dislocations in Fig.\,S8d.

In marked contrast to this behavior, zero dislocations are observed across a $226$-unit-cell-wide field of view in Fig.\,S8b.  Furthermore, although a significantly higher density of dislocations is present across the $204$-unit-cell-wide field of view in Fig.\,S8d, there are nearly equal numbers of edge dislocations having Burgers vectors of $-\mathbf{c} \text{ (orange)}$ and $+\mathbf{c} \text{ (green)}$, respectively, and the dislocations are rather uniformly distributed throughout the entire thickness of the film.  These observations imply that throughout a sizable volume fraction of the superconducting film, the crystal structure is, on average, much closer to the commensurately-strained limit than to the fully-relaxed limit.  We note that this agrees well with the distribution of x-ray scattering intensities plotted in the RSMs for this sample and others in Figs.\,S9-S10.  Based on these data, we suggest that it is appropriate to consider the local strain gradients that inevitably accompany the nucleation of dislocations in \ROA to be sample-dependent ``perturbations'' to significantly larger average components of the substrate-imposed strain fields that are present throughout all films shown in the manuscript.  Because superconductivity is an essentially mean-field phenomenon, we believe that the latter average components of the strain fields in \ROA are the key ingredients for stabilizing superconductivity with transition temperatures at least an order of magnitude larger than in bulk \ROnospace; finer details of the local strain gradients probably determine finer details of the superconductivity, such as the exact sample-dependent \TCnospace s measured by non-bulk-sensitive probes of superconductivity, such as resistivity.  Since our platform for applying strain enables scanning-probe measurements of the superconducting condensate, future experiments may be able to provide direct experimental evidence to support these general expectations of mesoscale or nanoscale strain inhomogeneity resulting in spatially inhomogeneous superconductivity~\cite{bachmann_spatial_2019}.

\section{Structural and electrical characterization data for \ROA films of different thicknesses}

\begin{figure*}
\begin{center}
\includegraphics[width=7in]{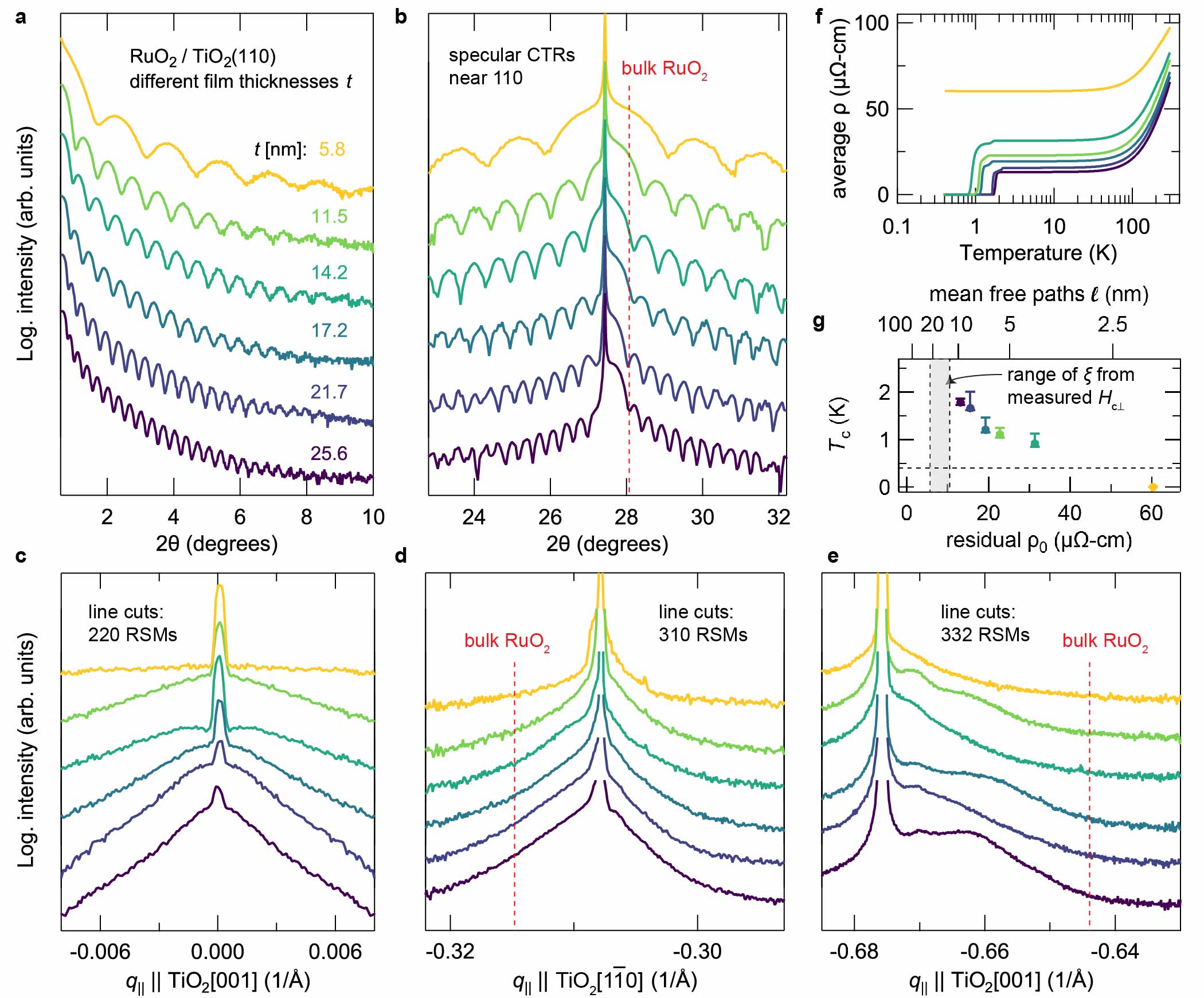}
\end{center}
\caption{\textbf{Evolution of crystal structure and electrical transport behavior as a function of film thickness for \RO / \TOnospace(110).} \textbf{a,} X-ray reflectivity and \textbf{b,} x-ray diffraction data along the specular CTR show comparable levels of flatness and crystalline order along the out-of-plane direction for all samples.  \textbf{c - e,} Average line cuts versus $q_{||}$ through the 220, 310, and 332 RSMs (fully $\mathbf{q}$-resolved data are shown in Fig.\,S10) indicate that all samples with $t > $ 5.8 nm exhibit partial strain relaxation.  \textbf{f - g,} Zero-field $\rho(T)$ data show that thinner films generally have higher residual resistivities $\rho_0$ and lower superconducting \TCnospace s.  The horizontal dashed line in \textbf{g} represents the base temperature attainable in our cryostat~($0.4\text{ K}$), and the gray-shaded region indicates the range of superconducting coherence lengths ($\xi = 12 - 22 \text{ nm}$) extracted from magnetoresistance measurements of the upper critical fields for ten \ROA thin films.  Comparison of these $\xi$ with the mean free paths $\ell$ corresponding to the measured $\rho_0$ shows that superconductivity persists even in the dirty limit $\ell < \xi$.}
\end{figure*}

\begin{figure*}
\begin{center}
\includegraphics[width=7in]{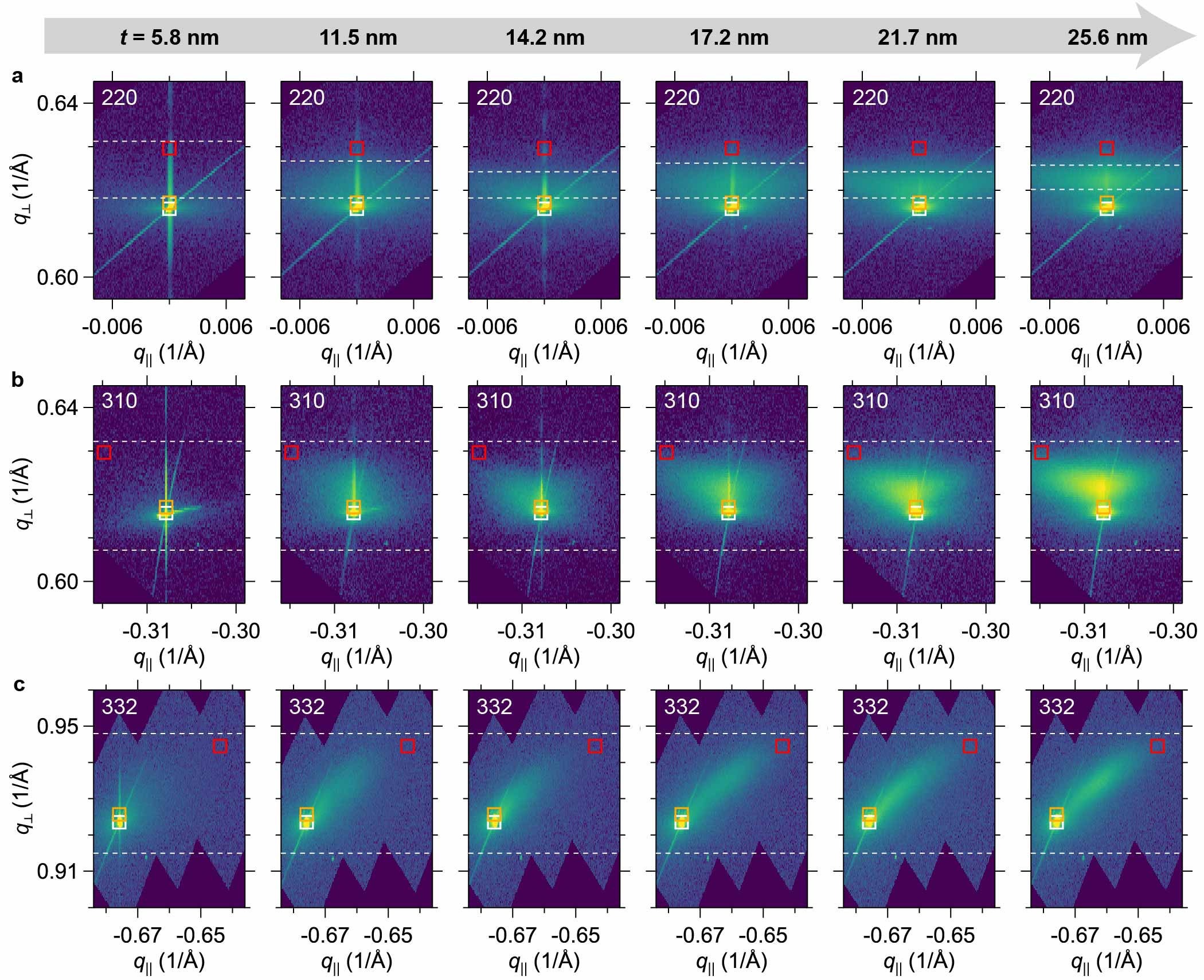}
\end{center}
\caption{\textbf{RSMs for \RO / \TOnospace(110) samples with different film thicknesses, \textit{t}.} \textbf{a,} RSMs along the specular CTR near the 220 Bragg reflections for films with increasing $t$, moving from left to right.  $q_{||}$ is aligned with $[001]$ in all panels, although the phenomenology is similar with $q_{||}$ along $[1\overline{1}0]$, \textit{cf.} Fig.\,S4.  \textbf{b,} Thickness-dependent RSMs near the off-specular 310 Bragg reflections where $q_{||}$ is purely along $[1\overline{1}0]$.  \textbf{c,} Same as \textbf{b}, but near the 332 Bragg reflections where $q_{||}$ is purely along $[001]$.  White, red, and orange squares represent the central peak positions expected for bulk \TOnospace, bulk \ROnospace, and commensurately strained \ROAnospace, as in Fig.\,S3.  The line cuts plotted in Fig.\,S9c-e are averaged over the ranges of $q_{\perp}$ of the RSMs between the horizontal white dashed lines.}
\end{figure*}

Figure 2b,c of the main text show that the superconducting \TCnospace s of \RO thin films synthesized on \TOnospace(110) substrates depend sensitively on the thickness of the films, $t$.  We do not purport to completely understand this empirical observation, but Figs.\,S9-S10 include several additional pieces of structural and electrical characterization data for this same thickness series of \ROA samples that constrain potential explanations.  

In Fig.\,S9a,b, we plot x-ray reflectivity (XRR) data taken at low incident angles, and XRD data taken near the 110 Bragg peaks of the film and substrate, both acquired along the specular CTRs using Cu-K$\alpha$ radiation.  Finite-thickness fringes are present over a wide range of angles in both data sets, evidencing (from reflectivity) atomically abrupt interfaces of the films with the substrates and with vacuum, and (from diffraction) comparable levels of crystallinity along the out-of-plane direction across samples.  Furthermore, the spacings between secondary maxima on either side of the primary film Bragg peaks match the spacings between the low-angle reflectivity fringes, suggesting that the crystal structures of all films are essentially homogeneous along the out-of-plane direction.  The film thicknesses $t$ quoted here and in the main text are obtained by directly fitting the XRR data in Fig.\,S9a using a genetic algorithm, which yields sub-nanometer roughnesses in all cases in the refined models.

Given that there are no obvious differences in film morphology or out-of-plane crystallinity between \ROA samples with different $t$, an alternative explanation that may account for the thickness-dependent superconducting \TCnospace s is the proliferation of misfit dislocations in thicker films that progressively relax the epitaxial---\textit{i.e.}, in-plane---strains; in this scenario, it may be that partially strain-relaxed \ROA films have higher (average) superconducting \TCnospace s compared with fully commensurately strained \ROA films.  To investigate this possibility, in Fig.\,S9c-e we plot line cuts of the intensity versus $q_{||}$ extracted from RSMs near the 220, 310, and 332 Bragg reflections of the \ROA thin films and \TO substrates.  The raw data for all RSMs are plotted in Fig.\,S10 using logarithmic false color scales; the line cuts in Fig.\,S9 are averaged over the ranges of $q_{\perp}$ between the dashed white lines in Fig.\,S10.  All of the samples except the thinnest film exhibit diffuse scattering surrounding the CTRs, indicating that partial strain relaxation onsets between film thicknesses of 5.8 nm and 11.5 nm for the growth conditions used in this work to synthesize \RO / \TOnospace(110) samples.  Since the in-plane lattice mismatches between \RO and \TO are highly anisotropic for the (110) orientation, it might also be expected that the substrate-imposed compressive strain along $[001]$ ($-4.7$\%) starts to relax at smaller film thicknesses than the tensile strain along $[1\overline{1}0]$ ($+2.3$\%)~\cite{kortan_structure_1999, biegalski_critical_2008}.  The off-specular RSMs in Fig.\,S10b,c qualitatively agree with this expectation, inasmuch as finite-thickness fringes can still be observed along the CTRs in the RSMs near 310 for films up to at least $t = 17.2 \text{ nm}$, whereas only the $t = 5.8 \text{ nm}$ film shows a contribution to the coherent CTR scattering in the RSMs near 332 that clearly rises above the contributions of the substrate.  

Although signatures of scattering from partially strain-relaxed \ROA are manifestly present in the data for all of the superconducting samples in Fig.\,S9---namely, broader distributions of intensity versus $q_{||}$ that asymmetrically gain weight towards the positions expected for bulk \RO as the film thickness increases---it remains somewhat ambiguous whether this data can be interpreted in a straightforward manner to gain insight into what levels of strain optimize the superconducting $T_c$s in \ROnospace.  Strain relaxation in oxide thin films often occurs inhomogeneously, with a mixture of commensurately strained and partially relaxed material~\cite{warusawithana_ferroelectric_2009}.  Indeed, examining the transport data for these same samples, it is tempting to ascribe the multi-stage behavior of the superconducting transitions to temperature-dependent Josephson coupling of regions of the films under different amounts of strain with correspondingly different ``local'' \TCnospace s; similar behavior has been described theoretically~\cite{glatz_self-organized_2011} and observed experimentally in patterned niobium islands on gold substrates~\cite{eley_approaching_2011}.  Nonetheless, because of the close proximity of the substrate peaks along $q_{\perp}$ ($d_{110} = 3.248$ \ang) with the positions expected for commensurately strained \ROA ($d_{110} = 3.241$ \ang), it is difficult to disentangle their respective contributions to the total scattering observed in XRD.  

Despite these complications in quantitatively analyzing the XRD results, we can use the values of $t$ obtained from XRR to plot the normalized resistance versus temperature curves from Fig.\,2b of the main text in terms of absolute resistivities, as shown in Fig.\,S9f.  From these data, a robust \textit{correlation} between the superconducting \TCnospace s and the residual resistivities $\rho_0$ immediately becomes apparent, as displayed in Fig.\,S9g.  As noted in the main text, this correlation may suggest that the primary effect of reducing $t$ is to enhance the relative importance of elastic scattering off disorder near the substrate-film interfaces, which is known to decrease \TC in numerous families of thin-film superconductors, both conventional~\cite{pinto_dimensional_2018} and unconventional~\cite{meyer_strain-relaxation_2015}.  It is largely outside the scope of this paper to contribute meaningful data to ongoing research efforts investigating the mechanism underlying thickness-induced suppressions of $T_c$ that are ubiquitously observed for superconducting films in the two-dimensional limit~\cite{kapitulnik_colloquium_2019}; however, in passing we note that the residual resistivity ($\rho_{0} = 60\,\mu\Omega\text{-cm}$) of the thinnest ($t = 5.8\text{ nm}$) non-superconducting ($T_{c} < 0.4 \text{ K}$) \ROA film shown in Fig.\,S9 corresponds to a sheet resistance of $R_{s} = \rho_{0}/t = 0.10 \,\, \text{k}\Omega/\square$.  This value of $R_s$ is about $60$ times \emph{less than} the Cooper pair quantum of resistance $h/(2e)^2 = 6.45 \,\, \text{k}\Omega/\square$ that was empirically noted to separate insulating from superconducting ground states in ultrathin films of numerous elemental metals~\cite{haviland_onset_1989,jaeger_onset_1989}; this indicates that quantitatively different physics is likely operative here in suppressing \TCnospace, which may place ultrathin films of \ROA in closer proximity to the ``anomalous metal'' regime that was shown to occur at weaker levels of disorder in Ref.\cite{steiner_approach_2008}.

We believe that identifying the exact mechanism underlying the strain-stabilized superconductivity in \ROA is also well beyond the scope of the current paper.  Phase-sensitive measurements of the superconducting order parameter—and/or momentum-resolved measurements of the superconducting gap magnitude—are notoriously challenging in multi-band materials with small (sub-meV) gaps, and a definitive answer to whether the pairing is (un)conventional will have to wait until such data become available.  With that qualification in mind, it is natural to consider whether the \TC versus $\rho_0$ behavior shown in Fig.\,S9g can shed any light on the answer to this question.  To address this possibility, we need to convert measured properties of the normal-metal and superconducting states into comparable characteristic length scales.  In unconventional low-temperature superconductors with sign-changing order parameters, such as Sr$_2$RuO$_4$~\cite{mackenzie_extremely_1998}, superconductivity is completely suppressed by non-magnetic impurity scattering whenever the normal-state mean free path $\ell$ is comparable with the clean-limit superconducting coherence length $\xi_0$---\textit{i.e.}, $T_{c} \rightarrow 0 \text{ K}$ if $\ell \approx \xi_0$.  By contrast, superconductivity persists in conventional superconductors even in the ``dirty limit'' $\ell << \xi_0$.  

On the top horizontal axis of Fig.\,S9g, we indicate approximate values of $\ell$ corresponding to the measured values of $\rho_0$ on the bottom horizontal axis; these numbers are computed following the analysis of Glassford \textit{et al}.~\cite{glassford_electron_1994}, who used the DFT-computed plasma frequencies and Fermi velocities for bulk \RO to obtain $\ell \, [\text{nm}]= 3.6 \cdot 35 / \rho \,[\mu\Omega\text{-cm}]$~\footnote{Because of various uncertainties implicit in this argument, we neglect any strain-dependent changes in Fermiology for \ROA that will, of course, quantitatively renormalize the actual relationship between $\ell$ and $\rho$.}.  To compare with $\ell$, we also include a gray-shaded region on Fig.\,S9g corresponding to the range of average in-plane superconducting coherence lengths $\xi$ we extracted experimentally from magnetoresistance measurements of the perpendicular upper critical magnetic fields $H_{c\perp}$ for ten different superconducting \ROA samples, following the procedures detailed in Fig.\,S2.  As noted previously, these values of $\xi$ likely represent a \emph{lower bound} for what the clean-limit $\xi_0$ would be in the absence of extrinsic defects in the films that impede vortex motion.  In any case, we find that superconductivity robustly persists in \ROA even when $\ell < \xi < \xi_0$; for example, the samples shown here with $T_{c} = 0.9 - 1.8 \text{ K}$ have measured residual resistivities corresponding to mean free paths $\ell = 4.0 - 9.6 \text{ nm}$, which are all less than the range of measured superconducting coherence lengths, $\xi = 12 - 22 \text{ nm}$.  Therefore, whatever the superconducting pairing mechanism is in \ROAnospace, these empirical considerations demonstrate that it is rather insensitive to defect scattering.  

\begin{figure*}
\begin{center}
\includegraphics[width=7in]{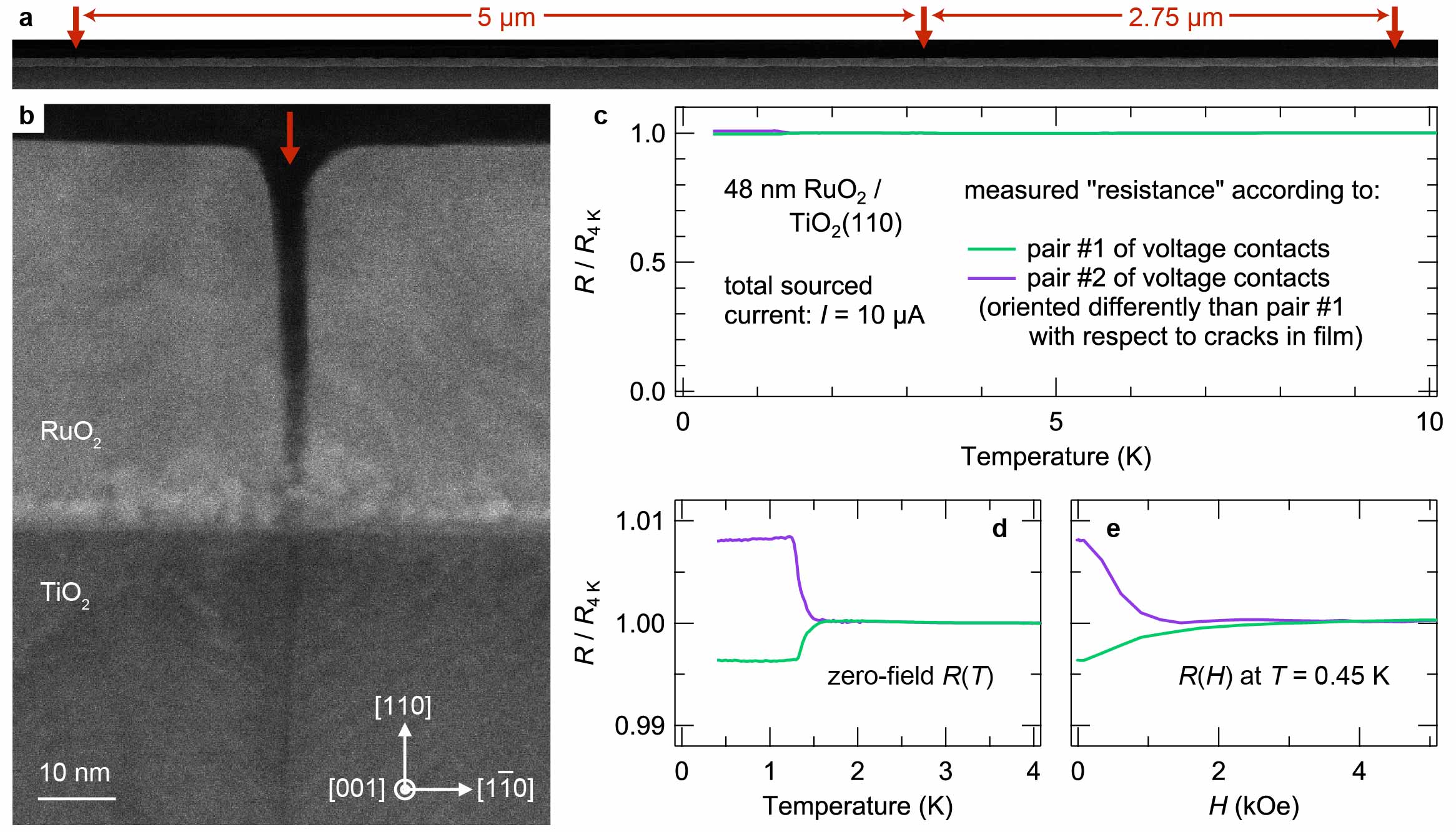}
\end{center}
\caption{\textbf{Structural and electrical characterization of a cracked 48 nm thick \RO / \TOnospace(110) sample.} \textbf{a,}~HAADF-STEM image of the sample over a wide field of view, showing an anisotropically oriented network of cracks (red arrows) each separated by a few microns.  Note that the cracks appear only in projection view along $[001]$---\textit{i.e.}, perpendicular to $[1\overline{1}0]$, which is the in-plane direction of the film under tensile strain.  \textbf{b,}~Magnified image of a crack from \textbf{a}.  \textbf{c,}~Zero-field ``resistance'' versus temperature data for two different pairs of voltage contacts oriented differently with respect to the network of cracks visualized in \textbf{a} and \textbf{b}.  \textbf{d,} Same $R(T)$ data as in \textbf{c}, plotted on a greatly expanded vertical scale.  \textbf{e,} ``Resistance'' versus externally applied magnetic field data for the pairs of voltage contacts from \textbf{c} and \textbf{d}, taken at fixed temperature $T = 0.45 \text{ K}$.}
\end{figure*}

Conceptually, perhaps the most straightforward test of our proposal that substrate-imposed strains are an essential ingredient in stabilizing superconductivity in \ROA would be to perform electrical transport measurements for a thick \ROA film where the epitaxial strains are almost completely relaxed.  In reality, however, such efforts are complicated by the observation that the $+2.3\%$ tensile strain along $[1\overline{1}0]$ is not released in sufficiently thick films of \ROA by the nucleation of misfit dislocations; instead, \emph{cracks} form in such samples (in our studies, for $t \gtrapprox 30 \text{ nm}$) that propagate through the entire thickness of the film and even tens of nanometers into the substrate.  

In Fig.\,S11a,b, we show HAADF-STEM images of such oriented micro-cracks for the same 48 nm thick \ROA film characterized by ARPES and LEED in Figs.\,3 and 4 of the main text and in Figs.\,S14-S15.  Although these images evidence strong interfacial bonding between film and substrate---which is certainly important to maintain high levels of strain throughout the thinner \ROA films we characterize elsewhere in the manuscript---the strongly anisotropic nature of the cracks makes the distribution of current flow through such samples extremely non-uniform.  Accordingly, putative measurements of electrical ``resistance'' $R(T, H)$ displayed in Fig.\,S11 should be more pedantically interpreted as the voltage difference measured between two voltage contacts, divided by the total current sourced through two other contacts placed elsewhere on the film.  Unsurprisingly, the results thus obtained for $R$ depend essentially on the orientation of the voltage contacts relative to the cracks in the film: in Fig.\,S11c,d we observe either a downturn (green trace) or upturn (purple trace) in the apparent $R$ as the temperature is decreased below $T_{c} = 1.3 - 1.5 \text{ K}$, followed by plateaus at lower temperatures.  In Fig.\,S11e, we show that these temperature-induced anomalies in $R$ can be suppressed in both cases by applying small magnetic fields $H_{c\perp} < 3 \text{ kOe}$ at fixed $T = 0.45 \text{ K}$, confirming that they result from (inhomogeneous) superconductivity.  Irrespective of whether the measured $R$ decreases or increases at $(T_{c}, H_{c})$, we emphasize that the fractional changes in resistance induced by superconductivity in this 48 nm thick \ROA film are less than $1\%$ of the residual normal-state resistances $R_{\text{4 K}}$, in marked contrast to the full $100\%$ drops to zero resistance exhibited by all thinner, more highly strained, non-cracked \ROA films shown in other figures.

\section{Density functional theory calculations}

\subsection*{Structural relaxations}

One of the central themes of this work is the exploration of strain-induced changes to the electronic structure in epitaxial thin films of \RO subject to biaxial epitaxial strains imposed by differently oriented rutile TiO$_2$ substrates.  To model this situation computationally within the framework of density functional theory (DFT), we started by using the Vienna Ab Initio Software Package~\cite{kresse_efficient_1996, kresse_ultrasoft_1999} to perform full structural relaxations (of lattice parameters and internal coordinates) to minimize the DFT + $U$-computed total energy of \RO in the ideal tetragonal rutile crystal structure (space group $\#136$, $P4_2/mnm$). Structural relaxations employed the same exchange-correlation functional and calculational parameters as for the DFT + $U$ ($U = 2 \text{ eV}$) calculations described in the Methods section of the main text, and forces were converged to $< 1\text{ meV}/\ang$.

Throughout the main text and supplemental information, we refer to DFT results for this minimum energy structure as ``bulk \ROnospace''.  The actual lattice parameters for this structure, $(a_{\text{bulk}}=4.517 \text{ \ang}, c_{\text{bulk}} =3.130 \text{ \ang})$, overestimate the experimentally measured lattice parameters at 295 K for \RO single crystals of $(a=4.492 \text{ \ang}, c=3.106 \text{ \ang})$ by $< 1\%$, due to well-established deficiencies of the generalized gradient approximation.  With the former as the bulk reference structure, we then simulated biaxial epitaxial strains to (110)-oriented \TO substrates by performing constrained structural relaxations for \RO in which the in-plane lattice parameters $c = (1 - 0.047) \times c_{\text{bulk}}$ and $d_{1\overline{1}0} = (1 + 0.023) \times d_{1\overline{1}0\text{, bulk}}$ were held fixed, while the out-of-plane lattice constant $d_{110}$ and all other internal coordinates of the structure were allowed to relax so as to minimize the total energy.  The fixed compression and expansion of $c$ and $d_{1\overline{1}0}$, respectively, correspond to the experimentally measured lattice mismatches between \TO and \RO single crystals at 295 K~\cite{berlijn_itinerant_2017, burdett_structural-electronic_1987}.  

Within this scheme, DFT + $U$ predicts that commensurately strained \ROA thin films will have an out-of-plane lattice constant $d_{110} = (1 + 0.017) \times d_{110\text{, bulk}}$, which compares reasonably well with the 2.0\% expansion of $d_{110}$ measured experimentally on a 5.8 nm thick \ROA film.  Because the splitting of $d_{110}$ and $d_{1\overline{1}0}$ in strained \ROnospace(110) breaks the non-symmorphic glide plane symmetry of the parent rutile structure, we used a base-centered orthorhombic structure (space group $\#65$, $Cmmm$) with lattice constants of $c \times 2d_{1\overline{1}0} \times 2d_{110}$ for DFT simulations of \ROnospace(110).  The primitive unit cell of this $Cmmm$ structure contains the same number of atoms as the parent rutile unit cell, so there is no apparent doubling and/or folding of the bands in spaghetti plots that compare the bandstructures of \ROA and bulk \ROnospace, such as in Fig.\,S12 or Fig.\,4a of the main text.

To simulate the electronic structure of commensurately strained (101)-oriented \RO thin films, we adopted a slightly different approach, since it is not straightforward to perform constrained structural relaxations with DFT in such a low-symmetry situation.  Specifically, we took the rutile $b$ axis to be under 2.3\% tension, $b = (1 + 0.023) \times b_{\text{bulk}}$, as dictated by the lattice mismatch of \RO with \TO along this direction.  On the other hand, the lengths of the rutile $a$ and $c$ axes are free to adjust their lengths, but are subject to the simultaneous constraints:

\begin{equation}
\sqrt{a^2 + c^2} = \sqrt{\left(a_{\text{TiO2}}\right)^2 + \left(c_{\text{TiO2}}\right)^2} = 5.464 \text{ } \ang
\end{equation}

\begin{equation}
\left|\mathbf{q}_{202}\right| = \sqrt{\left(\frac{2}{a}\right)^2 + \left(\frac{2}{c}\right)^2} = 0.792 \text{ } \ang^{-1}
\end{equation}

\begin{equation}
\left|\mathbf{q}_{103}\right| = \sqrt{\left(\frac{1}{a}\right)^2 + \left(\frac{3}{c}\right)^2} = 0.998 \text{ } \ang^{-1}
\end{equation}

\begin{equation}
 \left|\mathbf{q}_{402}\right| = \sqrt{\left(\frac{4}{a}\right)^2 + \left(\frac{2}{c}\right)^2} = 1.114 \text{ } \ang^{-1}.
\end{equation}

\noindent Equation~(3) ensures that the film is lattice matched to the substrate along $[\overline{1}01]$, and equations (4) - (6) ensure that the $d$-spacings for the $HKL=202$, 103, and 402 Bragg reflections reproduce the values measured experimentally for strained \ROB films (a proper lattice constant refinement would, of course, include data for many more reflections).  Note that in deriving these equations, we assumed for simplicity that the angle between the rutile $a$ and $c$ axes remains $90^{\circ}$ in epitaxially strained films; small deviations away from this limit should be expected in reality, since this orthogonality is not guaranteed by any symmetry or constraint of the system.  Nonetheless, finding the best-fit solution to equations (3) - (6) gives lattice constants of $(a = 4.501 \text{ } \ang$, $c = 3.077 \text{ } \ang)$ in absolute units; dividing through by the experimentally measured lattice constants of bulk \RO yields $a = (1 + 0.002) \times a_{\text{bulk}}$ and $c = (1 - 0.009) \times c_{\text{bulk}}$ as appropriately scaled inputs for DFT simulations.  With $a \neq b \neq c$ and all angles between the primitive unit cell translations equal to $90^{\circ}$, the crystal structure for \ROB DFT simulations was taken as the primitive orthorhombic space group \#58, $Pnnm$.  Table 1 summarizes all parameters of the crystal structures used in DFT simulations for bulk \ROnospace, \ROAnospace, and \ROBnospace.
  
\begin{table*}[ht]
\caption{Crystal structures used in DFT simulations (all units of length are given in \ang)}% title of Table
\centering % used for centering table
\begin{tabular}{c c c c c c c c c}% centered columns (4 columns)
\hline\hline                        %inserts double horizontal lines
& & & & & & & Ru - apical O & Ru - equatorial O \\ 
Name & Space group & \;$a_{\text{rutile}}$ \; & \; $b_{\text{rutile}}$ \; & \; $c_{\text{rutile}}$ \; & \; $d_{1\overline{1}0}$ \; & \; $d_{110}$ \; & bond length(s) & bond length(s) \\ [1ex]
\hline               % inserts single horizontal line
bulk \RO - expt. & \#136 & 4.492 & 4.492 & 3.106 & 3.176 & 3.176 & 1.941 & 1.984  \\
bulk \RO - DFT & \#136 & 4.517 & 4.517 & 3.130 & 3.194 & 3.194 & 1.945 & 2.002  \\% inserting body of the table
\ROA - DFT & \#65 & 4.606 & 4.606 & 2.982 & 3.266 & 3.249 & 1.946, 1.957 & 1.980, 1.984 \\
\ROB - DFT & \#58 & 4.525 & 4.618 & 3.101 & 3.233 & 3.233 & 1.969 & 2.000 \\ [1ex]      % [1ex] adds vertical space
\hline%inserts single line
\end{tabular}
\label{table:DFT_cryst_structs}% is used to refer this table in the text
\end{table*}

\subsection*{Effects of adding $+U$}

\begin{figure*}
\begin{center}
\includegraphics[width=7in]{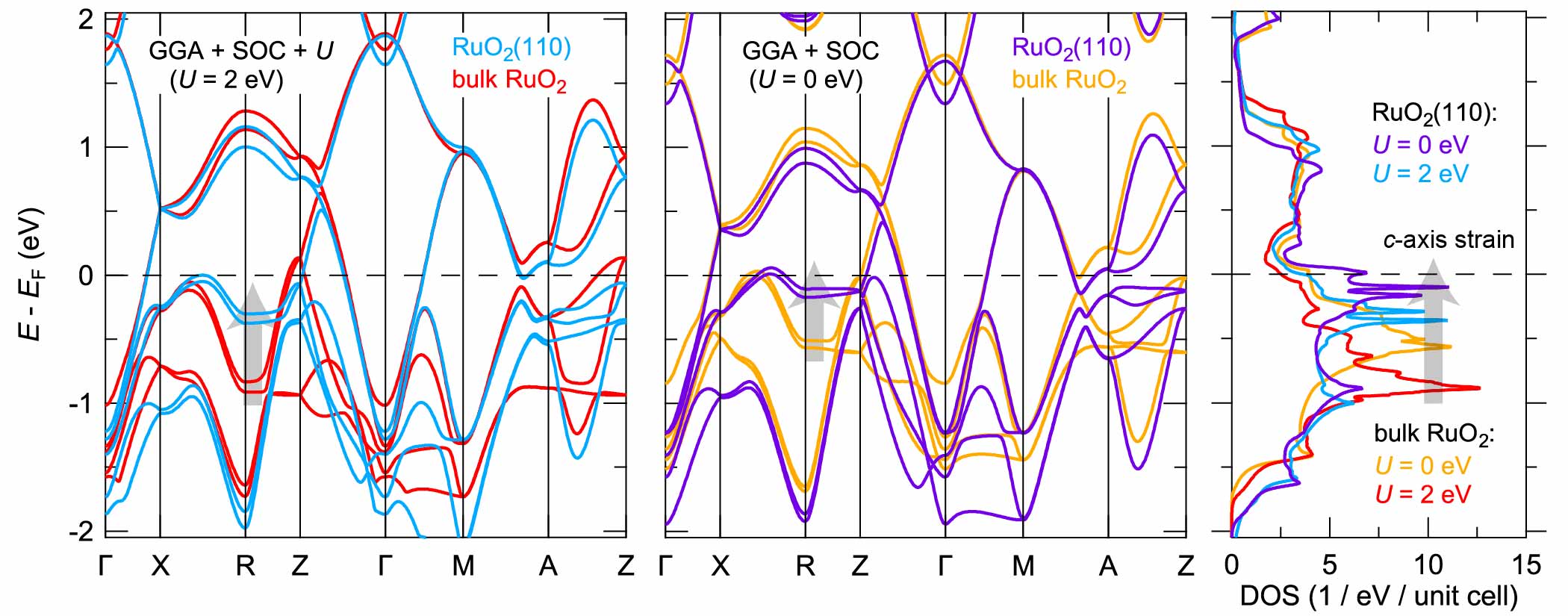}
\end{center}
\caption{\textbf{Strain dependence of the electronic structure of \ROnospace, according to DFT (+ $U$).}}
\end{figure*}

In Fig.\,S12, we show the effects of including an \textit{ad hoc} static mean-field $+U$ term on the Ru sites in DFT calculations. Adding such a phenomenological term to the Kohn-Sham Hamiltonian shifts the bands relative to each other (up/down in energy) so as to force the orbital occupancies towards integer fillings, rather than also shrinking the bandwidths, as would occur in a more realistic theory that includes dynamical electron-electron interactions.  The blue and red traces are reproduced from the non-magnetic DFT + $U$ results presented in Fig.\,4a of the main text, where we noted that the most prominent effect of the epitaxial strains present in the \ROA simulations is to push the near-$E_F$ \dparnospace-derived flat bands (concentrated around the $k_{001} = \pi / c$ plane) closer to $E_F$, which enhances the density of states (DOS) near $E_F$.  The purple and orange traces are the results of repeating GGA + SOC calculations for the same \ROA and bulk \RO crystal structures, but now setting $U = 0$.  The same shift of the \dpar flat bands towards $E_F$ observed in the $U = 2\text{ eV}$ calculations and concomitant enhancement of the DOS near $E_F$ are observed here when the $c$-axis compressive strain is increased upon going from bulk \RO to \ROAnospace.  While these strain-dependent \textit{trends} in the electronic structure are robust against details of the calculations, Fig.\,S12 also suggests that the calculated energy positions of the peaks in the DOS and the exact value of the DOS near $E_F$ should not be taken too seriously, as there are considerable theoretical uncertainties in these quantities.  We leave a more complete treatment of the effects of $\vec{Q}_{\text{AFM}} = (1\,0\,0)$ antiferromagnetic spin-density wave order~\cite{berlijn_itinerant_2017, zhu_anomalous_2019} on the electronic structure to future studies~\cite{ahn_antiferromagnetism_2019}, because it is not possible in purely DFT-based approaches to stabilize self-consistent solutions with the small values of the ordered moment that are measured experimentally~\cite{berlijn_itinerant_2017}.

\section{Determination of out-of-plane momenta probed by angle resolved photoemission spectroscopy}

\begin{figure*}
\begin{center}
\includegraphics[width=7in]{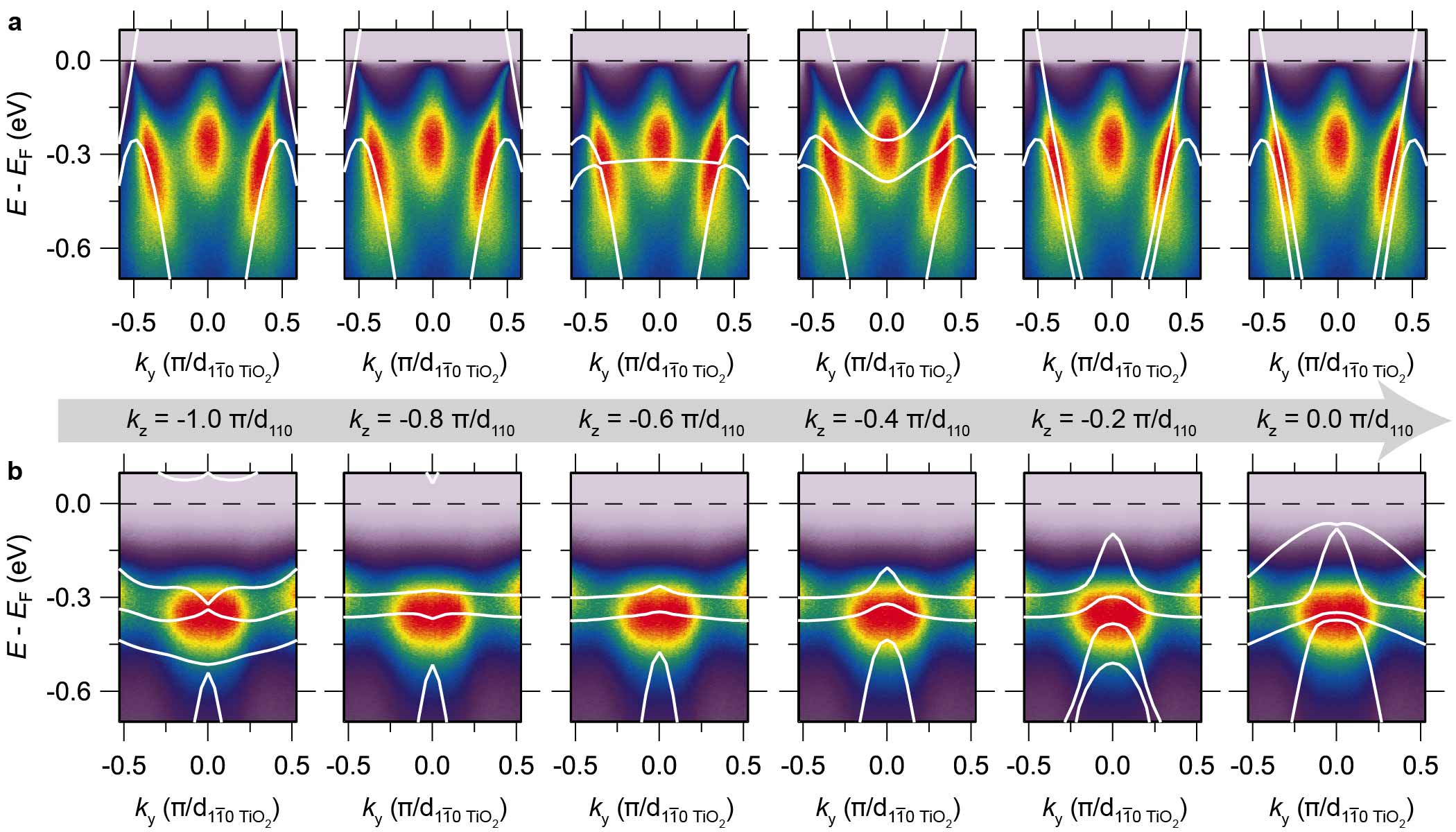}
\end{center}
\caption{\textbf{Determination of $k_z$ probed by ARPES for \ROAnospace.}  Experimental ARPES spectra for a 7 nm thick \RO / \TOnospace(110) sample reproduced from \textbf{a,} Fig.\,3d of the main text and \textbf{b,} Fig.\,3c of the main text, overlaid with the band dispersions from DFT + $U$ ($U = 2\text{ eV}$) calculations in white.  For the data in \textbf{a,} $k_x = k_{001}$ is fixed at zero and in \textbf{b,} $k_x = \pi/c$.  Moving from left to right, $k_z$ is incremented in the DFT simulations in steps of 0.2 $\pi/d_{110}$ starting from -1.0 $\pi/d_{110}$.}
\end{figure*}

Figure 3 of the main text compares the electronic structure of a 7 nm thick \RO / \TOnospace(110) sample experimentally measured by angle-resolved photoemission spectroscopy (ARPES) with the results of DFT + $U$ simulations.  To make this comparison, it is necessary to determine what range of out-of-plane momenta $k_z = k_{110}$ in the initial state are probed by ARPES at a given final-state kinetic energy and momentum.  This was established by plotting the DFT-computed $E(\vec{k})$ dispersions on top of the experimentally measured spectra along several one-dimensional cuts through momentum space measured over a small range of kinetic energies corresponding to near-$E_F$ states at the given photon energy (21.2 eV), and allowing $k_z$ to vary in the calculations so as to best match the experimental data.  

Figure S13 shows representative examples of this procedure for experimental spectra taken along the one-dimensional cuts shown in Fig.\,3d (top row, $k_x = k_{001}$ fixed at zero) and in Fig.\,3c (bottom row, $k_x = \pi/c$) of the main text.  Moving from left to right, $k_z$ is incremented in the DFT simulations in steps of 0.2 $\pi/d_{110}$ starting from -1.0 $\pi/d_{110}$.  For the panels in Fig.\,S13a, the $k_F$s and electron-like character of the band crossing $E_F$ are best fitted by calculations with $k_z$ in the range $-0.2 \rightarrow 0.0 \text{ } \pi/d_{110}$.  Likewise, for the panels in Fig.\,S13b, $k_z$ values in the range $-0.6 \rightarrow -0.3 \text{ } \pi/d_{110}$ best reproduce the measured spectra, although the results here are more ambiguous because of the insensitivity of the flat-band energies to the precise value of $k_{110}$.  Therefore, we took the range of reduced initial-state out-of-plane momenta probed at normal emission ($k_x = k_y = 0$) to be $k_{z,i} = -0.1 \pm 0.1 \text{ } \pi / d_{110}$.  Assuming a free-electron-like model of final states, the final-state $k_{z,f}$ is given by the expression

\begin{equation}
k_{z,f} = \sqrt{\frac{2 m_e (E_k\cos^2\theta + V_0) }{\hbar^2}} = \frac{2\pi}{2d_{110}}N + k_{z,i}\;\;,
\end{equation}

\noindent where $m_e$ is the free electron mass, $E_k$ is the kinetic energy of the photoelectrons, $\theta$ is the emission angle relative to the surface normal, $V_0$ is the inner potential, and $2d_{110}$ is the spacing between equivalent lattice points along the out-of-plane direction ($N$ can adopt any integer value).  Substituting $E_k = 16.6 \pm 0.3 \text{ eV}$, $\theta = 0^{\circ}$, $k_{z,i} = -0.1 \pm 0.1 \text{ } \pi / d_{110}$, and $d_{110} = 3.23 \text{ \normalfont\AA}$ into equation (7), we find that an inner potential of $13.7 \, \pm \, 2.3 \text{ eV}$ is compatible with our determination of $k_{z,i}$.  Taking this same value of $V_0$ and setting $\theta = 30 - 35^{\circ}$ in equation (7)---as is appropriate for the experimental data in the panels displayed in Fig.\,S13b---yields $k_{z,i} = -0.35 \pm 0.17 \text{ } \pi/d_{110}$; visual inspection of the DFT bands for this range of $k_{z,i}$ show that the calculations also reasonably reproduce the experimental spectrum in this region of the Brillouin zone.  The curved green planes drawn in the Brillouin zone schematic in Fig.\,3 of the main text are constructed by evaluating equation (7) with $V_0 = 13.7 \text{ eV}$ and $N = 3$ for all $(k_x, k_y)$, and accounting for an intrinsic uncertainty of $\approx 0.2 \text{ } \pi/d_{110}$ in $k_{z}$ owing to the finite elastic escape depth of photoelectrons, which we take to be $\approx 5 \text{ \normalfont\AA}$.

\section{Surface lattice constant refinement}

\begin{figure*}
\begin{center}
\includegraphics[width=7in]{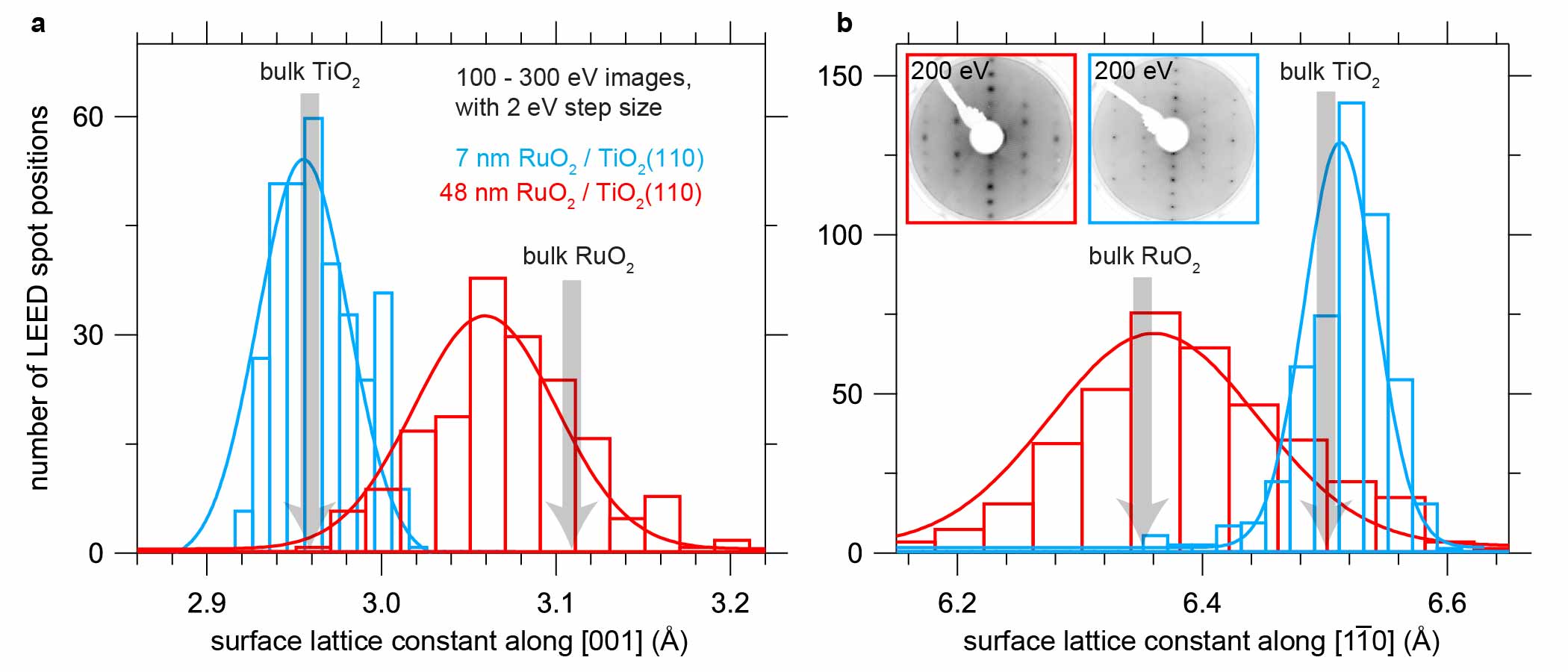}
\end{center}
\caption{\textbf{Surface lattice constant refinement by LEED spot position analysis.}  Results are shown for two \ROA films of different thicknesses, 7 nm (blue) and 48 nm (red).  The histograms in panels \textbf{a} and \textbf{b} summarize the measured values for $c$ and $2d_{1\overline{1}0}$, respectively.  For reference, the gray arrows indicate the values expected for a commensurately strained film (bulk \TOnospace) and a fully strain relaxed film (bulk \ROnospace).}
\end{figure*}

In Fig.\,S14, we present the results of a surface lattice constant refinement for two different \RO / \TOnospace(110) films of different thicknesses, 7 nm and 48 nm.  For both samples we acquired many low-energy electron diffraction (LEED) images at normal incidence using incident energies ranging from 100 - 300 eV in 2 eV steps; for examples of the raw data, the insets contain representative images taken at 200 eV.  For each image, we located the positions of all visible spots and indexed the spots according to their in-plane momentum transfer values $\mathbf{q_{||}} = 2\pi(H/c, K/2d_{1\overline{1}0})$, where $H$ and $K$ are integers and, by our convention, $H$ defines the magnitude of $\mathbf{q_{||}}$ along [001] (nearly horizontal in the images in Fig.\,S14), and $K$ defines the magnitude of $\mathbf{q_{||}}$ along $[1\overline{1}0]$ (nearly vertical).  We then calculated the distance of all spots from the specular $\mathbf{q_{||}} = (0, 0)$ reflection and converted these image distances $D$ (in pixel space) to scattering angles $\sin(\theta)$ (where $\theta$ is the angle of each diffracted electron beam relative to the surface normal) based on $D \rightarrow \sin(\theta)$  calibrations that were independently determined from reference measurements on SrTiO$_3$(001) surfaces having a known lattice constant.  Note that these calibrations absorb the overall scaling factor that depends on the working distance between the LEED screen and the sample (and the camera image magnification factor), as well as some higher-order distortions of the spot patterns that result from the sample not being positioned precisely at the center of curvature of the LEED screen (and the screen itself being slightly aspherical).

From these values of $\sin(\theta)$, the electron energies $E$ at which each LEED pattern was recorded, and the $(H, K)$ indices, we compiled lists of lattice constants corresponding to each fitted spot position.  For simplicity in analysis, we restricted our attention to spots having $\mathbf{q_{||}}$ purely aligned with [001] or [$1\overline{1}0$].  Elastic scattering and conservation of momentum modulo translations of the surface reciprocal lattice require that:   

\begin{equation}
|\mathbf{q_{||}}| = k \sin(\theta) = \sqrt{2 m_e E}\sin(\theta)/\hbar \;\;,
\end{equation}

\noindent which for Bragg spots of the type $\mathbf{q_{||}} = 2\pi(H/c, 0)$ and $2\pi(0, K/2d_{1\overline{1}0})$, reduces to:

\begin{equation}
c = \frac{2 \pi \hbar H}{\sqrt{2 m_e E}\sin(\theta)} \;\; \text{ and } \;\; 2d_{1\overline{1}0} = \frac{2 \pi \hbar K}{\sqrt{2 m_e E}\sin(\theta)}\;\; .
\end{equation}

\noindent Here $c$ and $2d_{1\overline{1}0}$ are the surface lattice constants along $[001]$ and $[1\overline{1}0]$ that are expected for an unreconstructed (110)-oriented rutile surface.  Histograms of the values obtained in this way for $c$ are displayed in Fig.\,S14a, and the results for $2d_{1\overline{1}0}$ are displayed in Fig.\,S14b.  For reference we also indicate by gray arrows the surface lattice constants expected for bulk-terminated \TO $(c=2.96 \text{ \ang}, 2d_{1\overline{1}0}= 6.50 \text{ \ang})$ and bulk-terminated \RO $(c=3.11 \text{ \ang}, 2d_{1\overline{1}0}= 6.35 \text{ \ang})$.  The surface lattice constants for the 7 nm thick \ROA sample in blue, $(c= 2.96 \pm 0.03 \text{ \ang}, 2d_{1\overline{1}0}= 6.51 \pm 0.05 \text{ \ang})$, match those of \TO within the $\approx 1.0\%$ resolution of the measurements, indicating that this film is (nearly) commensurately strained to the substrate along both $[001]$ and $[1\overline{1}0]$.  By contrast, the 48 nm thick \ROA sample in red shows broader LEED spots, indicating lower surface crystallinity than the 7 nm thick sample, which results in wider distributions of the extracted lattice constants.  Furthermore, the centers of mass of the red distributions, $(c= 3.07 \pm 0.06 \text{ \ang}, 2d_{1\overline{1}0}= 6.39 \pm 0.11 \text{ \ang})$, are displaced away from the blue distributions towards the values expected for bulk \ROnospace, which is why in the main text we suggest that the surface electronic structure of this sample measured by ARPES---which probes the film over a comparable depth to LEED, within $< 1 \text{ nm}$ from the surface---should be more representative of ``bulk'' \ROnospace.

\section{Extracting the near-$E_F$ density of states from ARPES measurements}

\begin{figure*}
\begin{center}
\includegraphics[width=7in]{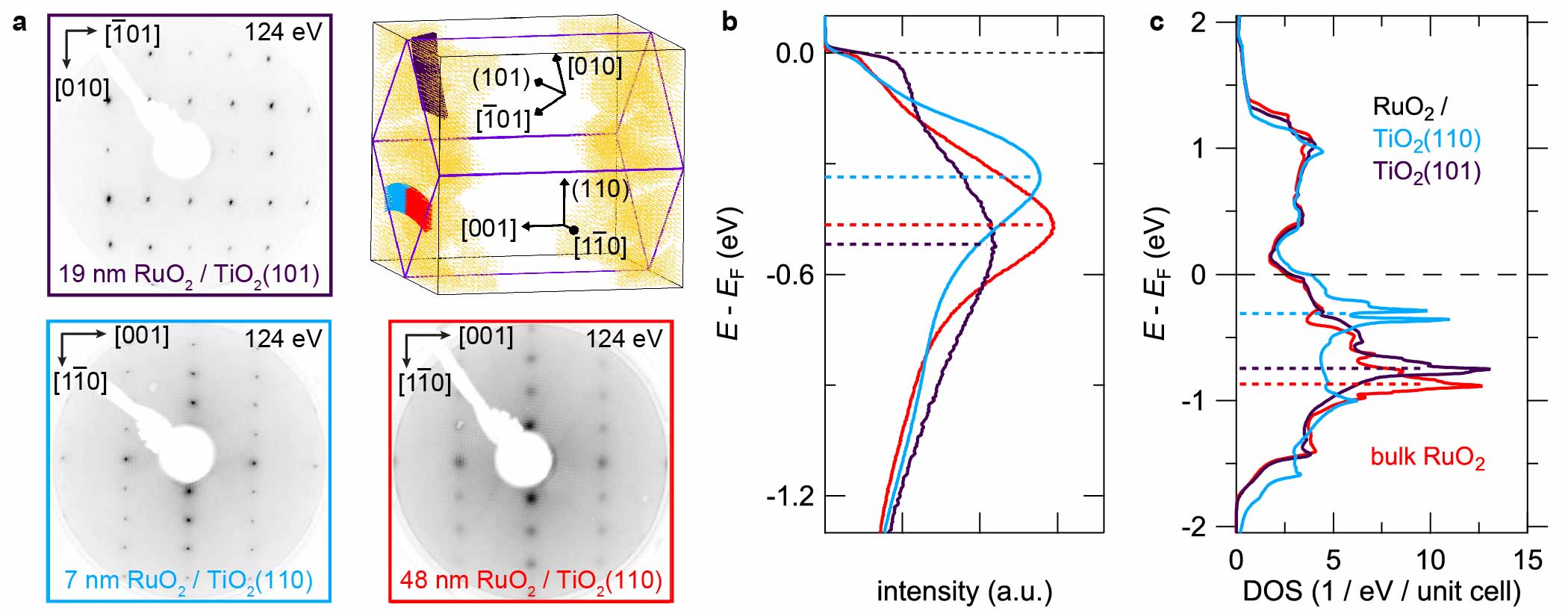}
\end{center}
\caption{\textbf{Strain dependence of the near-$E_F$ density of states.} \textbf{a,} LEED patterns recorded at 124 eV for a 19 nm thick \ROBnospace, a 7 nm thick \ROAnospace, and a 48 nm thick \ROA sample, along with a schematic of the Brillouin zone showing the regions of $k$-space over which we integrated the ARPES data for these samples to obtain the average energy distribution curves in \textbf{b}.  Yellow regions of the Brillouin zone indicate where the near-$E_F$ Kohn-Sham wavefunctions have $> 90\%$ \dpar orbital character.  \textbf{b,} The colored dashed lines represent the peak positions of the energy distribution curves.  By locating the \dparnospace-derived flat bands in this way, trends in how the DOS($E_F$) evolves with strain can be more reliably extracted from ARPES data than by directly reading off the measured photoemission intensity at $E_F$.  \textbf{c,} Expected strain-dependent changes in the total DOS according to DFT + $U$ simulations, reproduced from Fig.\,4a of the main text.}
\end{figure*}

In Fig.\,S15 we present LEED and ARPES data taken on three different \RO thin-film samples: 19 nm thick \ROBnospace, 7 nm thick \ROAnospace, and 48 nm thick \ROAnospace.  In Fig.\,S15a, all of the samples show LEED spot patterns with the periodicities expected for unreconstructed (101)- and (110)-oriented rutile surfaces, respectively; furthermore, the sharpness of the patterns suggest high degrees of surface crystallinity, such that in-plane momentum should be a nearly conserved parameter in photoemission.  Given that strained \ROA samples superconduct at measurable \TCnospace s, while \ROB samples and bulk \RO do not, the question we wanted to address using ARPES was: how does the density of states (DOS) near the Fermi level ($E_F$) evolve between these samples?  Recall that based on the LEED lattice constant analysis described in Fig.\,S14, most of the epitaxial strains are relaxed at the surface of the 48 nm thick \ROA sample, such that its electronic structure is a reasonable proxy for that of bulk \ROnospace.  Specifically, the quantity of interest as it relates to the low-energy physics is:

\begin{equation}
\text{DOS}(E_F) = \int_{E_F - \delta}^{E_F + \delta} \int_{\text{BZ}} A(\vec{k}, \omega) d\vec{k} d\omega \;\;,
\end{equation}

\noindent where $A(\vec{k},\omega)$ is the single-particle spectral function, integrated over all momenta $\vec{k}$ in the Brillouin zone (BZ) and over some limited range of energies $\omega$ near $E_F$ ($\delta$ is some small parameter).

Two separate factors make it extremely challenging to quantitatively extract the total DOS($E_F$) directly from data taken with our lab-based ARPES equipment.  First, our inability to continuously vary the photon energy---or equivalently, the kinetic energy of the photoelectrons at $E_F$---implies that only regions of the Brillouin zone with specific $k_z$ can be probed, \textit{cf.} equation~(7).  Therefore the full integration over $\vec{k}$ in equation~(10) cannot be performed in a lab-based ARPES setup, which is especially problematic in a material such as \RO that has a highly three-dimensional electronic structure depending strongly on $k_z$ (\textit{cf.} Fig.\,S13).  Second, even if the entire Brillouin zone could be mapped exhaustively, the intensity measured in ARPES is not the initial-state spectral function $A(\vec{k}, \omega)$, but rather this quantity multiplied by probabilities (matrix elements) for photoemission, which are difficult to calculate theoretically.

With these qualifications in mind, there is a route to answering the simpler question of whether the DOS($E_F$) \textit{increases} in strained \ROA compared with strained \ROB or bulk \ROnospace:  we simply must determine where the flat bands with \dpar orbital character are located in energy relative to $E_F$.  DFT calculations suggest that if these bands move closer to (further away from) $E_F$, the total DOS($E_F$) will increase (decrease).  To approximately determine the positions of these bands experimentally, we integrated the photoemission intensity over the color-coded slabs in the Brillouin zone schematic in Fig.\,S15a, plotted the resulting energy distribution curves (EDCs) in Fig.\,S15b, and found the maxima in the EDCs as indicated by the dashed lines.  The regions colored yellow in the Brillouin zone denote where the near-$E_F$ wavefunctions have greater than 90\% \dpar orbital character, according to our DFT + Wannier90 calculations; since all slabs lie in this region, we expect that the dominant contributions to the measured EDCs are from \dpar initial states.  Note that the region of $k_{z}=k_{110}$ probed by ARPES with He-I$\alpha$ (21.2 eV) light on the (110)-oriented samples is well-constrained by analysis of the $E(\vec{k})$ dispersions as outlined in Fig.\,S13; however, for the (101)-oriented sample the region of $k_{z}=k_{101}$ probed by ARPES with He-II$\alpha$ (40.8 eV) light is merely calculated from the free-electron final state model in equation (7), using the same inner potential as for \ROAnospace, and thus is subject to greater experimental uncertainties.  Nonetheless, the results of this anlaysis are in qualitative agreement with the strain-dependent trends anticipated by DFT shown in Fig.\,S15c:  in highly strained \ROA films, the flat bands move closer to $E_F$ compared with either more strain-relaxed \ROA films or commensurately strained \ROB films.  This modification of the effective $d$ orbital degeneracies boosts the total DOS near $E_F$, which---as argued in the main text---likely contributes to the enhanced superconducting \TCnospace s observed in highly strained \ROAnospace. 

\section{Migdal-Eliashberg calculations of \TCnospace}

\begin{figure*}
\begin{center}
\includegraphics[width=7in]{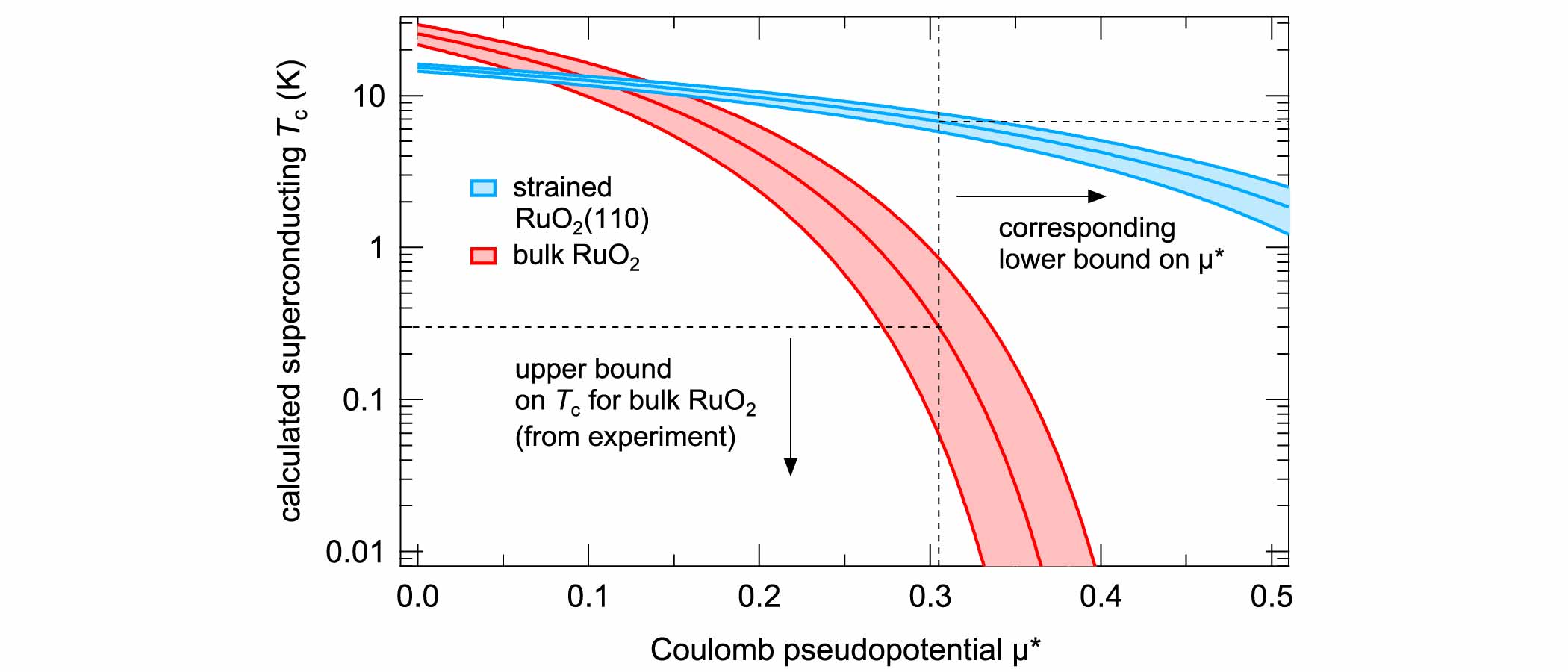}
\end{center}
\caption{\textbf{Calculated \TCnospace s for bulk \RO and strained \ROAnospace, assuming an electron-phonon mediated mechanism of superconductivity.} \TC is evaluated from the McMillan-Allen-Dynes formula using DFT-based calculations of $\lambda_{\text{el-ph}}$ and $\omega_{\log}$, and allowing $\mu^{*}$ to vary.  As indicated by the vertical and horizontal dashed lines, $\mu^{*} > 0.30$ is the region of parameter space consistent with the reported non-superconducting behavior of bulk \RO \cite{lin_low_2004}, which in turn places an upper bound on the strain-enhanced \TCnospace s expected for \ROAnospace.  Shaded regions indicate the changes in \TC that would result from $\pm10\%$ errors in the calculated $\lambda_{\text{el-ph}}$.}
\end{figure*}

As described in the Methods section of the main text, we performed first-principles DFT-based electron-phonon coupling calculations of the isotropic Eliashberg spectral function $\alpha^2 F(\omega)$ and total electron-phonon coupling constant $\lambda_{\text{el-ph}}$ (integrated over all phonon modes and wavevectors) for bulk \RO and commensurately strained \ROAnospace.  From these quantities, we estimated the superconducting transition temperature using the semi-empirical McMillan-Allen-Dynes formula:

\begin{equation}
T_c = \frac{\omega_{\log}}{1.2} \exp\left[ - \frac{1.04 (1 + \lambda_{\text{el-ph}} ) } { \lambda_{\text{el-ph}} - \mu^{*} (1 + 0.62 \lambda_{\text{el-ph}} ) } \right]  
\end{equation}

McMillan first obtained a formula resembling equation (11) by numerically solving the equations of finite-temperature Migdal-Eliashberg theory using the experimentally measured spectral function of niobium~\cite{mcmillan_transition_1968}; Allen and Dynes improved the agreement of McMillan's formula with experimentally measured $\alpha^2 F(\omega)$ and \TCnospace s for a variety of conventional superconductors by introducing an appropriately weighted average over $\alpha^2 F(\omega)$ in the prefactor of the exponential, rather than using the Debye temperature~\cite{allen_transition_1975}.  In essence, equation~(11) can be considered an extension of equation~(1) from the main text that identifies phonons as the bosonic modes which mediate Cooper pairing (\textit{i.e.}, $\lambda = \lambda_{\text{el-ph}}$), and which remains valid even in the limit of stronger couplings (\textit{i.e.}, $\lambda > 1$) by virtue of Migdal's theorem for the electron-phonon interaction.

For bulk \RO (strained \ROAnospace, respectively), we obtained $\lambda_{\text{el-ph}} = 0.685$ and $\omega_{\log} = 34.1 \text{ meV}$ ($\lambda_{\text{el-ph}} = 1.97$ and $\omega_{\log} = 7.59 \text{ meV}$).  The large strain-induced enhancement in $\lambda_{\text{el-ph}}$ and shift of $\omega_{\log}$ to lower frequencies is caused by substantial phonon softening that occurs under $c$-axis compression in the rutile structure.  In fact, we found that some of the calculated zone-boundary phonon frequencies even become imaginary under strain in \ROAnospace, possibly indicating an incipient structural instability.  The full details of this phenomenon will be described in a future publication; for the purposes of this work, we omitted such phonon modes in subsequent electron-phonon coupling calculations, and neglected any errors this may cause in $\lambda_{\text{el-ph}}$ and $\omega_{\log}$. \\

To convert the calculated values of $\lambda_{\text{el-ph}}$ and $\omega_{\log}$ to \TCnospace s via equation~(11) requires knowledge of the appropriate value(s) of the screened Coulomb interaction $\mu^{*}$ between quasiparticles.  Typically $\mu^{*}$ is chosen in an \emph{ad hoc} fashion to match the experimentally measured \TCnospace.  Because bulk \RO is not known to be superconducting at experimentally accessible temperatures, we cannot employ such a prescription here; nevertheless, we can use the experimentally measured \emph{least upper bound} on $T_{c} < 0.3\text{ K}$ to place a \emph{lower bound} on $\mu^{*} > 0.30$, as illustrated in Fig.\,S16.  For this range of $\mu^{*}$, inserting the values of $\lambda_{\text{el-ph}}$ and $\omega_{\log}$ calculated for \ROA into equation~(11) predicts $T_{c} < 7\text{ K}$, which agrees reasonably well with the experimentally measured values.  Because several uncontrolled approximations enter into these estimates of \TCnospace, we consider this level of agreement as suggestive, although not conclusive, evidence for a phonon-mediated mechanism of superconductivity; in any case, it is clear that reducing the axial ratio $c/a$ in strained \RO robustly boosts $\lambda_{\text{el-ph}}$, in excellent agreement with expectations based on steric trends for other rutile compounds~\cite{Eyert_2002, hiroi_structural_2015}.  Any effects of strain on $\mu^*$ are ignored for the purposes of this work.

\newpage